\documentclass{aa}  

\usepackage{graphicx}
\usepackage[utf8]{inputenc}
\usepackage{amssymb}
\usepackage{amsmath}
\usepackage{graphicx} 
\usepackage{enumitem}
\usepackage{url}
\usepackage{multirow}
\setlist{nosep} %
\usepackage{sectsty}
\usepackage[colorlinks=true,linkcolor=blue,citecolor=blue]{hyperref}
\usepackage{ulem}

\sectionfont{\fontsize{12}{12}\selectfont}
\subsectionfont{\fontsize{11}{11}\selectfont}

\usepackage{txfonts}
\newcommand{\simpropto}{\mathrel{\vcenter{
  \offinterlineskip\halign{\hfil$##$\cr
    \propto\cr\noalign{\kern2pt}\sim\cr\noalign{\kern-2pt}}}}}

\newcommand{\lpop}{\vec\lambda_\mathrm{pop}}

\newcommand{\lsource}{\vec\lambda_\mathrm{src}}

\newcommand{\thv}{\theta_\mathrm{v}}
\newcommand{\thvi}{\theta_{\mathrm{v},i}}
\newcommand{\Ep}{E_\mathrm{p}}
\newcommand{\Epi}{E_{\mathrm{p},i}}

\defcitealias{Mandel2019}{M19}
\defcitealias{Meegan2009}{M09}
\defcitealias{vonKienlin2020}{vK20}
\defcitealias{Ghirlanda2016}{G16}
\defcitealias{Wanderman2015}{W15}

\begin{document}

   \title{The short gamma-ray burst population in a quasi-universal jet scenario}
   \titlerunning{SGRB population in a quasi-universal jet scenario}

   \author{{Om Sharan} Salafia\inst{1,2}, {Maria Edvige} Ravasio\inst{3,1}, Giancarlo Ghirlanda\inst{1,2} and Ilya Mandel\inst{4,5}}

   \institute{INAF -- Osservatorio Astronomico di Brera, via Emilio Bianchi 46, I-23807 Merate (LC), Italy
        \and
             INFN -- sezione di Milano-Bicocca, Piazza della Scienza 3, I-20126 Milano (MI), Italy
        \and
            Department of Astrophysics/IMAPP, Radboud University, 6525
            AJ Nijmegen, The Netherlands
        \and School of Physics and Astronomy, Monash University, Clayton, Victoria 3800, Australia
        \and ARC Center of Excellence for Gravitational Wave Discovery -- OzGrav
             }

    \authorrunning{O.~S.\ Salafia, M.~E.\ Ravasio, G.\ Ghirlanda \& I.\ Mandel}
  
   \date{Received xxx; accepted xxx}

 
 \abstract{We describe in detail a model of the short gamma-ray burst (SGRB) population under a `quasi-universal jet' scenario in which jets can differ somewhat in their on-axis peak prompt emission luminosity $L_\mathrm{c}$, but share a universal angular luminosity profile $\ell(\theta_\mathrm{v})=L(\theta_\mathrm{v})/L_\mathrm{c}$ as a function of the viewing angle $\theta_\mathrm{v}$. The model is fitted, through a Bayesian hierarchical approach inspired by gravitational wave (GW) population analyses, to three observed SGRB samples simultaneously: the \textit{Fermi}/GBM sample of SGRBs with spectral information available in the catalogue (367 events); a flux-complete sample of 16 \textit{Swift}/BAT SGRBs that are also detected by GBM, with a measured redshift; and a sample of SGRBs with a binary neutron star (BNS) merger counterpart, which only includes GRB~170817A at present.  Particular care is put in modelling selection effects. The resulting model, which reproduces the observations, favours a narrow jet `core' with half-opening angle $\theta_\mathrm{c}= 2.1_{-1.4}^{+2.4}\,\mathrm{deg}$ (uncertainties hereon refer to 90\% credible intervals from our fiducial `full sample' analysis) whose peak luminosity, as seen on-axis, is distributed as a power law $p(L_\mathrm{c}) \propto L_\mathrm{c}^{-A}$ with $A=3.2_{-0.4}^{+0.7}$ above a minimum isotropic-equivalent luminosity $L_\mathrm{c}^\star = 5_{-2}^{+11}\times 10^{51}\,\mathrm{erg\,s^{-1}}$. For viewing angles larger than $\theta_\mathrm{c}$, the luminosity profile scales as a single power law $\ell\propto \theta_\mathrm{v}^{-\alpha_L}$ with $\alpha_L=4.7_{-1.4}^{+1.2}$, with no evidence for a break, despite the model allowing for it. While the model implies an intrinsic `Yonetoku' correlation between $L$ and the peak photon energy $E_\mathrm{p}$ of the spectral energy distribution (SED), its slope is somewhat shallower $E_\mathrm{p}\propto L^{0.4\pm 0.2}$ than the apparent one, and the normalization is offset towards larger $E_\mathrm{p}$, due to selection effects. The implied local rate density of SGRBs (regardless of the viewing angle) is between about one hundred up to several thousands of events per cubic Gpc per year, in line with the binary neutron star (BNS) merger rate density inferred from GW observations. Based on the model, we predict 0.2 to 1.3 joint GW+SGRB detections per year by the Advanced GW detector network and \textit{Fermi}/GBM during the O4 observing run. }


   \maketitle
%

\section{Introduction}


Two main clusters of gamma-ray bursts (GRBs) have long been identified in the two-dimensional plane of duration \textit{versus} hardness ratio\footnote{In X-ray and $\gamma$-ray astronomy, detectors typically identify `events' (interactions between photons and the active part of the detector) in different energy channels. The hardness ratio is generally defined as the ratio of the event counts in a higher-energy (`harder') channel (or group of channels) to that in a lower-energy (`softer') channel.} of the large sample collected by the Burst Alert and Transient Source Experiment (BATSE) onboard the Compton Gamma-Ray Observatory \citep{Kouveliotou1993}. The bimodality in the BATSE GRBs is apparent also when considering the durations only (more precisely, the time $T_{90}$ over which 5\% to 95\% of the background-subtracted counts are collected), with the histogram of the logarithms of the durations featuring two peaks separated by a valley at around $T_{90}=2\,\mathrm{s}$. This has since become the customary separation between the `long' (LGRB) and the `short' (SGRB) events in the GRB class, although the actual position of the valley varies somewhat across different detectors \citep[e.g.][]{Bromberg2013}. The evidence accumulated during the following decades painted a picture of two different progenitor systems: starting with GRB~980425 \citep{Galama1998}, a number of LGRBs have been firmly associated with type Ib/c core-collapse supernovae \citep[e.g.][]{Bloom2002,Malesani2004,Mirabal2006,Kann2011,Cano2014,DElia2018,Hu2021}, securing the scenario of a massive star progenitor \citep{Woosley1993}; the progenitors of SGRBs remained elusive for a longer time, while all hints consistently pointed \citep[e.g.][]{Nakar2007,Fong2013b,Berger2014,DAvanzo2015} to a compact binary merger progenitor \citep{Eichler1989,Mochkovitch1993}. Such a progenitor has been confirmed by the astounding association \citep{Abbott2017_GW170817_and_GRB170817A,Abbott2017multimessenger} of the short GRB~170817A \citep{Goldstein2017} -- detected by the Gamma-ray Burts Monitor (GBM, \citealt{Meegan2009}) onboard the \textit{Fermi} spacecraft and by \textit{INTEGRAL} \citep{Savchenko2017} -- with the first-ever binary neutron star (BNS) merger detected by humankind, GW170817 \citep{Abbott2019_GW170817_properties}, whose gravitational wave (GW) signal was captured on the 17$^\mathrm{th}$ of August 2017 by the Advanced Laser Interferometer Gravitational wave Observatory (aLIGO, \citealt{Aasi2015}) and localized also thanks to Advanced Virgo \citep{Acernese2015}.

The properties of SGRBs -- especially their shorter duration and harder spectrum with respect to LGRBs \citep{Ghirlanda2009,Calderone2015} -- make for arduous detection with current facilities: of more than 280 GRBs revealed by \textit{Fermi}/GBM every year, only about 40 are SGRBs. The softer sensitivity band of the Burst Alert Telescope (BAT) onboard the Neil Gehrels \textit{Swift} Observatory \citep{Gehrels2004}, together with its smaller field of view, allows it to identify and localize only a handful of SGRBs per year. Moreover, even when localized by \textit{Swift}, the fraction of SGRBs that end up with a secure redshift determination is relatively low, due to a combination of a fainter X-ray `afterglow' \citep{Costa1997} -- whose detection is a requirement for precise localization -- and a larger typical offset from the host galaxy \citep{Fong2013a,Fong2013b,DAvanzo2014,Berger2014,Fong2022} with respect to LGRBs, which renders the host galaxy identification ambiguous in cases where multiple galaxies stand at similar offsets from the position of the afterglow. As a result, the intrinsic properties of the SGRB population are much more uncertain than for LGRBs. Indeed, a variety of attempts at constraining the properties of the SGRB population throughout the years, in some cases with very different methodologies and reference samples, has yielded varied results (e.g.\ \citealt{Schmidt2001,Guetta2005,Guetta2006,Virgili2011,Yonetoku2014,Wanderman2015,Shahmoradi2015,Ghirlanda2016,Zhang2018,Paul2018,Tan2020}). Some of these results are in clear tension with each other: in this work, we consider \citealt{Wanderman2015} (hereafter \citetalias{Wanderman2015}) and \citealt{Ghirlanda2016} (hereafter \citetalias{Ghirlanda2016}), which reach very different conclusions, as our benchmarks.

Among the challenges along the path towards unveiling the intrinsic properties of SGRBs, a prominent one is the uncertainty about which processes shape the luminosity function, that is, the probability distribution from which the luminosity of each event in the population is sampled. \citet{Lipunov2001} and \citet{Rossi2002} were the first to realize that the highly relativistic nature of GRB jets would make their angular structure an important factor in determining the luminosity function, in addition to the intrinsic spread in luminosities. Indeed, if the energy density and the typical Lorentz factor of a GRB jet are functions of the angular separation from the jet axis, then the apparent energetics are viewing-angle-dependent by virtue of relativistic aberration effects \citep[e.g.][]{Salafia2015}. Since jets are isotropically oriented in space, this naturally produces a large spread in the apparent luminosities, with a well-defined dependence on the angular structure \citep{Pescalli2015}. In the presence of a narrow distribution of intrinsic jet luminosities, the luminosity function is then mainly shaped by the angular structure \citep[e.g.][]{Salafia2020,Tan2020}.

An angular profile in the jet properties arises naturally in essentially any physically viable jet formation scenario (see \citealt{Salafia2022_structjet} for a recent review). Moreover, a non-trivial jet energy density angular profile is required \citep[e.g.][]{Mooley2018,Lamb:2018,Lamb:2018obs,Ghirlanda2019,Takahashi2020,Takahashi2021,Beniamini2022} to explain the observed properties of the non-thermal afterglow of the SGRB associated with GW170817. Hence, the question whether the observed distribution of SGRB properties can be traced back, at least in part, to the differing viewing angles is of particular relevance, as it would provide a route to a unification of these sources and a way to disentangle the intrinsic diversity in their properties (and hence those of their progenitor) from the apparent diversity due to extrinsic factors, particularly the viewing angle.

In the future, joint GW-GRB observations will provide direct information on the structure of SGRB jets, thanks to the measurements of the inclination of the merging binary's angular momentum (which is most likely a proxy of the jet viewing angle) that can be inferred from the GW analysis \citep{Hayes2020,Farah2020,Biscoveanu2020}. Still, such information must be combined self-consistently with that encoded in the SGRB population observed in gamma-rays only. In this work, we describe our investigation of the population properties of SGRBs within a `quasi-universal jet' scenario. We assumed that all SGRB jets share the same angular profile of luminosity as a function of the viewing angle, while we allowed for a spread in the on-axis luminosities \citep[similarly as in, e.g.,][]{Tan2020,Hayes2023}. This produces a particular parametrization of the SGRB population properties, which we derive and describe in detail in \S\ref{sec:population_model} and \S\ref{sec:struct_to_lumfunc}. In order to constrain the parameters of this model, we considered the sample of SGRBs detected by \textit{Fermi} and carefully modelled the underlying selection effects (\S\ref{sec:sample_and_sel_effects}). This allowed us to fit the model to the data through a hierarchical Bayesian approach (\S\ref{sec:inference}). In \S\ref{sec:results} we describe in detail the results of the fit, and in \S\ref{sec:discussion} we discuss several implications.

Throughout this work, we assume a flat Friedmann-Lema\^itre-Robertson-Walker cosmology with \citet{Planck2016} parameters, that is, $H_0=67.74\,\mathrm{km\,Mpc^{-1}\,s^{-1}}$ and $\Omega_\mathrm{m,0}=0.3075$.

\section{Methodology}\label{sec:methodology}

\subsection{Apparent versus intrinsic structure}\label{sec:apparent_vs_intrinsic}

The dominant form of energy in GRB jets and the processes that dissipate such energy leading to the observed `prompt' gamma-ray emission are still a matter of debate \citep[see][for a recent review]{Kumar2015}. Still, regardless of the particular dissipation and emission process, the observed emission is affected by relativistic aberration effects in a way that depends on the Lorentz factor profile and the viewing angle \citep[e.g.][]{Woods1999,Salafia2015}. For example, an observer looking, from a viewing angle $\theta_\mathrm{v}$ (angle between the line of sight and the jet axis), at an axisymmetric jet with a bulk Lorentz factor profile $\Gamma(\theta)$ (where $\theta$ is the angle from the jet axis) that radiates an energy per unit solid angle $\mathrm{d}E/\mathrm{d}\Omega(\theta)$, would measure an isotropic-equivalent gamma-ray energy \citep{Salafia2015}
\begin{equation}
 E_\mathrm{iso}(\theta_\mathrm{v})=\int_0^{\pi/2}\int_0^{2\pi}\frac{\delta^3(\theta,\phi,\theta_\mathrm{v})}{\Gamma(\theta)}\frac{\mathrm{d}E}{\mathrm{d}\Omega}(\theta)\,\mathrm{d}\phi\,\sin\theta\,\mathrm{d}\theta,
 \label{eq:Eiso_thv}
\end{equation}
where $\phi$ is the azimuthal angle of a spherical coordinate system whose $z$-axis coincides with the jet axis, $\delta(\theta,\phi,\thv)=\Gamma^{-1}(\theta)\left[1-\vec\beta(\theta,\phi)\cdot \vec e_\mathrm{LoS}(\thv)\right]^{-1}$ is the Doppler factor, $\vec \beta(\theta,\phi)$ is the local jet bulk velocity vector -- with a magnitude $\beta(\theta)=\sqrt{1-1/\Gamma^{2}(\theta)}$ -- and $\vec e_\mathrm{LoS}$ is a unit vector pointing to the observer.
Hence, when considering the emitted energy in gamma-rays, the \textit{apparent structure} $E_\mathrm{iso}(\theta_\mathrm{v})$ depends on the \textit{intrinsic structure} $\left(\Gamma(\theta),\mathrm{d}E/\mathrm{d}\Omega(\theta)\right)$ through a functional that is not invertible in general.
The situation is even more nuanced when considering the luminosity, as the effective angular profile $L_\mathrm{iso}(\theta_\mathrm{v})$ depends also on the degree of overlap between different pulses \citep{Salafia2016}, and hence on the intrinsic variability.

For these reasons, in the absence of strong theoretical constraints on the intrinsic jet structure and on the dissipation and emission processes, the most straightforward approach -- which we adopt in this work -- is that of parametrizing directly the apparent luminosity structure $L(\theta_\mathrm{v})$ (we drop the `$\mathrm{iso}$' suffix from here on for simplicity), which also reduces the number of parameters. An assessment of the intrinsic jet structures that are compatible with a given apparent luminosity profile can be then carried out \textit{a posteriori}.

\subsection{Statistical model of a SGRB population with a quasi-universal jet}\label{sec:population_model}

Within our framework, we describe each SGRB by four physical quantities, namely its viewing angle $\thv$, its peak isotropic-equivalent luminosity $L$, the photon energy $\Ep$ at the peak of the spectral energy distribution (SED, i.e.\ the $\nu F_\nu$ spectrum) and its redshift $z$. These collectively represent what we refer to as the source parameter vector, $\lambda_\mathrm{src}^\star=(\thv,L,\Ep,z)$. For most events, the viewing angle is unknown and we will therefore consider a reduced source parameter vector $\lsource=(L,\Ep,z)$. We assumed the diversity in the latter parameters to be the combined result of intrinsic heterogeneity in the physical properties of jets within the population and extrinsic diversity induced by the differing viewing angles under which these jets are observed.

In order to represent the intrinsic heterogeneity in SGRB jets, we opted for parametrising the joint probability density $P(L_\mathrm{c},E_\mathrm{p,c})$ of their on-axis (`core') peak luminosity $L_\mathrm{c}$ and peak SED photon energy $E_\mathrm{p,c}$ as follows. We assumed $L_\mathrm{c}$ to be distributed as a power law with index $-A$, with a lower exponential cutoff below $L_\mathrm{c}^\star$, namely
\begin{equation}
 P(L_\mathrm{c}\,|\,A,L_\mathrm{c}^\star) = \frac{A}{\tilde\Gamma(1-1/A)L_\mathrm{c}^\star}\exp\left[-\left(\frac{L_\mathrm{c}^\star}{L_\mathrm{c}}\right)^A\right]\left(\frac{L_\mathrm{c}}{L_\mathrm{c}^\star}\right)^{-A},
 \label{eq:P(Lc)}
\end{equation}
where $\tilde\Gamma$ indicates a Gamma function, and the integrated probability is normalized to unity. Such probability density is defined in such a way that it peaks at $L_\mathrm{c}^\star$ regardless of the value of $A$. The probability distribution on $E_\mathrm{p,c}$, conditional on $L_\mathrm{c}$, was assumed log-normal and centered at a $L_\mathrm{c}$-dependent value $\tilde E_\mathrm{p,c}=E_\mathrm{p,c}^\star (L_\mathrm{c}/L_\mathrm{c}^\star)^y$, where $y$ sets the slope of the relation. Hence
\begin{equation}
 P(E_\mathrm{p,c}\,|\,L_\mathrm{c},E_\mathrm{p,c}^\star,y,\sigma_\mathrm{c}) = \frac{\exp\left[-\frac{1}{2}\left(\frac{\ln(E_\mathrm{p,c})-\ln(\tilde E_\mathrm{p,c})}{\sigma_\mathrm{c}}\right)^2\right]}{E_\mathrm{p,c}\sqrt{2\pi\sigma_\mathrm{c}^2}},
 \label{eq:P(Epc)}
\end{equation}
where $\sigma_\mathrm{c}$ sets the dispersion of $E_\mathrm{p,c}$ around $\tilde E_\mathrm{p,c}$. This assumption allows for a `Yonetoku'  correlation\footnote{The name follows from the apparent correlation between $\log(L)$ and $\log(\Ep)$ in long GRBs originally found by \citet{Yonetoku2004}.} with slope $y$ between the logarithms of the on-axis peak SED photon energy and the luminosity, which may be induced, for example, by the underlying emission process. The case with no correlation is represented by $y=0$ and it is therefore naturally included. The joint probability density of the core quantities is
\begin{equation}
P(L_\mathrm{c},E_\mathrm{p,c}\,|\,\lpop) = P(E_\mathrm{p,c}\,|\,L_\mathrm{c},\lpop)P(L_\mathrm{c}\,|\,\lpop),
\end{equation}
where the population parameter vector $\lpop$ contains $L_\mathrm{c}^\star$, $A$, $E_\mathrm{p,c}^\star$, $y$, $\sigma_\mathrm{c}$ and all other parameters that fully specify the SGRB population model.

The next, key assumption of the model is that all jets share a universal `structure' that specifies the dependence of $L$ and $E_\mathrm{p}$ on the viewing angle $\thv$. In practice, we assumed the viewing-angle-dependent luminosity and SED peak photon energy to be expressed as
\begin{equation}
\begin{split}
 L(\thv,\lpop)= & \,L_\mathrm{c}\,\ell(\thv,\lpop),\,\text{and}\\
 E_\mathrm{p}(\thv,\lpop)= & \,E_\mathrm{p,c}\,\eta(\thv,\lpop),
\end{split}
\label{eq:universal_structure}
\end{equation}
where $\ell$ and $\eta$ are functions of the viewing angle and of some parameters included in the $\lpop$ vector. These functions, which we assumed to be redshift-independent, define the universal apparent structure of the jet.

For a population of isotropically-oriented jets, whose viewing angle probability distribution is $P(\thv)=\sin\thv$, we have
\begin{equation}
\begin{split}
 & P(\thv,L,E_\mathrm{p}\,|\,\lpop) = P(L,\Ep\,|\,\thv,\lpop) P(\thv) = \\
 & \delta\left(L-L_\mathrm{c}\ell(\thv,\lpop)\right)\delta\left(\Ep-E_\mathrm{p,c}\eta(\thv,\lpop)\right) P(L_\mathrm{c},E_\mathrm{p,c}\,|\,\lpop)\sin\thv.
\end{split}
\end{equation}

The induced joint luminosity and peak photon energy distribution, given the isotropic viewing angles, is then given by
\begin{equation}
\begin{split}
 & P(L,E_\mathrm{p}\,|\,\lpop)=\int_0^{\pi/2} P(L,E_\mathrm{p}\,|\,\thv,\lpop) \sin\thv\,\mathrm{d}\thv =\\
 & \int_0^{\pi/2} \frac{\exp\left\lbrace-\left[y\ln(\ell(\thv) L_\mathrm{c}^\star/ L) - \ln(E_\mathrm{p,c}^\star \eta(\thv)/E_\mathrm{p}) \right]^2/2\sigma_\mathrm{c}^2\right\rbrace}{E_\mathrm{p}\sqrt{2\pi\sigma_\mathrm{c}^2}}\times\\
 &\times \frac{A}{\tilde\Gamma(1-1/A)\ell(\thv) L_\mathrm{c}^\star}\exp\left[-\left(\frac{\ell(\thv) L_\mathrm{c}^\star}{L}\right)^A\right]\left(\frac{L}{\ell(\thv) L_\mathrm{c}^\star}\right)^{-A}\sin\thv\,\mathrm{d}\thv.
\end{split}
 \label{eq:P(L Ep)}
\end{equation}
In general this must be solved numerically, but in the next section we analyze two cases where the intrinsic dispersion is negligible (i.e.\ $\sigma_\mathrm{c}\to 0$ and $A\to\infty$) and $\ell$ and $\eta$ take simple forms, so that an analytical integration is possible: this will help in demonstrating the main features of the $P(L,\Ep)$ distribution induced by such a quasi-universal structure scenario.

As a consequence of the assumption of redshift independence of the jet structure parameters, the probability distribution of the population source parameters is $P_\mathrm{pop}(L,\Ep,z\,|\,\lpop)=P(L,\Ep\,|\,\lpop)P(z\,|\,\lpop)$, where the redshift probability distribution can be expressed as
\begin{equation}
 P(z\,|\,\lpop)\propto \frac{\dot\rho(z,\lpop)}{1+z}\frac{\mathrm{d}V}{\mathrm{d}z}.
\end{equation}
Here $\mathrm{d}V/\mathrm{d}z$ is the differential comoving volume and $\dot\rho(z,\lpop)$ is the SGRB rate density at redshift $z$. We parametrize the latter as a smoothly broken power law, namely
\begin{equation}
 \dot\rho(z,\lpop)=R_0 \frac{(1+z)^a}{1 + \left(\frac{1+z}{1+z_p}\right)^{b+a}}
 \label{eq:rhoz}
\end{equation}
where $a$, $b$ and $z_p$ are free parameters and $R_0$ is the local rate density of SGRBs with any viewing angle.

\subsection{Apparent jet structure models and the implied luminosity-peak energy distribution}\label{sec:struct_to_lumfunc}

\subsubsection{Luminosity function}

A simple and widely adopted parametric form for the jet structure is a Gaussian one, namely
\begin{equation}
 \begin{split}
  &\ell = \exp\left(-\frac{1}{2}\left(\frac{\thv}{\theta_\mathrm{c}}\right)^2\right),\\
  &\eta = \exp\left(-\frac{1}{2}\left(\frac{\thv}{\theta_{\mathrm{c},E_\mathrm{p}}}\right)^2\right).
 \end{split}
\label{eq:avgstruct_Gaussian}
\end{equation}
In absence of a dispersion in the core quantities, which formally corresponds to the limit $\sigma_\mathrm{c}\to 0$ and $A\to\infty$, and in the case where $\ell$ and $\eta$ are monotonic, Eq.~\ref{eq:P(L Ep)} reduces to a change of variables from $\thv$ to either $L$ or $\Ep$ applied to the viewing angle probability $P(\thv)=\sin\thv$, that is \citep{Pescalli2015,Salafia2022_structjet}
\begin{equation}
\begin{split}
 & P(L,\Ep) = \left[\left(L_\mathrm{c}\frac{\partial \ell}{\partial \thv}\right)^{-1}\sin\thv\,\delta\left(\Ep - \eta(\thv)E_\mathrm{p,c}\right)\right]_{\thv = \ell^{-1}(L/L_\mathrm{c})} =\\
 &=\left[\left(E_\mathrm{p,c}\frac{\partial \eta}{\partial \thv}\right)^{-1}\sin\thv\,\delta\left(L-\ell(\thv)L_\mathrm{c}\right)\right]_{\thv = \eta^{-1}(\Ep/E_\mathrm{p,c})},
\end{split}
\end{equation}
where $\ell^{-1}$ is the inverse function of $\ell$ and $\eta^{-1}$ is the inverse function of $\eta$. In what follows, we will show results using the first of the above equalities, which highlights the dependence on $L$, but the results using the second equality are entirely analogous and can be obtained by exchanging $L\leftrightarrow\Ep$, $\theta_\mathrm{c}\leftrightarrow\theta_{\mathrm{c},\Ep}$ and $L_\mathrm{c}\leftrightarrow E_\mathrm{p,c}$.
In the Gaussian apparent structure case this yields, for $L<L_\mathrm{c}$ and $\Ep<E_\mathrm{p,c}$,
\begin{equation}
\begin{split}
& P(L,\Ep)=\frac{\theta_\mathrm{c}^2}{L}\frac{\sin\left(\theta_\mathrm{c}\sqrt{2\ln(L_\mathrm{c}/L)}\right)}{\theta_\mathrm{c}\sqrt{2\ln(L_\mathrm{c}/L)}}\delta\left(\Ep-E_\mathrm{p,c}\left(\frac{L}{L_\mathrm{c}}\right)^{\theta_\mathrm{c}^2/\theta_{\mathrm{c},\Ep}^2}\right)\sim\\
& \sim \frac{\theta_\mathrm{c}^2}{L}\delta\left(\Ep-E_\mathrm{p,c}\left(\frac{L}{L_\mathrm{c}}\right)^{\theta_\mathrm{c}^2/\theta_{\mathrm{c},\Ep}^2}\right),
\end{split}
 \label{eq:PLEp_Gaussian}
\end{equation}
where the last approximate equality is valid for $\thv\ll \pi/2$, which corresponds to $L\gg L_\mathrm{c}\exp(-1/2\theta_\mathrm{c}^2)$ (or $E_\mathrm
{p}\gg E_\mathrm{p,c}\exp(-1/2\theta_{\mathrm{c},E_\mathrm{p}}^2)$). For typical values $\theta_\mathrm{c}\ll 1$ (or $\theta_{\mathrm{c},E_\mathrm{p}}\ll 1$), the exponential factor is tiny and hence the approximation applies to essentially all relevant luminosities and $\Ep$'s. The luminosity function induced by a Gaussian universal apparent structure is therefore uniform in $\log(L)$, and the same applies to the $\Ep$ distribution.

\begin{figure*}
 \includegraphics[width=\textwidth]{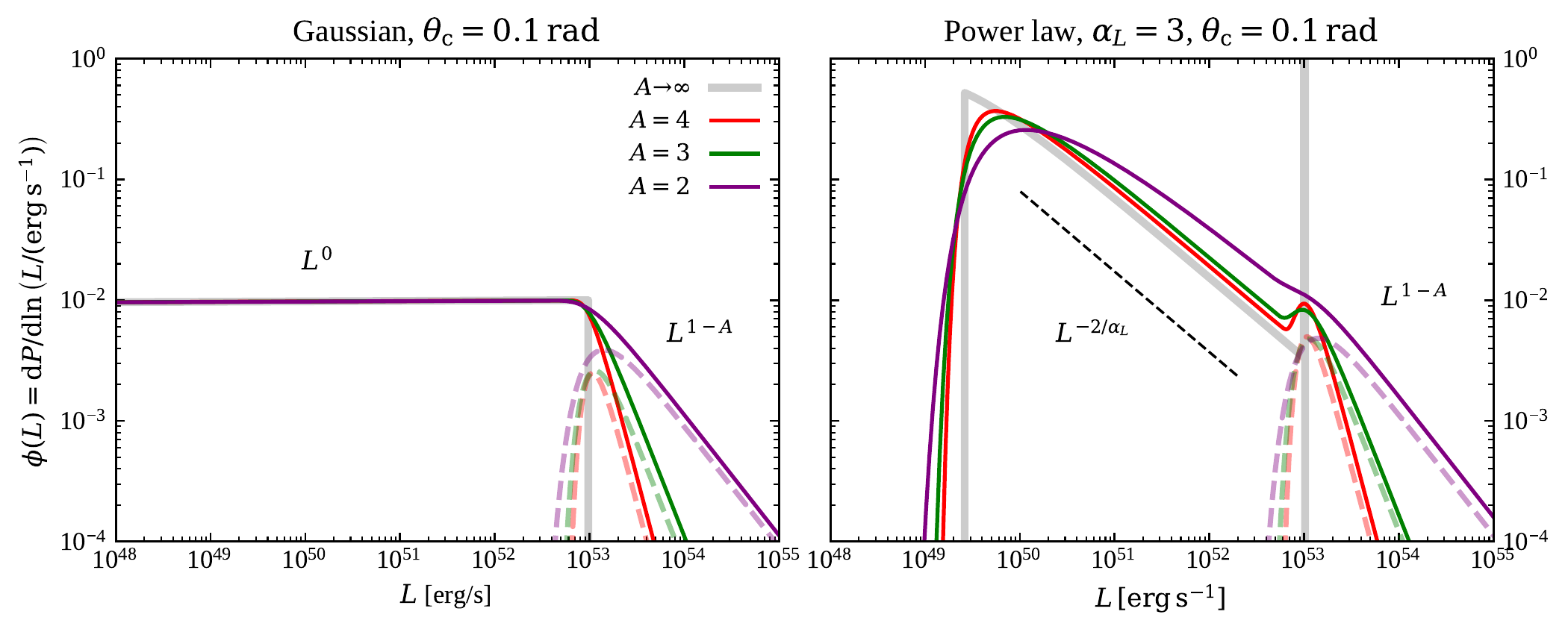}
 \caption{Example luminosity functions $\phi(L)=\mathrm{d}P/\mathrm{d}\ln(L)$ induced by a Gaussian apparent structure model (Eqs.~\ref{eq:avgstruct_Gaussian}, left-hand panel) and a power law apparent structure model (Eqs.~\ref{eq:avgstruct_plaw}) with a luminosity profile slope $\alpha=3$ (right-hand panel), both with a core half-opening angle $\theta_\mathrm{c}=0.1\,\mathrm{rad}$ and $L_\mathrm{c}^\star=10^{53}\,\mathrm{erg\,s^{-1}}$. In each panel, the luminosity function corresponding to a population with zero core dispersion ($A\to\infty$ in Eq.~\ref{eq:P(Lc)}) is shown in grey, while three cases with different core luminosity distributions (slopes reported in the legend) are shown with colored lines. The corresponding $L_\mathrm{c}$ distributions (arbitrarily scaled) are shown with dashed lines. In the right-hand panel, the black dashed line shows the trend $\phi(L)\propto L^{-2/\alpha_L}$ expected at intermediate luminosities.}
\label{fig:appstruct_lumfunc_examples}
\end{figure*}

The effect of a non-zero dispersion in the core quantities is equivalent to that of convolving the zero-dispersion $P(L,\Ep)$ probability density distribution with the probability density distribution of the core luminosity and peak SED photon energy $P(L_\mathrm{c},E_\mathrm{p,c})$. In that sense, the probability density distribution of the core quantities acts essentially as a smoothing kernel: in the Gaussian apparent structure case, it introduces a smooth transition to a power law fall-off above $L_\mathrm{c}^\star$ and $E_\mathrm{p,c}^\star$. In more physical terms, and for typical parameters relevant to the quasi-universal jet scenario, the high-end of the luminosity function is shaped by the distribution of core luminosities, while below $L_\mathrm{c}^\star$ the luminosity function is set by the jet (apparent) structure. The left-hand panel in Figure \ref{fig:appstruct_lumfunc_examples} shows the luminosity function $\phi(L)=\mathrm{d}P/\mathrm{d}\ln(L)=L\int P(L,\Ep)\,\mathrm{d}\Ep$ for an example Gaussian apparent structure case, demonstrating the effect of introducing a core luminosity dispersion with three different values of the slope $A$.

We can get some further insight by adopting a power law apparent structure model with a `uniform core' within $\thv\leq \theta_\mathrm{c}$, namely
\begin{equation}
 \begin{split}
  & \ell = \min\left[1,\left(\frac{\thv}{\theta_\mathrm{c}}\right)^{-\alpha_L}\right]\\
  & \eta = \min\left[1,\left(\frac{\thv}{\theta_\mathrm{c}}\right)^{-\alpha_{\Ep}}\right],
 \end{split}
\label{eq:avgstruct_plaw}
\end{equation}
again with no dispersion. The change of variables approach can be applied to $\thv>\theta_\mathrm{c}$, where the structure is monotonic; for $\thv\leq \theta_\mathrm{c}$ it is sufficient to note that all observers see $L=L_\mathrm{c}$ and $\Ep=E_\mathrm{p,c}$, and the probability of having a viewing angle in this range is $1-\cos\theta_\mathrm{c}$. Hence, the $L-\Ep$ distribution is
\begin{equation}
\begin{split}
 & P(L,\Ep)=\frac{\theta_\mathrm{c} L_\mathrm{c}^{1/\alpha_L}}{\alpha_L}\frac{\sin\left[\theta_\mathrm{c}\left(\frac{L_\mathrm{c}}{L}\right)^{1/\alpha_L}\right]}{L^{1+1/\alpha_L}}\delta\left(\Ep-E_\mathrm{p,c}\left(\frac{L}{L_\mathrm{c}}\right)^{\alpha_{\Ep}/\alpha_{L}}\right) \sim \\
 & \sim \frac{\theta_\mathrm{c}^2 L_\mathrm{c}^{2/\alpha_L}}{\alpha_L}L^{-1-2/\alpha_L}\delta\left(\Ep-E_\mathrm{p,c}\left(\frac{L}{L_\mathrm{c}}\right)^{\alpha_{\Ep}/\alpha_{L}}\right)
\end{split}
\label{eq:PLEp_plaw}
\end{equation}
for $L_\mathrm{min}^\star\leq L<L_\mathrm{c}$ and $E_\mathrm{p,min}^\star\leq \Ep<E_\mathrm{p,c}$, where $L_\mathrm{min}^\star=L_\mathrm{c}(2\theta_\mathrm{c}/\pi)^{\alpha_L}$ and $E_\mathrm{p,min}^\star=E_\mathrm{p,c}(2\theta_\mathrm{c}/\pi)^{\alpha_{\Ep}}$, and the last approximate equality is valid for $L\gg L_\mathrm{min}^\star$ and $\Ep\gg E_\mathrm{p,min}^\star$. Again, the alternate form which highlights the dependence on $\Ep$ can be obtained by the substitutions $L\leftrightarrow\Ep$, $\alpha_{L}\leftrightarrow\alpha_{\Ep}$ and $L_\mathrm{c}\leftrightarrow E_\mathrm{p,c}$. For $L\geq L_\mathrm{c}$, $\Ep\geq E_\mathrm{p,c}$, we have
\begin{equation}
 P(L,\Ep)=(1-\cos\theta_\mathrm{c})\delta(L-L_\mathrm{c})\delta(\Ep-E_\mathrm{p,c}).
\end{equation}
While in this limiting zero-dispersion case the jet core clearly extends over a zero-measure region of the $(L,\Ep)$ plane, a non-zero dispersion spreads this over a more physically sound, finite region of the plane. Hence, a power law apparent structure induces a power law luminosity function with a slope $L^{-2/\alpha_L}$ in the logarithm of $L$ \citep{Pescalli2015}. Similar conclusions apply to the induced $\Ep$ distribution. The slope parameter $\alpha_L$, together with the core half-opening angle $\theta_\mathrm{c}$, control the extent of the luminosity function as a consequence of the limited physical viewing angle range $0\leq \thv \leq \pi/2$. The right-hand panel in Figure \ref{fig:appstruct_lumfunc_examples} shows the luminosity function in an example power law case with $\alpha_L=3$.

The above derivation also shows that in a quasi-universal structured jet scenario $\phi(L)$ can never attain a positive slope, except at the low-luminosity end, or in a narrow luminosity range close to the `core' for small dispersions (unless $\alpha_L<0$, which however does not seem a likely physical possibility). This is a feature of quasi-universal jet models.

\subsubsection{$L-\Ep$ correlation induced by the jet structure}

Since both $\ell$ and $\eta$ are functions of the viewing angle, the quasi-universal structured jet model inherently implies a `Yonetoku' \citep{Yonetoku2004} correlation between $L$ and $\Ep$ within the observed population \citep[e.g.][]{Salafia2015}, in addition to any intrinsic correlation that may hold between the core quantities $L_\mathrm{c}$ and $E_\mathrm{p,c}$. The shape of the induced correlation can be obtained by eliminating $\thv$ in the average apparent structure functions, to obtain $\eta(\ell)$, and is already apparent in the Dirac-delta functions in Eqs.~\ref{eq:PLEp_Gaussian} and \ref{eq:PLEp_plaw}. In the Gaussian average apparent structure case, it yields $(\Ep/E_\mathrm{p,c})=(L/L_\mathrm{c})^{\theta_\mathrm{c}^2/\theta_{\mathrm{c},\Ep}^2}$, so that the slope of the correlation is set by the ratio of the scale parameters $\theta_\mathrm{c}$ and $\theta_{\mathrm{c},\Ep}$ over which $L$ and $\Ep$ decay with the viewing angle. Similarly, in the power law case the relation is $(\Ep/E_\mathrm{p,c})=(L/L_\mathrm{c})^{\alpha_{\Ep}/\alpha_L}$. In general, the induced correlation is a power law (or a collection of power law branches) whenever the functional forms of $\ell$ and $\eta$ are the same, and its slope is set by the ratio of the decay rates of $\ell$ and $\eta$.

\subsubsection{Average apparent structure model adopted in this work}

\begin{figure*}
 \centering
 \includegraphics[width=\textwidth]{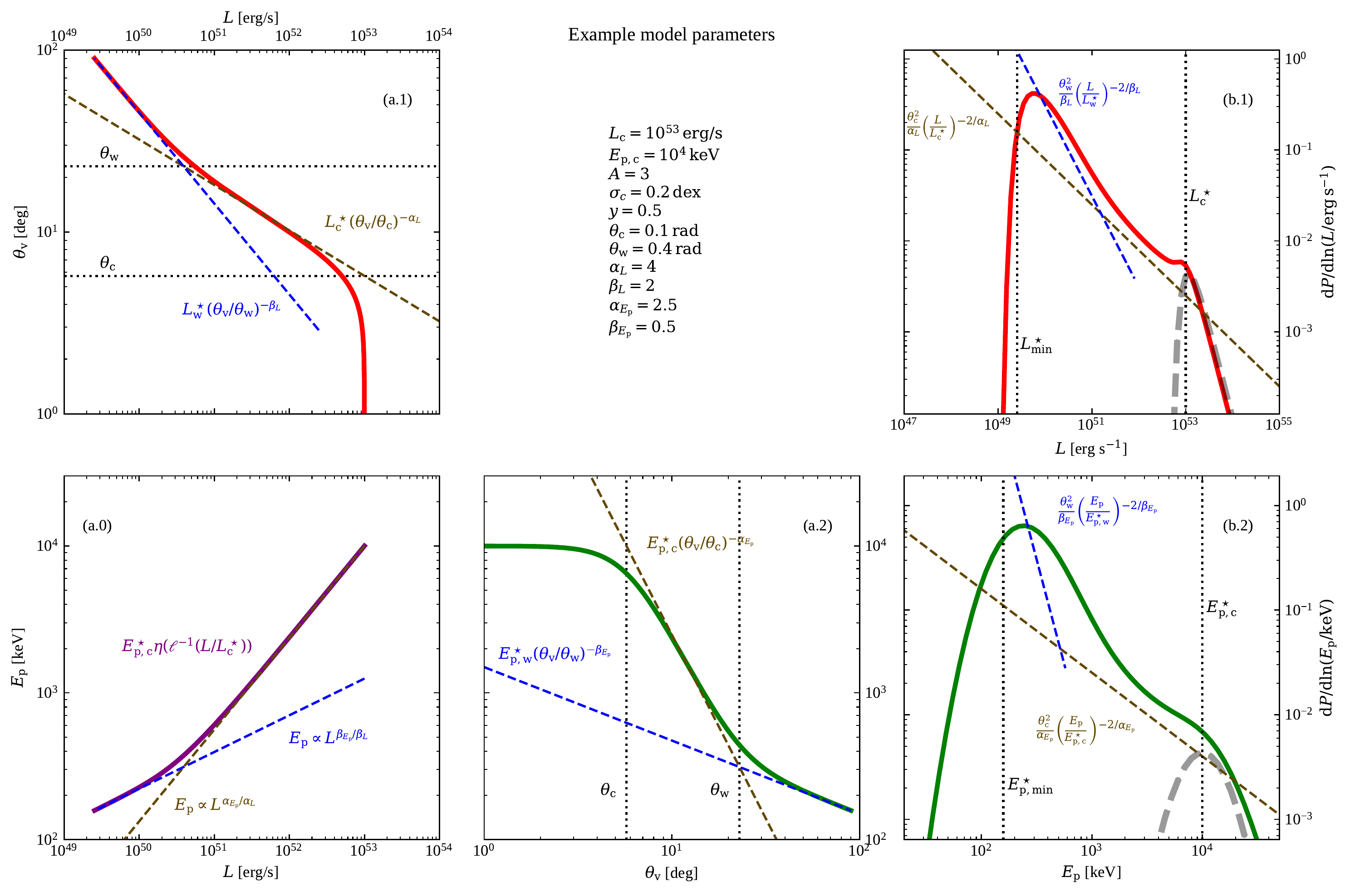}
 \caption{Example double smoothly broken power law jet apparent structure model and induced $L$ and $\Ep$ probability distributions. Panels a.1 and a.2 show the apparent jet structure functions $L_\mathrm{c}^\star \ell (\thv)$ and $E_\mathrm{p,c}^\star\eta(\thv)$ (red and green solid line, respectively) adopting the parametrization defined in Eq.~\ref{eq:avgstruct_dbpl} and using the parameter values reported in the middle panel of the figure. Panel a.0 shows the induced $E_\mathrm{p}(L)$ relation due to the structure (purple line).  In panels a.1 and a.2, the black dotted lines show the transition angles $\theta_\mathrm{c}$ and $\theta_\mathrm{w}$. The dashed lines demonstrate the analytical trends discussed in the text. Panel b.1 shows the induced luminosity function $\phi(L)=\mathrm{d}P/\mathrm{d}\ln L = L\int P(L,\Ep)\,\mathrm{d}\Ep$ (red line), with dashed lines demonstrating approximate analytical trends (significant departures are due to the core dispersion), and with the core luminosity distribution shown with a dark grey dashed line, centered at $L_\mathrm{c}$ and with its integral normalized to $1-\cos\theta_\mathrm{c}$. Panel b.2 is the analogue of b.1, but showing the $\mathrm{d}P/\mathrm{d}\ln\Ep$ probability distribution (green line).}
 \label{fig:dbpl_example}
\end{figure*}

In this study, in order to endow the average apparent structure model with a high degree of flexibility (in the absence of strong theoretical constraints on the expected shape), we adopted a double smoothly broken power law model with a nearly constant  `core' within $\thv<\theta_\mathrm{c}$ and a break at a wider angle $\theta_\mathrm{w}$, that is
\begin{equation}
 \begin{split}
  & \ell(\thv) = \left[1+\left(\frac{\thv}{\theta_\mathrm{c}}\right)^{s}\right]^{-\alpha_L/s}\left[1+\left(\frac{\thv}{\theta_\mathrm{w}}\right)^{s}\right]^{-(\beta_L-\alpha_L)/s}\\
  & \eta(\thv) = \left[1+\left(\frac{\thv}{\theta_\mathrm{c}}\right)^{s}\right]^{-\alpha_{\Ep}/s}\left[1+\left(\frac{\thv}{\theta_\mathrm{w}}\right)^{s}\right]^{-(\beta_{\Ep}-\alpha_{\Ep})/s},
 \end{split}
\label{eq:avgstruct_dbpl}
\end{equation}
where the `smoothness' parameter is set to $s=4$, which makes the transitions between the power law branches relatively sharp, to compensate the fact that the intrinsic dispersion of core quantities tends to smooth out the induced breaks in the luminosity function. The implied $L-\Ep$ correlation has two branches with slopes $\alpha_{\Ep}/\alpha_L$ and $\beta_{\Ep}/\beta_L$, the break being around $L\sim L_\mathrm{w}^\star= L_\mathrm{c}^\star(\theta_\mathrm{w}/\theta_\mathrm{c})^{-\alpha_L}$ and $\Ep\sim E_\mathrm{p,w}^\star=E_\mathrm{p,c}^\star(\theta_\mathrm{w}/\theta_\mathrm{c})^{-\alpha_{\Ep}}$.  Figure~\ref{fig:dbpl_example} demonstrates the features of such model in detail, including the induced $L$ and $\Ep$ probability distributions, based on an example choice of parameters.

With this choice of average apparent jet structure functions, the model features a total of 14 free parameters. We list these parameters in Table \ref{tab:parameters}, along with brief definitions and with information on the priors adopted on each of them in the analysis, discussed later in the text.

\begin{table*}
 \caption{Model parameters. The table lists the parameters that constitute the components of the $\lpop$ vector in our analysis.}
 \label{tab:parameters}
 \centering
 \renewcommand*{\arraystretch}{1.2}
 \begin{tabular}{cll}
  Parameter & Prior & Description  \\
  \hline
  $\theta_\mathrm{c}$ & \multirow{2}{*}{Isotropic$^a$, $0.01\,\mathrm{rad}\leq\theta_\mathrm{c}<\theta_\mathrm{w}\leq\pi/2\,\mathrm{rad}$} & `core' half-opening angle\\
  $\theta_\mathrm{w}$ &  & `break' angle \\
  $L_\mathrm{c}^\star$ & Uniform-in-log, $3\times 10^{51}$ erg s$^{-1}$ $\leq L_\mathrm{c}^\star\leq 10^{55}$ erg s$^{-1}$ & Low-end cutoff luminosity for on-axis observers ($\thv=0$) \\
  $\alpha_{L}$ & Uniform, $0\leq\alpha_\mathrm{L}\leq6$ & Slope of $\ell$ at intermediate viewing angles $\theta_\mathrm{c}<\thv<\theta_\mathrm{w}$ \\
  $\beta_{L}$  & Uniform, $-3\leq\beta_\mathrm{L}\leq6$ & Slope of $\ell$ at large viewing angles $\thv>\theta_\mathrm{w}$ \\
  $E_\mathrm{p,c}^\star$  & Uniform-in-log, $10^{2}\,\mathrm{keV}\leq E_\mathrm{p,c}^\star\leq 10^{5}\,\mathrm{keV}$ & Low-end cutoff $\Ep$ for on-axis observers ($\thv=0$)  \\
  $\alpha_{E_\mathrm{p}}$  & Uniform, $0\leq \alpha_{\Ep}\leq 6$ & Slope of $\eta$ at intermediate viewing angles $\theta_\mathrm{c}<\thv<\theta_\mathrm{w}$ \\
  $\beta_{E_\mathrm{p}}$  & Uniform, $-3\leq \beta_{\Ep}\leq 10$ & Slope of $\eta$ at large viewing angles $\thv>\theta_\mathrm{w}$ \\
  $A$  & Uniform, $1.5\leq A\leq 5$ & Slope of $L_\mathrm{c}$ distribution \\
  $\sigma_{c}$  & Uniform-in-log, $0.3\leq \sigma_\mathrm{c}\leq 3$ & $E_\mathrm{p}$ dispersion around $\tilde E_\mathrm{p} = E_\mathrm{p,c}^\star(L_\mathrm{c}/L_\mathrm{c}^\star)^y$ \\
  $y$  & Uniform, $-3\leq y \leq 3$ & $L_\mathrm{c}-E_\mathrm{p,c}$ `on-axis Yonetoku' correlation slope \\
  $a$  & Uniform, $-1\leq a \leq 5$ & Rate density evolution slope at low redshift, $\dot\rho\propto(1+z)^a$\\
  $b$  & Uniform, $1\leq b \leq 10$ & Rate density decay slope at high redshift, $\dot\rho\propto(1+z)^{-b}$\\
  $z_\mathrm{p}$  & Uniform, $0.1\leq z_\mathrm{p}\leq 3$ & Redshift of rate density peak\\
  \hline
 \end{tabular}\\
 \flushleft
 \footnotesize{$^a$Isotropic $=$ uniform in the subtended solid angle, i.e.\ $\pi(\theta)=\sin\theta$.}
\end{table*}

\subsection{Sample definition and selection effect modelling}\label{sec:sample_and_sel_effects}

In order to compare a population model with an observed sample, the selection effects that shape the latter must be taken into account. Here we describe our sample choice and the procedure that we employed to model the underlying selection effects. We adopted a description of the SGRB photon spectrum as a cut-off power law \citep{Ghirlanda2004}, $\mathrm{d}\dot N/\mathrm{d}E (E,E_\mathrm{p,obs},\alpha) \propto E^{\alpha} \exp[-(2+\alpha)E/E_\mathrm{p,obs}]$, 
where $\dot N$ represents the rate of photons hitting the detector, $E$ is the photon energy, and $E_\mathrm{p,obs}=\Ep/(1+z)$. We set the low-energy photon index to $\alpha=-0.4$, which is the median of the values reported in the \textit{Fermi}/GBM online catalog for SGRBs. The peak photon flux in the $E_0-E_1$ keV band is then defined as
\begin{equation}
 p_{[E_0,E_1]}(L,\Ep,z) = \frac{1}{4\pi d_\mathrm{L}^2(z)}\frac{L}{\mathcal{E}_{[E_0,E_1]}(\Ep,z)}
 \label{eq:pflux}
\end{equation}
where $d_\mathrm{L}$ is the luminosity distance and
\begin{equation}
 \mathcal{E}_{[E_0,E_1]}(\Ep,z) = \frac{\int_{0.1\,\mathrm{keV}/(1+z)}^{10^7\,\mathrm{keV}/(1+z)}\mathrm{d}\dot N/\mathrm{d}E (E,E_\mathrm{p,obs},\alpha)\,E\,\mathrm{d}E}{\int_{E_0\,\mathrm{keV}}^{E_1\,\mathrm{keV}}\mathrm{d}\dot N/\mathrm{d}E (E,E_\mathrm{p,obs},\alpha)\,\mathrm{d}E}.
\end{equation}
We stress here again that we extended the customary $1-10^4$ keV pseudo-bolometric rest-frame band to the wider $0.1-10^7$ keV to include possible cases with very high $E_\mathrm{p}$. In practice, we pre-computed\footnote{While \textit{Fermi}/GBM is sensitive over a larger energy band, and the results in the catalog usually refer to the 10-1000 keV band, the 50-300 keV band is where most of the online GRB trigger algorithms look for excess \citep{vonKienlin2020}, so that the flux in that band is the most relevant for what concerns the modelling of the GBM detection -- see Appendix \ref{sec:pdet_construction}.} $\mathcal{E}_{[50-300]}$ for \textit{Fermi}/GBM and $\mathcal{E}_{[15-150]}$ for \textit{Swift}/BAT over a uniformly-spaced two dimensional grid in $(\log\Ep,\log z)$ and then used two-dimensional linear interpolation to recover it and obtain the photon flux from Eq.~\ref{eq:pflux} (or equivalently to obtain $L$ from $p$ and $E_\mathrm{p,obs}$, by inverting the equation).

\begin{figure}
 \centering
 \includegraphics[width=\columnwidth]{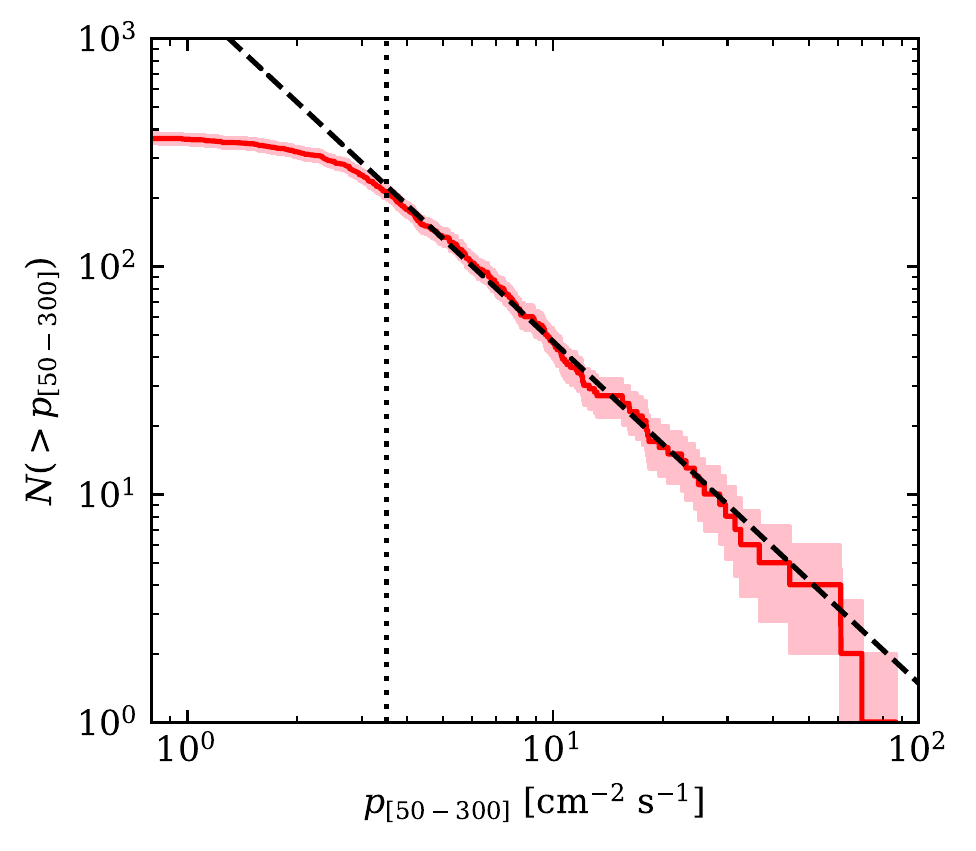}
 \caption{Inverse cumulative distribution of \textit{Fermi}/GBM SGRB peak photon fluxes. The red solid line shows the inverse cumulative number of SGRBs detected by GBM with spectral information available in the catalog, as a function of the peak photon flux measured on a 64 ms timescale in the 50-300 keV band. The pink band shows the one-sigma-equivalent Poisson error. The dashed black line shows a power law $p_{[50-300]}^{-3/2}$ for reference. The vertical dotted line shows the chosen value of the flux cut for the flux-limited sample, $p_\mathrm{lim,GBM}=3.5\,\mathrm{cm^{-2}\,s^{-1}}$.}
 \label{fig:logN_logS}
\end{figure}

We considered three reference short GRB samples: (i) SGRBs detected by \textit{Fermi}/GBM, with available spectral information in the public catalog; (ii) SGRBs detected by both \textit{Swift}/BAT and \textit{Fermi}/GBM, with a number of additional cuts to reach a high completeness in redshift; (iii) SGRBs detected by \textit{Fermi}/GBM with a gravitational wave counterpart, which currently includes only GRB~170817A/GW170817. In what follows, we describe in detail the selection cuts of each sample and our approach to the modelling of selection effects.

\subsubsection{Observer-frame sample: \textit{Fermi}/GBM SGRBs with spectral information}\label{sec:obsframe_sample}

Our ``observer-frame'' sample includes GRBs detected by the GBM onboard \textit{Fermi} from the start of the mission up to mid July 2018, after which no spectral information is present in the online catalog \citep{vonKienlin2020} at the time of writing. Among these, we selected 367 events that are nominally `short', that is, their duration $T_{90}<2\,\mathrm{s}$, as our initial raw sample. As a first approach, we considered working with the sub-sample of bursts whose peak photon flux is larger than a `completeness' threshold $p_{[50-300]}>p_\mathrm{lim,GBM}$, where the GBM flux in the 50-300 keV band is measured with 64 ms binning to best approximate the actual peak photon flux. By visual inspection we found that above $p_\mathrm{lim,GBM}=3.5\,\mathrm{cm^{-2}\,s^{-1}}$ the observed $N(>p_{[50-300]})$ distribution looks like a single power law (see Figure~\ref{fig:logN_logS}), and hence we selected this as our completeness threshold. This is a practical approach historically employed to construct flux-complete samples. The corresponding detection probability can be modelled simply as
\begin{equation}
 P_\mathrm{det,GBM}(L,\Ep,z)=\Theta\left(p_{[50-300]}(L,\Ep,z)-p_\mathrm{lim,GBM}\right),
 \label{eq:pdet_simple}
\end{equation}
where $\Theta$ is the Heaviside step function, that is $\Theta(x)=1$ if $x>0$ and $\Theta(x)=0$ otherwise.

This selection criterion significantly reduces the sample size: out of a total of 367 SGRBs in the raw sample, only 212 have $p_{[50-300]}>3.5\,\mathrm{cm^{-2}\,s^{-1}}$. Most importantly, the discarded events possibly probe the luminosity function down to lower luminosities, which is where most of the useful information on the jet structure resides in a quasi-universal jet scenario. Last, but not least, the only SGRB with reliable viewing angle information, that is GRB~170817A, is not included in this flux-limited sample. In order to access the flux-incomplete part of the \textit{Fermi}/GBM SGRB sample, we carefully constructed a detection probability $P_\mathrm{det}(L,E_\mathrm{p},z)$ by simulating the response of the GBM NaI detectors to SGRBs with a broad range of characteristics, as described in detail in Appendix~\ref{sec:pdet_construction}. In this study, we compare the results obtained by using either of the strategies in modelling the selection effects: hereafter, we refer to the reduced sample of \textit{Fermi}/GBM SGRBs with $p_{[50-300]}>3.5\,\mathrm{cm^{-2}\,s^{-1}}$ as the `flux-limited sample', and to the analysis adopting the associated simplified selection effect model as the `flux-limited sample analysis'. Conversely, the analysis performed adopting the simulated \textit{Fermi}/GBM detection efficiency is referred to as the `full sample analysis'.

As a further countermeasure against possible biases, we applied an additional quality cut to both the above samples: we removed events with best-fit $E_\mathrm{p,obs}>10\,\mathrm{MeV}$ or $E_\mathrm{p,obs}<50\,\mathrm{keV}$, which fall outside the spectral range where the effective area of the GBM detectors is optimal: this removes 2 events whose uncertainty on $E_\mathrm{p,obs}$ is very large, reducing the flux-limited sample to 210 events. In the full sample analysis, we also removed another 10 events with a low best-fit peak 64-ms photon flux $p_{[50-300]}<1\,\mathrm{cm^{-2}\,s^{-1}}$, all of which have very large errors on both $p_{[50-300]}$ and $E_\mathrm{p,obs}$. This reduces the `full' sample to 355 events. To reflect these further quality cuts, we updated our model detection probabilities by multiplying them by $\Theta(p_{[50-300]}-1\,\mathrm{cm^{-2}\,s^{-1}})\Theta(E_\mathrm{p,obs}-50\,\mathrm{keV})\Theta(10\,\mathrm{MeV}-E_\mathrm{p,obs})$.

\subsubsection{Rest-frame sample: Flux-complete sample of \textit{Swift}/BAT SGRBs observed also by \textit{Fermi}/GBM}\label{sec:restframe_sample}

\begin{figure*}
 \includegraphics[width=\textwidth]{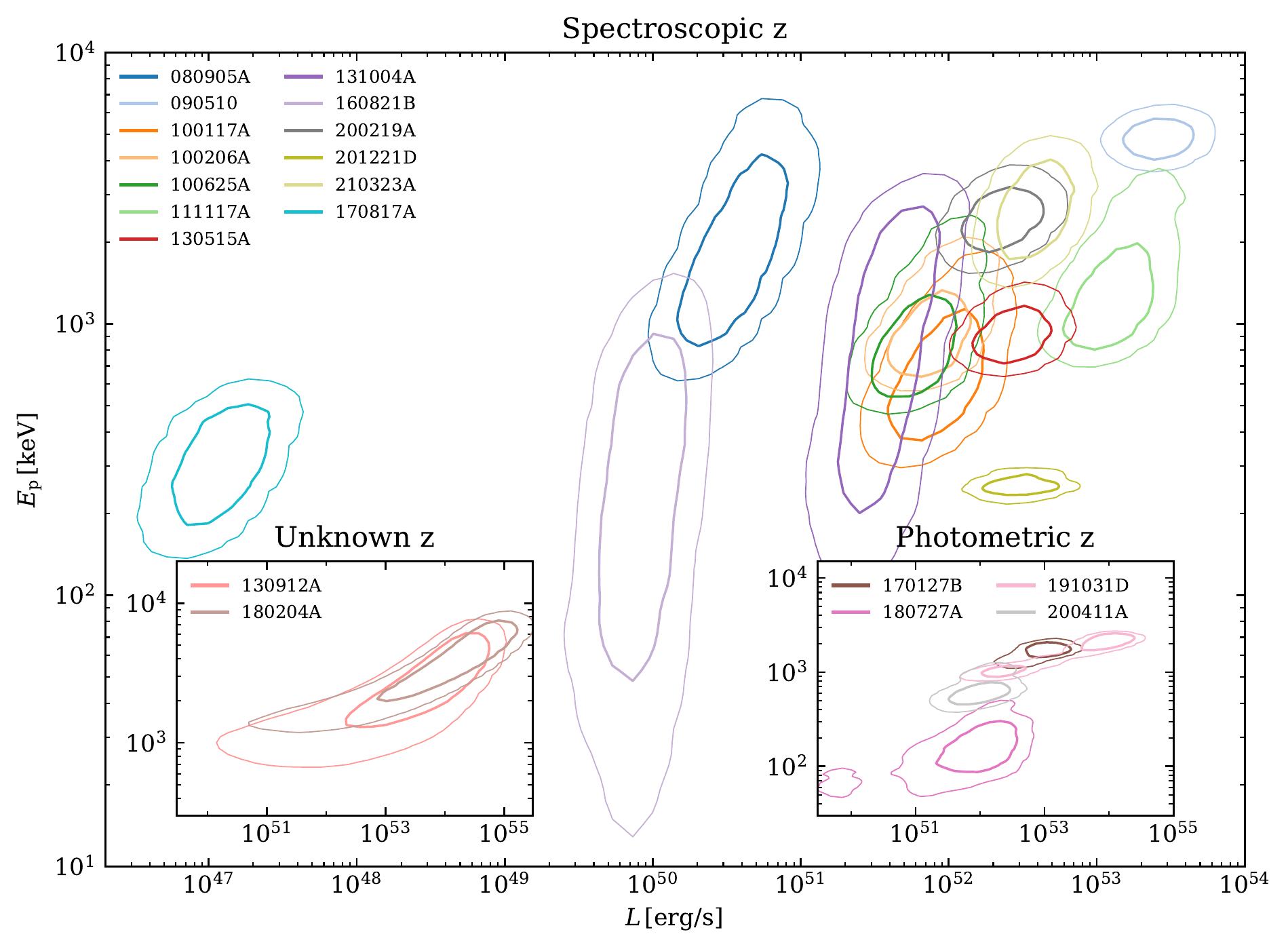}
 \caption{$L-\Ep$ contours for our sub-sample of \textit{Fermi}/GBM SGRBs detected also by \textit{Swift}/BAT with a 64 ms peak flux $p_\mathrm{[15-150]}> 3.5\,\mathrm{cm^{-2}\,s^{-1}}$ (see the text for the full sample selection criteria). The main panel shows the contours that contain 50\% (thick lines) and 90\% (thin lines) of the posterior probability on $(L,\Ep)$ for bursts with a spectroscopically measured redshift, plus GRB~170817A (which does not belong to the GBM+BAT sample); the right-hand inset shows the corresponding contours for four events with a photometric redshift measurement (from \citealt{Fong2022}); the left-hand inset shows the contours for the remaining two SGRBs with an unknown redshift, constructed assuming a uniform prior on $z$ in the range $(10^{-4},4)$. The latter two events were not included in the analysis.}
 \label{fig:GBM_SGRBs_with_z}
\end{figure*}

For some of the events in our \textit{Fermi}/GBM observer-frame sample, redshift information is available and can be used to constrain the population parameters better than can be done using the observer-frame information $(p_{[50-300]},E_\mathrm{p,obs})$ only.
In order to avoid biases, on the other hand, any additional selection effects at play in the sub-sample with known redshift must be accounted for in the inference, that is, in the $P_\mathrm{det}$ model. The redshift determination is a complex process that involves multiple facilities and depends not only on the prompt emission properties, but also on those of the afterglow. Therefore, modelling the associated selection effects is prohibitive. On the other hand, \citet{Salvaterra2012} and \citet{DAvanzo2014} showed that it is possible to construct a sample with a selection that is easier to model, but that leads to a high redshift completeness. The selection involves two cuts that do not bias the redshift distribution, namely (i) a cut on the foreground interstellar dust extinction $A_V$ and (ii) a cut on the \textit{Swift}/XRT slew time; plus a cut on the BAT 64 ms peak flux in the 15-150 keV band, $p_{[15-150]}>3.5\,\mathrm{cm^{-2}\,s^{-1}}$, to ensure flux completeness. The original short GRB sample constructed in this way, known as the S-BAT4 \citep{DAvanzo2014}, included 16 events, 11 of which had a measured redshift. Thanks to a considerable effort spent by the community, and in particular by \citet{Fong2022} and \citet{Nugent2022}, in identifying SGRB host galaxies and measuring their redshifts, it has been recently possible to construct an extended ``S-BAT4ext'' sample (Matteo Ferro et al., submitted; D'Avanzo et al., in preparation) with a more than doubled size and an increased redshift completeness. For this work, we adopt the S-BAT4ext sub-sample of 18 events that have been jointly detected by \textit{Fermi}/GBM with $T_{90}<2\,\mathrm{s}$ (for consistency with our observer-frame sample). Of these, 16 have either a spectroscopic (12 events) or photometric (4 events) redshift measurement, as listed in Table \ref{tab:GBM_BAT_sample}. For these SGRBs we performed an independent analysis of their peak spectra, based on publicly available \textit{Fermi}/GBM data, as described in Appendix \ref{sec:gbm_analysis}, obtaining posterior samples of their bolometric flux $F$ and observed peak photon energy $E_\mathrm{p,obs}$, adopting a prior $\pi(F,E_\mathrm{p,obs})\propto F^{-1}$. For the 12 events with a spectroscopic redshift, these were converted into samples of $P(L,\Ep,z\,|\,d_i)$ by simply fixing the redshift at the best fit value. For the four events with photometric redshift, we obtained samples of $P(L,\Ep,z\,|\,d_i)$ by computing $L_{l,m}=4\pi d_{\mathrm{L},l}^2 F_m$ and $E_{\mathrm{p},l,m}=(1+z_l)E_{\mathrm{p,obs},m}$, where $\lbrace z_l \rbrace_{l=1}^{N_\mathrm{s}}$ are $N_\mathrm{s}$ photometric redshift posterior samples from the BRIGHT catalog \citep{Nugent2022}, $\lbrace d_{\mathrm{L},l} \rbrace_{l=1}^{N_\mathrm{s}}$ are the corresponding luminosity distances under our assumed cosmology and $\lbrace (F_m,E_{\mathrm{p,obs},m}) \rbrace_{m=1}^{M_\mathrm{s}}$ are $M_\mathrm{s}$ posterior samples from the spectral analysis. In both cases, the effective prior on the source parameters is $\pi(L,\Ep,z)\propto (1+z)^{-1}L^{-1}$. The two events with unknown redshift were not included in the analysis.

\renewcommand{\arraystretch}{1.2}
\begin{table*}
\caption{``Rest-frame sample'' of SGRBs detected by \textit{Swift}/BAT and \textit{Fermi}/GBM which belong to the S-BAT4ext complete sample and have $T_{90}<2\,\mathrm{s}$ in the GBM catalog.}

    \centering
    \begin{tabular}{lccccccc}
\hline
GRB		&    GBM name & $T_{90}/\mathrm{s}$ & z & Redshift source$^\star$ \\
\hline
080905A & GRB080905499 & $0.960\pm 0.345$   & $0.1218 \pm 0.0003$          & \citealt{Fong2022} (Gold) \\
090510	& GRB090510016 & $0.960\pm 0.138$	& $0.903 \pm 0.002$	   & \citet{Rau2009GCN,Levan2009GCN} \\
100117A	& GRB100117879 & $0.256\pm 0.834$	& $0.914 \pm 0.0004$  & \citealt{Fong2022} (Gold)   \\
100206A	& GRB100206563 & $0.176\pm 0.072$	& $0.40 \pm 0.002$	   & \citealt{Fong2022} (Gold)   \\
100625A	& GRB100625773 & $0.240\pm 0.276$	& $0.452_\pm 0.002$	   & \citealt{Fong2022} (Silver) \\
111117A	& GRB111117510 & $0.432\pm 0.082$	& $2.211 \pm 0.001$    & \citealt{Fong2022} (Silver) \\
130515A	& GRB130515056 & $0.256\pm 0.091$	& $0.80 \pm 0.01$	   & \citealt{Fong2022} (Silver) \\
130912A	& GRB130912358 & $0.512\pm 0.143$	&               N/A            &                             \\
131004A	& GRB131004904 & $1.152\pm 0.590$	& $0.717 \pm 0.002$	   & \citealt{Fong2022} (Silver) \\
160821B	& GRB160821937 & $1.088\pm 0.977$	& $0.1619 \pm 0.0002$ & \citealt{Fong2022} (Silver) \\
170127B	& GRB170127634 & $1.728\pm 1.346$	& $2.28 \pm 0.14$ $^\dagger$	   & \citealt{Fong2022} (Silver) \\
180204A	& GRB180204109 & $1.152\pm 0.091$	&               N/A            &        \\
180727A	& GRB180727594 & $0.896\pm 0.286$   &   $1.95_{-0.58}^{+0.50}$ $^\dagger$    & \citealt{Fong2022} (Gold)   \\
191031D	& GRB191031891 & $0.256\pm 0.023$	& $1.93_{-1.44}^{+0.22}$ $^\dagger$      & \citealt{Fong2022} (Silver) \\
200219A	& GRB200219317 & $1.152\pm 1.032$	& $0.48 \pm 0.02$	   & \citealt{Fong2022} (Gold)   \\
200411A	& GRB200411187 & $1.440\pm 0.506$   & $0.82_{-0.17}^{+0.18}$ $^\dagger$	   & \citealt{Fong2022} (Bronze) \\
201221D	& GRB201221963 & $0.144\pm 0.066$   & $1.046 \pm 0.002$	   & \citet{deUgartePostigo2020GCN}       \\
210323A	& GRB210323918 & $0.960\pm 0.781$	& $0.733 \pm 0.001$	   & \citealt{Fong2022} (Gold)   \\
\hline
    \end{tabular}\\
    \label{tab:GBM_BAT_sample}
    \footnotesize{$^\star$ \citet{Fong2022} host galaxy class in parentheses, where available, based on the probability $P_\mathrm{cc}$ of a chance association with the SGRB. Gold: $P_\mathrm{cc}\leq 0.02$; Silver: $0.02 <P_\mathrm{cc}\leq 0.1$;  Bronze: $0.1< P_\mathrm{cc}\leq 0.2$.}\\
    \footnotesize{$^\dagger$ Photometric redshift posterior samples retrieved from the BRIGHT online catalog at \url{https://bright.ciera.northwestern.edu/}.}
\end{table*}

The careful selection adopted to construct this sample allowed us to model the underlying selection effects by taking the product of the GBM detection efficiency $P_\mathrm{det,GBM}$ times the simple BAT detection efficiency
\begin{equation}
P_\mathrm{det,BAT} = \Theta\left(p_{[15-150]}(L,\Ep,z)-p_\mathrm{lim,BAT}\right),
\end{equation}
where $p_\mathrm{lim,BAT}=3.5\,\mathrm{cm^{-2}\,s^{-1}}$ is numerically identical to $p_\mathrm{lim,GBM}$ by pure chance.

\subsubsection{Viewing angle sample: GRB~170817A / GW170817}\label{sec:viewing_angle_sample}

\begin{figure}
 \includegraphics[width=\columnwidth]{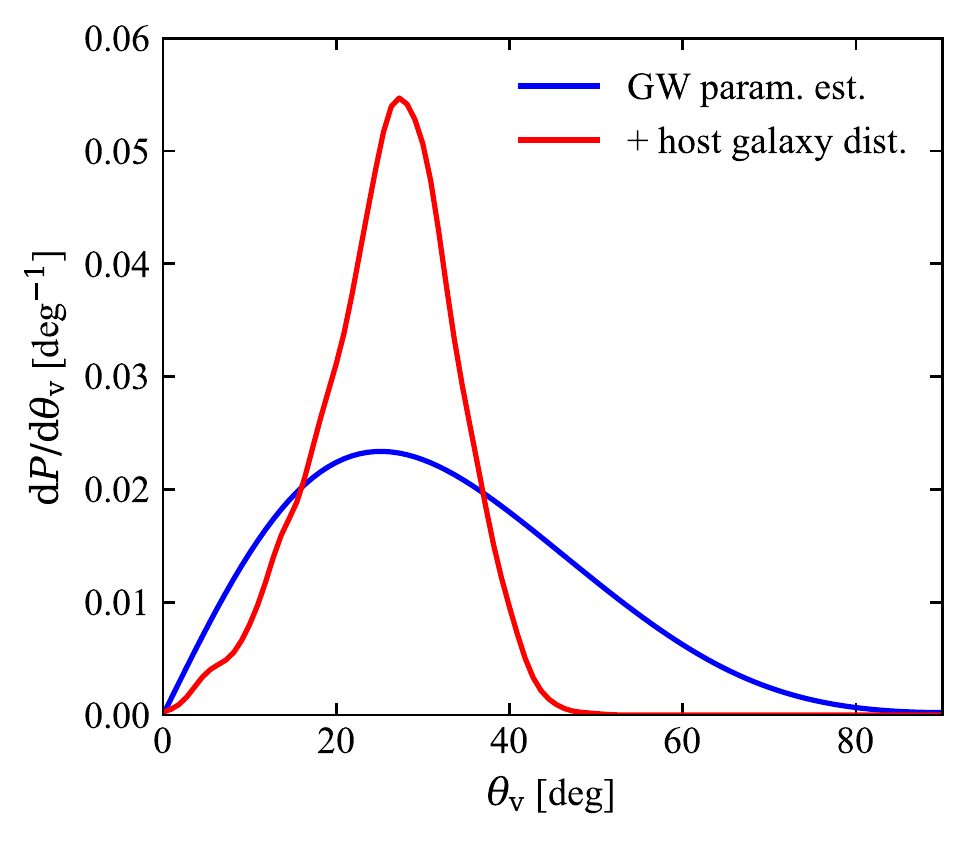}
 \caption{GRB 170817A viewing angle posterior probability distribution, assuming the jet to be aligned with the GW170817 binary total angular momentum. The blue line shows the posterior probability constructed using the posterior samples from the low-spin-prior GW analysis \citep{Abbott2019_GW170817_properties}, while the red line shows the result of conditioning on the host galaxy distance \citep{Cantiello2018} as explained in the text.}
 \label{fig:GW170817_theta_view}
\end{figure}

The third and last sample we considered is that of \textit{Fermi}/GBM SGRBs which have a GW counterpart produced by the inspiral of a BNS merger, from which a measurement of $\thv$ can be obtained under the assumption that the jet is launched along the direction of the total angular momentum. At present, the sample clearly consists of the single event GRB~170817A with its counterpart GW170817. Let us indicate with $d_\mathrm{G17}$ the available data regarding the prompt emission and the GW signal of the event, and with $d_\mathrm{HOST}$ the available data of the host galaxy NGC4993 \citep{Coulter2017,Hjorth2017,Cantiello2018}. We took the host galaxy spectroscopic redshift $z_\mathrm{H}=0.009783$ \citep{Coulter2017,Hjorth2017} as the redshift of the source, neglecting the small uncertainty on the actual cosmological redshift (i.e.\ corrected for the galaxy proper motion), hence $P(z\,|\,d_\mathrm{HOST})=\delta(z-z_\mathrm{H})$. We obtained the posterior $P(L,\Ep\,|\,d_\mathrm{G17},z_\mathrm{H})$, whose 50\% and 90\% containment contours are shown in the left-hand panel of Figure \ref{fig:GBM_SGRBs_with_z}, from our own analysis of the peak spectrum (Appendix \ref{sec:gbm_analysis}). For what concerns the viewing angle, in order to break the distance-inclination degeneracy inherent in the BNS inspiral GW analysis, we proceeded similarly as in \citet[][but keeping the cosmological parameters fixed, differently from them]{Mandel:2018}, as follows: we approximated the host galaxy luminosity distance $r$ uncertainty as
\begin{equation}
 P(r\,|\,d_\mathrm{HOST}) = \frac{\exp\left[-\frac{1}{2}\left(\frac{r-\mu_r}{\sigma_r}\right)^2\right]}{\sqrt{2\pi \sigma_r^2}},
 \label{eq:PrNGC4993}
\end{equation}
with $\mu_r=40.7\,\mathrm{Mpc}$ and $\sigma_r=2.36\,\mathrm{Mpc}$ (mean and square sum of statistical and systematic errors from the measurements performed by \citealt{Cantiello2018}). We then computed the posterior on the viewing angle as
\begin{equation}
 P(\thv\,|\,d_\mathrm{G17},d_\mathrm{HOST}) = \int_0^\infty P(\thv,r\,|\,d_\mathrm{G17}) P(r\,|\,d_\mathrm{HOST})\,\mathrm{d}r,
 \label{eq:Pthv170817}
\end{equation}
where $P(\thv,r\,|\,d_\mathrm{G17})$ is the joint posterior on $\thv$ and $r$ from the GW analysis.  In practice, we obtained samples of the posterior probability density in Eq.~\ref{eq:Pthv170817} by re-sampling the publicly available posterior samples from the low-spin prior analysis of GW170817 performed by \citet{Abbott2019_GW170817_properties} with a weight equal to the right-hand side of Eq.~\ref{eq:PrNGC4993} evaluated at the luminosity distance of each sample. The viewing angle posterior probability density obtained from a kernel density estimate on the resulting samples is shown in Figure \ref{fig:GW170817_theta_view}.

We modelled the selection effects acting on this sample as the product of $P_\mathrm{det,GBM}(L,\Ep,z)$ times a GW detection efficiency $P_\mathrm{det,GW}(\thv,z)$. Since the time-volume surveyed by aLIGO and Advanced Virgo so far is dominated by that of the third observing run O3 \citep{Abbott2021_GWTC3}, we constructed the latter detection efficiency assuming for simplicity the GW network sensitivity of O3, neglecting periods with a lower sensitivity. To do so, we retrieved the dataset made publicly available by the \citet{O3_search_sensitivity} that contains information on the response of online GW search pipelines to a large number of simulated signals injected into O3 data. We re-weighted the BNS merger injections in that dataset to reflect a population with a primary mass distribution $p(m_1)\propto m_1^{-2}$ between $m_\mathrm{1,min}=1.2\,\mathrm{M_\odot}$ and $m_\mathrm{1,max}=2.1\,\mathrm{M_\odot}$ (similar to the preferred distribution from the GWTC-3 population analysis, \citealt{Abbott2021_GWTC3_pop}) and a flat secondary mass distribution $p(m_2\,|\,m_1) \propto \Theta(m_1-m_2)\Theta(m_2-m_\mathrm{1,min})$. Assuming the inclination $\iota$ and the jet viewing angle $\thv$ to be related by
\begin{equation}
 \thv(\iota) = \left\lbrace\begin{array}{lr}
                            \iota & 0\leq \iota < \pi/2 \\
                            \pi-\iota & \pi/2 \leq \iota < \pi
                           \end{array}\right.,
\end{equation}
we binned the injected signals into a number of two-dimensional bins in $(\thv,z)$ space centered at $(\theta_{\mathrm{v},i},z_j)$. Calling $w_k$ and $\rho_k$ the weight and network signal-to-noise ratio (SNR) associated to the $k$-th injected signal, we estimated $P_\mathrm{det,GW}(\thv,z)$ at the center of each bin as
\begin{equation}
 P_\mathrm{det,GW}(\theta_{\mathrm{v},i},z_j)\sim \frac{\sum_{k\in \mathcal{I}_{i,j}}w_{k}\Theta(\rho_{k}\geq 12)}{\sum_{k\in \mathcal{I}_{i,j}}w_{k}},
\end{equation}
where $\mathcal{I}_{i,j}$ represents the set of indices $k$ of injections whose viewing angle and redshift fall into the bin centered at $(\theta_{\mathrm{v},i},z_j)$. The GW detection efficiency on the rest of the $(\thv,z)$ space was obtained by two-dimensional linear interpolation of the resulting values. The relatively high cut $\rho\geq 12$ includes GW170817 \citep{Abbott2017_GW170817_and_GRB170817A} and ensures that the detection can be represented by a simple cut in SNR, in analogy with the flux completeness cuts discussed previously.

\subsection{Inference on the population properties}\label{sec:inference}

Within a Bayesian hierarchical approach, the posterior probability on the population parameters $\lpop$ can be written as \citep[eqs. 7 and 8 in][hereafter \citetalias{Mandel2019}]{Mandel2019}
\begin{equation}
 P(\lpop\,|\,\vec d)=\frac{\pi(\vec\lambda_\mathrm{pop})}{P(\vec{d})}\prod_{i=1}^{N}\frac{\int P(d_i\,|\,\lsource)P_\mathrm{pop}(\lsource\,|\,\lpop)\,\mathrm{d}\lsource}{\int P_\mathrm{pop}(\lsource\,|\,\lpop)P_\mathrm{det}(\lsource)\,\mathrm{d}\lsource},
 \label{eq:Plpop}
\end{equation}
where the index $i$ runs over the $N$ events in the sample, $d_i$ (the $i$-th element of $\vec d$) represents the corresponding data in the detectors (i.e.\ \textit{Fermi}/GBM, plus either \textit{Swift}/BAT or the aLIGO and Advanced Virgo detectors in our case), $\pi(\vec\lambda_\mathrm{pop})$ is the prior on the population parameters, and $P(\vec{d})=\prod_{i=1}^N P(d_i)$ is a normalization factor. We indicate with $\mathcal{N}_i=\mathcal{N}_i(d_i,\vec\lambda_\mathrm{pop})$ the numerator of the fraction after the product symbol in the above equation, that is
\begin{equation}
 \mathcal{N}_i = \int P(d_i\,|\,\lsource)P_\mathrm{pop}(\lsource\,|\,\lpop)\,\mathrm{d}\lsource,
\end{equation}
and with $\mathcal{D}=\mathcal{D}(\vec\lambda_\mathrm{pop})$ the denominator, namely
\begin{equation}
 \mathcal{D} = \int P_\mathrm{pop}(\lsource\,|\,\lpop)P_\mathrm{det}(\lsource)\,\mathrm{d}\lsource.
\end{equation}
The latter contains the detection efficiency $P_\mathrm{det}(\lsource)$, which must be chosen consistently with the selection effects acting on each of the considered samples.

\subsubsection{Numerator $\mathcal{N}_i$ for observer-frame sample events}

For events in our observer-frame sample, which have unknown redshift, we approximated the posterior probability on the source parameters $P(L,E_\mathrm{p},z\,|\,d_i) = P(d_i\,|\,\vec\lambda_\mathrm{src})\pi(L,E_\mathrm{p})\pi(z)/P(d_i)$ by neglecting the uncertainty on the measured peak photon flux $p_i=p_{[50-300],i}$ and peak photon energy $E_\mathrm{p,obs,i}$ (which is justified by the large sample size and by the quality cuts discussed in the previous section), and by assuming the posterior to simply reflect the prior $\pi(z)$ along the $z$ axis. In other words, we assumed $P(L,E_\mathrm{p},z\,|\,d_i)\propto \pi(z)$ when keeping $L$ and $E_\mathrm{p}$ fixed: this is equivalent to stating that no redshift information is available. This leads to
\begin{equation}
 P(d_i\,|\,\vec\lambda_\mathrm{src})\sim \frac{\delta\left(L-4\pi d_\mathrm{L}^2\mathcal{E} p_i\right)\delta\left(E_\mathrm{p}-(1+z)E_{\mathrm{p,obs},i}\right)}{\pi(L,E_\mathrm{p})}P(d_i),
\end{equation}
The appropriate prior $\pi(L,E_\mathrm{p})$ was obtained by applying a coordinate transform to a uniform prior in both $p_i$ and $E_\mathrm{p,obs}$ for these events, that is
\begin{equation}
 \pi(L,E_\mathrm{p})=\begin{array}{|cc|}
                      \frac{\partial p_i}{\partial L} & \frac{\partial p_i}{\partial E_\mathrm{p}}\\
                      \frac{\partial E_{\mathrm{p,obs},i}}{\partial L} & \frac{\partial E_{\mathrm{p,obs},i}}{\partial E_\mathrm{p}}
                     \end{array}\,\pi(p_i,E_{\mathrm{p,obs},i}) \propto \frac{p_i}{ (1+z)\hat L_i(z)},
\label{eq:prior_noredshift}
\end{equation}
where $|X|$ represents the determinant of matrix $X$, and $\hat L_i = 4\pi d_\mathrm{L}^2(z)\mathcal{E}_{[50-300]}\left((1+z)E_{\mathrm{p,obs},i},z\right)p_i$.
The integral in the numerator of Eq.~\ref{eq:Plpop} is then
\begin{equation}
\mathcal{N}_i \sim  P(d_i)\int_0^\infty (1+z)\frac{\hat L_i(z)}{p_i} P_\mathrm{pop}\left(\hat L_i(z),(1+z)E_{\mathrm{p,obs},i},z\,|\,\vec\lambda_\mathrm{pop}\right)\,\mathrm{d}z,
\end{equation}
and we note that the $P(d_i)$ term eventually cancels out in Eq.~\ref{eq:Plpop} (as in \citetalias{Mandel2019}).

\subsubsection{Numerator $\mathcal{N}_i$ for rest-frame sample events}

For events in the rest-frame sample, in order to carry out the integral over $z$, $L$ and $\Ep$ in $\mathcal{N}_i$, we employed the same Monte Carlo approximation as \citetalias{Mandel2019}, that is
\begin{equation}
\mathcal{N}_i \sim \frac{P(d_i)}{N_\mathrm{s}}\sum_{j=1}^{N_\mathrm{s}}\frac{P_\mathrm{pop}(L_{i,j},E_{\mathrm{p},i,j},z_{i,j}\,|\,\lpop)}{\pi(L_{i,j},E_{\mathrm{p},i,j})},
\label{eq:numerator_knownredshift}
\end{equation}
where $\lbrace L_{i,j} , E_{\mathrm{p},i,j}, z_{i,j}\rbrace_{j=1}^{N_\mathrm{s}}$ represent a total of $N_\mathrm{s}$ samples of the luminosity, peak photon energy and redshift posterior of the GRB. In cases with a photometric redshift, the latter was obtained by combining the results from our analysis of the spectrum at the peak of the GRB with the redshift posterior samples from \citet{Nugent2022}, obtained from the BRIGHT online catalog\footnote{\url{https://bright.ciera.northwestern.edu/}}. In cases with a spectroscopic redshift, we simply kept $z_{i,j}$ fixed at the best-fit redshift.
The prior $\pi(L,E_{\mathrm{p}})$ in Eq.~\ref{eq:numerator_knownredshift} is proportional to $(1+z)^{-1}L^{-1}$, as discussed in \S\ref{sec:restframe_sample}.

\subsubsection{Numerator $\mathcal{N}_i$ for viewing angle sample events}

For the viewing angle sample, that is, for GRB~170817A, it was necessary to proceed differently between the full sample and flux-limited sample analyses: the peak photon flux $p_{[50-300]}=2.02\,\mathrm{cm^{-2}s^{-1}}$ of the GRB is below the completeness cut $p_\mathrm{lim,GBM}=3.5\,\mathrm{cm^{-2}s^{-1}}$ adopted in the flux-limited sample analysis, hence in that case the simple treatment of GBM selection effects is not adequate. In the full sample analysis, instead, this is not a problem as the GBM detection efficiency model from Appendix \ref{sec:pdet_construction} accounts for the smooth decrease of the detection efficiency at low peak photon fluxes.  In the latter case, the correct form of $\mathcal{N}_i$ can be obtained by reformulating our population model by including $\thv$ among the source parameters, $\lsource^\star=(L,\Ep,z,\thv)$. The population probability then becomes $P_\mathrm{pop}^\star(L,\Ep,z,\thv\,|\,\lpop)=P(L,\Ep\,|\,\thv,\lpop)\sin\thv P(z\,|\,\lpop)$, and the integral over $\thv$ is deferred to $\mathcal{N}_i$, namely
\begin{equation}
\begin{split}
 & \mathcal{N}_{i} = \iiiint P(d_i\,|\,L,E_\mathrm{p},z,\thv)\times \\
 & \times P(L,E_\mathrm{p}\,|\,\thv,\lpop)\sin\thv\, P(z\,|\,\lpop)\,\mathrm{d}L\,\mathrm{d}E_\mathrm{p}\,\mathrm{d}z\,\mathrm{d}\thv.
\end{split}\label{eq:N_i_GW170817}
\end{equation}
It is straightforward to verify that whenever $d_i$ does not contain information on the viewing angle this leads to exactly the same definition of $\mathcal{N}_i$ as before.

For GRB~170817A, we have (see \S\ref{sec:viewing_angle_sample})
\begin{equation}
\begin{split}
& P(d_\mathrm{G17},d_\mathrm{HOST}\,|\,L,\Ep,z,\thv) = \frac{P(\thv\,|\,d_\mathrm{G17},d_\mathrm{HOST})}{\pi(\thv)}\times\nonumber\\
& \frac{P(L,\Ep\,|\,d_\mathrm{G17},z)}{\pi(L,\Ep,z)}\delta(z-z_\mathrm{H})P(d_\mathrm{G17},d_\mathrm{HOST}),
 \end{split}
\end{equation}
where the prior $\pi(\thv)=\sin\thv$ corresponds to that used in the GW analysis. This leads to
\begin{equation}
 \begin{split}
 & \frac{\mathcal{N}_\mathrm{G17}}{P(d_\mathrm{G17},d_\mathrm{HOST})} = \iiint P(\thv\,|\,d_\mathrm{G17},d_\mathrm{HOST})\frac{P(L,E_\mathrm{p},z_\mathrm{H}\,|\,d_\mathrm{G17})}{\pi(L,E_\mathrm{p},z_\mathrm{H})}\times\nonumber\\
 & \times P(L,E_\mathrm{p}\,|\,\thv,\lpop)P(z_\mathrm{H}\,|\,\lpop)\,\mathrm{d}L\,\mathrm{d}E_\mathrm{p}\,\mathrm{d}\thv.
 \end{split}
\label{eq:N_GW170817_final}
\end{equation}
Approximating again the integral with a Monte Carlo sum, we obtained
\begin{equation}
  \frac{\mathcal{N}_\mathrm{G17}}{P(d_\mathrm{G17},d_\mathrm{HOST}) }\sim \frac{P(z_\mathrm{H}\,|\,\lpop)}{N_\mathrm{g}N_\mathrm{s}}\sum_{j=1}^{N_\mathrm{g}}\sum_{k=1}^{N_\mathrm{s}}\frac{P(L_k,E_{\mathrm{p},k}\,|\,\theta_{\mathrm{v},j},\lpop)}{\pi(L_k,E_{\mathrm{p},k},z_\mathrm{H})},
\end{equation}
where $\lbrace (L_k,E_{\mathrm{p},k})\rbrace_{k=1}^{N_\mathrm{s}}$ are a total of $N_\mathrm{s}$ samples from our analysis of the peak spectrum of GRB~170817A and $\lbrace \theta_{\mathrm{v},j}\rbrace_{j=1}^{N_\mathrm{g}}$ are $N_\mathrm{g}$ posterior samples of the probability in Eq.~\ref{eq:Pthv170817}. Clearly, the term $P(d_\mathrm{G17},d_\mathrm{HOST})$ eventually cancels out in Eq.~\ref{eq:Plpop} as before.

In the flux-limited sample analysis, the information on the viewing angle, luminosity and peak photon energy of GRB~170817A/GW170817 can still be used to condition the apparent structure parameters \textit{before} the flux-limited sample analysis is performed, because the structure must be consistent with what has been observed in that event. The starting point is
\begin{equation}
\begin{split}
 & P(\lpop\,|\,d_\mathrm{G17},d_\mathrm{HOST})=\\
 & \iiint P(\lpop\,|\,L,\Ep,\thv)P(L,\Ep,\thv\,|\,d_\mathrm{G17},d_\mathrm{HOST})\,\mathrm{d}L\,\mathrm{d}\Ep\,\mathrm{d}\thv.
\end{split}
\label{eq:lpop_GW17}
\end{equation}
Application of Bayes' theorem gives
\begin{equation}
 P(\lpop\,|\,L,\Ep,\thv) = \frac{P(L,\Ep,\thv\,|\,\lpop)\pi(\lpop)}{\pi(\thv)\pi(L,E_\mathrm{p})},
\end{equation}
but we have $P(L,\Ep,\thv\,|\,\lpop)=P(L,\Ep\,|\,\thv,\lpop)\sin\thv$ and $\pi(\thv)=\sin\thv$, which therefore leads to $P(\lpop\,|\,L,\Ep,\thv) = P(L,\Ep\,|\,\thv,\lpop)\pi(\lpop)/\pi(L,E_\mathrm{p})$. Substitution of this into Eq.~\ref{eq:lpop_GW17} leads to
\begin{equation}
 \begin{split}
 & P(\lpop\,|\,d_\mathrm{G17},d_\mathrm{HOST}) = \pi(\lpop) \iiint P(L,\Ep,\thv\,|\,d_\mathrm{G17},d_\mathrm{HOST}) \times \\
 & \times \frac{P(L,\Ep\,|\,\thv,\lpop)}{\pi(L,\Ep)} \,\mathrm{d}L\,\mathrm{d}\Ep\,\mathrm{d}\thv.
 \end{split}
\end{equation}
This is similar to $\mathcal{N}_\mathrm{G17}$, but the redshift information is not used. This can be again approximated by Monte Carlo integration over samples drawn from the posterior $P(L,\Ep,\thv\,|\,d_\mathrm{G17},z_\mathrm{H})$, namely
\begin{equation}
 P(\lpop\,|\,d_\mathrm{G17},d_\mathrm{HOST})\sim \frac{\pi(\lpop)}{N_\mathrm{s}N_\mathrm{g}} \sum_{k=1}^{N_\mathrm{s}}\sum_{j=1}^{N_\mathrm{g}}\frac{P(L_k,E_{\mathrm{p},k}\,|\,\theta_{\mathrm{v},j},\lpop)}{\pi(L_k,E_{\mathrm{p},k},z_\mathrm{H})}.
\end{equation}
This result can then be used as a prior in the analysis of the observer-frame and rest-frame samples. The final posterior on $\lpop$ then becomes
\begin{equation}
\begin{split}
& P_\mathrm{flux-limited}(\lpop\,|\,\vec d,d_\mathrm{G17},d_\mathrm{HOST})\sim\frac{\pi(\vec\lambda_\mathrm{pop})}{P(\vec{d})}\left(\prod_{i=1}^{N}\frac{\mathcal{N}_i}{\mathcal{D}}\right)\times\nonumber\\
& \times\frac{1}{N_\mathrm{s}N_\mathrm{g}}\sum_{k=1}^{N_\mathrm{s}}\sum_{j=1}^{N_\mathrm{g}}\frac{P(L_k,E_{\mathrm{p},k}\,|\,\theta_{\mathrm{v},j},\lpop)}{\pi(L_k,E_{\mathrm{p},k},z_\mathrm{H})}.
\end{split}
 \label{eq:Plpop+GW17-fluxlimited}
\end{equation}
Hence, the posterior takes a similar form as in the full-sample analysis case, the difference being in a missing $P(z_\mathrm{H}\,|\,\lpop)/\mathcal{D}$ factor.

\subsubsection{Denominator $\mathcal{D}$ for events in the three samples}

The denominator $\mathcal{D}$ in the above expressions represents the fraction of events in the population that pass the sample selection criteria \citepalias{Mandel2019}, i.e.\ the `accessible' events according to the selection effects model. For the observer-frame sample events, the denominator takes the form
\begin{equation}
 \mathcal{D}(\lpop)=\iiint P_\mathrm{pop}\left(L,E_\mathrm{p},z\,|\,\vec\lambda_\mathrm{pop}\right)P_\mathrm{det,GBM}\left(L,\Ep,z\right)\mathrm{d}L\,\mathrm{d}E_\mathrm{p}\,\mathrm{d}z;
\end{equation}
For the rest-frame sample, it is
\begin{equation}
\begin{split}
& \mathcal{D}(\lpop)=\iiint P_\mathrm{pop}\left(L,E_\mathrm{p},z\,|\,\vec\lambda_\mathrm{pop}\right)\times \\
& \times P_\mathrm{det,GBM}\left(L,\Ep,z\right)P_\mathrm{det,BAT}\left(L,\Ep,z\right)\mathrm{d}L\,\mathrm{d}E_\mathrm{p}\,\mathrm{d}z;
\end{split}
\end{equation}
For the viewing-angle sample, it is a four-dimensional integral, namely
\begin{equation}
\begin{split}
& \mathcal{D}(\lpop)=\iiiint P_\mathrm{pop}^\star\left(L,E_\mathrm{p},z,\thv\,|\,\vec\lambda_\mathrm{pop}\right)\times \\
& \times P_\mathrm{det,GBM}\left(L,\Ep,z\right)P_\mathrm{det,GW}\left(\thv,z\right)\mathrm{d}L\,\mathrm{d}E_\mathrm{p}\,\mathrm{d}z\,\mathrm{d}\thv.
\end{split}
\end{equation}

\subsection{Choice of priors}\label{sec:priors}

Our Bayesian inference approach requires the definition of a prior probability density over the `hyper' parameter space the $\lpop$ vector belongs to. We chose independent priors on all parameters, except for the pair $(\theta_\mathrm{c},\theta_\mathrm{w})$ for which we enforce $\theta_\mathrm{w}>\theta_\mathrm{c}$, hence
\begin{equation}
 \pi(\lpop)=\pi(\theta_\mathrm{c},\theta_\mathrm{w})\pi(L_\mathrm{c}^\star)\pi(\alpha_\mathrm{L})\,...\,\pi(z_\mathrm{p}).
\end{equation}

As shown in Table \ref{tab:parameters}, we selected mildly informative priors on most parameters, typically adopting a uniform or uniform-in-log prior over a relatively wide range that encompasses what we consider reasonable values. There are two exceptions: for the core half-opening angle $\theta_\mathrm{c}$ and the break angle $\theta_\mathrm{w}$ we adopted a prior that is uniform in the subtended solid angle, $\pi(\theta)\propto\sin\theta$, since their role is that of setting the probability for a jet to be observed within the core or within the `wing' where the break occurs; for the low-end cutoff of the core luminosity $L_\mathrm{c}^\star$, we set a relatively high lower limit $L_\mathrm{c}^\star\geq 3\times 10^{51}\,\mathrm{erg\,s^{-1}}$, consistently with our quasi-universal structured jet scenario, where relatively low luminosities are produced by off-axis jets and not by intrinsically underluminous jets.

\subsection{Stochastic sampling of the population posterior}
In order to realise our inference in practice, we sampled the posterior probability density on $\lpop$ in Eq.~\ref{eq:Plpop} using the publicly available \texttt{python} package \texttt{emcee} \citep{emcee}, which constitutes an efficient and flexibile implementation of the \citet{Goodman2010} affine-invariant ensemble sampler. Numerical integrals were performed using the trapezoidal rule over a four-dimensional grid with a resolution of 1000 linearly-spaced points in $\thv$ between 0 and $\pi/2$ and 60 logarithmically-spaced points in each of the remaining axes, in the domain $L\in (10^{44},10^{56})\,\mathrm{erg\,s^{-1}}$, $\Ep\in(10^{-1},10^{7})\,\mathrm{keV}$ and $z\in(10^{-3},10^{1})$. We ensured that this resolution was sufficient by comparing the value of the posterior at a number of points in the parameter space with those obtained with a higher resolution of 100 points over each axis, and found a negligible difference as long as $A<10$ and $\sigma_\mathrm{c}\gtrsim 0.05$ dex (our prior limits fall well within these requirements).

We set up the sampler to employ $14\times 2=28$ walkers, and ran it for $10^4$ iterations for each analysis, for a total of $2.8\times 10^5$ posterior samples. The autocorrelation length of the resulting chains, averaged over all walkers, is around 650. A corner plot showing the density of the samples in the 14-dimensional parameter space is shown in Fig.~\ref{fig:corner_full}, while summary statistics for each parameter are reported in Table \ref{tab:fit_results_table}. The results presented in the next section are constructed using 1000 random posterior samples from these chains, after discarding the first half as burn-in.


\begin{table*}
\caption{Constraints on population model parameters from our analysis. For each parameter, and for each analysis set-up (full sample; flux-limited sample) we report the median of the marginalized posterior probability density and the symmetric 90\% credible interval. A horizontal line separates actual model parameters from derived ones.  The latter are: the typical luminosity at the `break' angle, $L_\mathrm{w}=\bar L(\theta_\mathrm{w})$; the slope $2/\alpha_\mathrm{L}$ of the luminosity function $\phi(L)=\mathrm{d}P/\mathrm{d}\ln L$ for jets seen between $\theta_\mathrm{c}$ and $\theta_\mathrm{w}$; the slope $2/\beta_\mathrm{L}$ for jets seen at larger viewing angles; the slope $\alpha_{\Ep}/\alpha_\mathrm{L}$ of the `Yonetoku' correlation for off-axis jets with $\theta_\mathrm{c}<\theta_\mathrm{v}<\theta_\mathrm{w}$; the slope of the `Yonetoku' correlation for $\theta_\mathrm{v}>\theta_\mathrm{w}$. }
\label{tab:fit_results_table}
\centering
\renewcommand*{\arraystretch}{1.2}
\begin{tabular}{ccc}
 \hline
 Parameter & Full-sample & Flux-limited sample \\
 \hline
$\theta_\mathrm{c}/\mathrm{deg}$  &  $2.1_{-1.4}^{+2.4}$  &  $3.0_{-2.1}^{+2.4}$ \\
$\theta_\mathrm{w}/\mathrm{deg}$  &  $63.4_{-51.5}^{+24.4}$  &  $64.5_{-42.5}^{+23.2}$ \\
$L_\mathrm{c}/\mathrm{10^{52}\,erg\,s^{-1}}$  &  $0.5_{-0.2}^{+1.1}$  &  $0.5_{-0.1}^{+0.7}$ \\
$\alpha_L$  &  $4.7_{-1.4}^{+1.2}$  &  $4.9_{-1.7}^{+1.0}$ \\
$\beta_L$  &  $1.6_{-4.3}^{+4.0}$  &  $1.9_{-4.4}^{+3.7}$ \\
$E_\mathrm{p,c}/\mathrm{MeV}$  &  $17.7_{-10.5}^{+39.3}$  &  $4.5_{-2.3}^{+6.3}$ \\
$\alpha_{E_\mathrm{p}}$  &  $1.9_{-0.9}^{+1.1}$  &  $1.5_{-0.9}^{+1.0}$ \\
$\beta_{E_\mathrm{p}}$  &  $1.1_{-3.6}^{+4.3}$  &  $1.5_{-4.1}^{+4.0}$ \\
$A$  &  $3.2_{-0.4}^{+0.7}$  &  $2.9_{-0.4}^{+0.7}$ \\
$\sigma_\mathrm{c}/\mathrm{dex}$  &  $0.4_{-0.1}^{+0.1}$  &  $0.4_{-0.1}^{+0.1}$ \\
$y$  &  $-0.3_{-0.4}^{+0.3}$  &  $-0.0_{-0.3}^{+0.3}$ \\
$a$  &  $4.6_{-0.8}^{+0.4}$  &  $3.8_{-1.1}^{+1.0}$ \\
$b$  &  $5.3_{-3.8}^{+4.2}$  &  $5.5_{-4.0}^{+4.1}$ \\
$z_\mathrm{p}$  &  $2.2_{-0.6}^{+0.6}$  &  $2.3_{-0.8}^{+0.6}$ \\
\hline
$\log(L_\mathrm{w}/\mathrm{erg\,s^{-1}})$  &  $45.5_{-2.0}^{+2.1}$  &  $45.7_{-1.4}^{+1.7}$ \\
$2/\alpha_L$  &  $0.4_{-0.1}^{+0.2}$  &  $0.4_{-0.1}^{+0.2}$ \\
$2/\beta_L$  &  $0.4_{-4.4}^{+4.1}$  &  $0.5_{-4.6}^{+4.2}$ \\
$\alpha_\mathrm{E_\mathrm{p}}/\alpha_L$  &  $0.4_{-0.2}^{+0.1}$  &  $0.3_{-0.2}^{+0.1}$ \\
$\beta_\mathrm{E_\mathrm{p}}/\beta_L$  &  $0.1_{-5.4}^{+5.2}$  &  $0.1_{-5.8}^{+5.9}$ \\
$R_0/\mathrm{Gpc^{-3}\,yr^{-1}}$ & $740_{-630}^{+3870}$ & $180_{-145}^{+660}$ \\
$R_0(L>10^{50}\,\mathrm{erg\,s^{-1}})/\mathrm{Gpc^{-3}\,yr^{-1}}$ & $3.6_{-2.5}^{+6.9}$ & $1.3_{-0.7}^{+2.0}$ \\
\end{tabular}

\end{table*}

\section{Results}\label{sec:results}

Generally speaking, the full sample and flux-limited sample analyses yielded quite similar results, as demonstrated by the detailed posterior probability density distributions (Figure \ref{fig:corner_full}) and the summary in Table \ref{tab:fit_results_table}. The main notable differences in the full sample analysis \textit{versus} the flux-limited sample analysis are a preference for a slightly larger on-axis SED peak photon energy $E_\mathrm{p,c}$; a steeper slope $a$ of the rate density evolution before peak; and a slightly better constrained redshift $z_\mathrm{p}$ of the peak of the rate density evolution. The posterior distributions from the two analyses show large overlaps for all parameters. In what follows, we present a thorough description of the results, further highlighting the differences in the results from the two analyses when relevant.

\subsection{Apparent structure}

\begin{figure}
 \centering
 \includegraphics[width=\columnwidth]{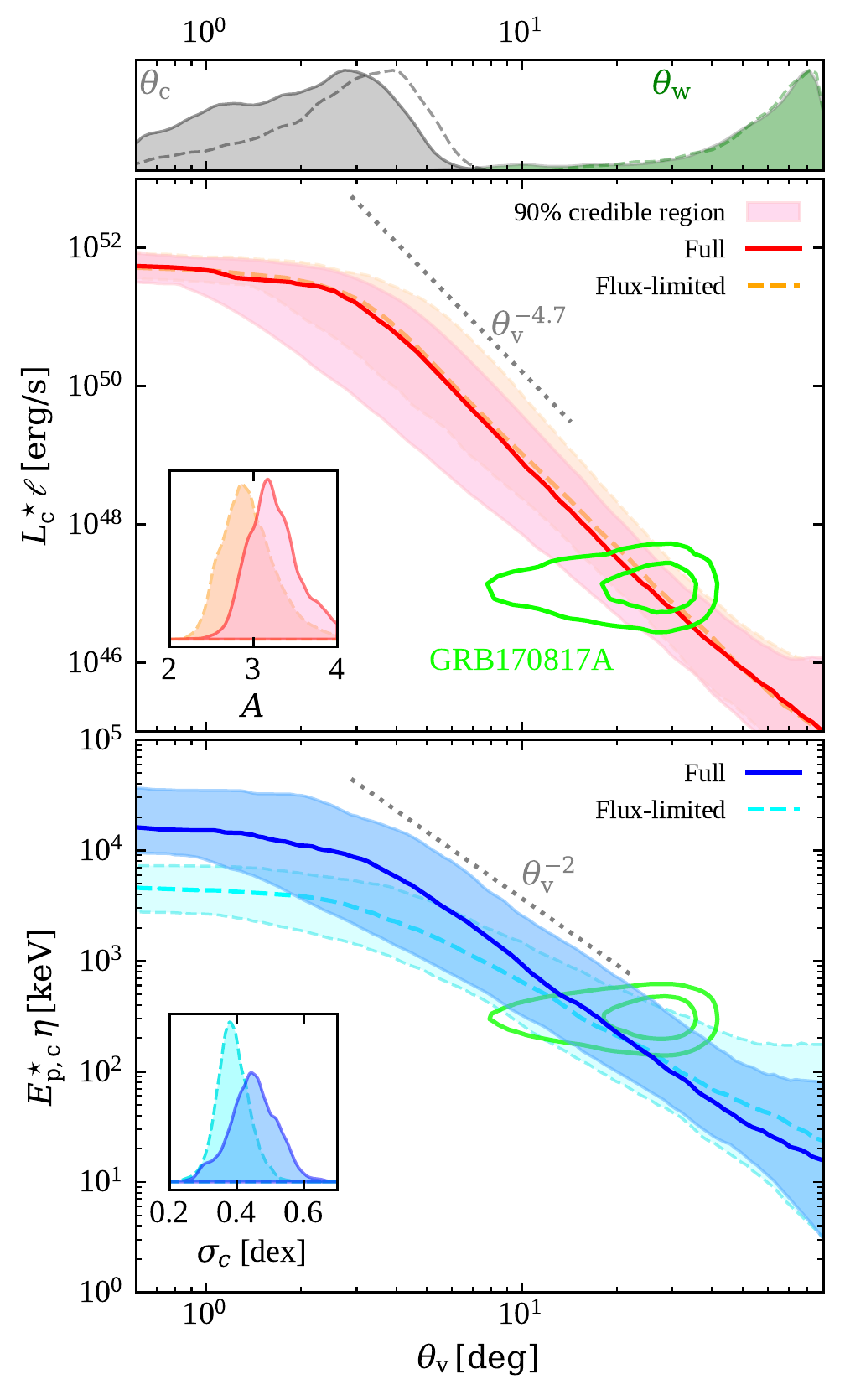}
 \caption{Average apparent jet structure. The two larger panels show the constraints on the SGRB average apparent structure obtained from our analysis: solid lines show the medians of the projected posterior distributions of the jet structure functions at fixed $\thv$ from the full sample analysis (red: $\bar L=L_\mathrm{c}^\star\ell(\thv)$; blue: $\bar E_\mathrm{p}=E_\mathrm{p,c}^\star \eta(\thv)$),
 while dashed lines refer to the flux-limited sample (orange: $\bar L$; cyan: $\bar E_\mathrm{p}$). The shaded region around each line encompasses the corresponding 90\% credible range at each fixed $\thv$. Dotted lines show reference power law trends. The light green contours show the 50\% and 90\% credible regions of $(L,\thv)$ (middle panel) and $(\Ep,\thv)$ (bottom panel)  for  GRB~170817A (\S\ref{sec:viewing_angle_sample}).  The insets show the posterior distributions on the parameters $A$ and $\sigma_c$ which determine the dispersion around the average structure within the population. The top smaller panel shows the posterior distributions on the logarithms of the transition angles, $\ln(\theta_\mathrm{c})$ (grey) and $\ln(\theta_\mathrm{w})$ (green), with the results for the full sample shown as filled areas, and those for the flux-limited sample shown as dashed lines.}
 \label{fig:tildeL_tilde_Ep}
\end{figure}

\begin{figure}
 \centering
 \includegraphics[width=\columnwidth]{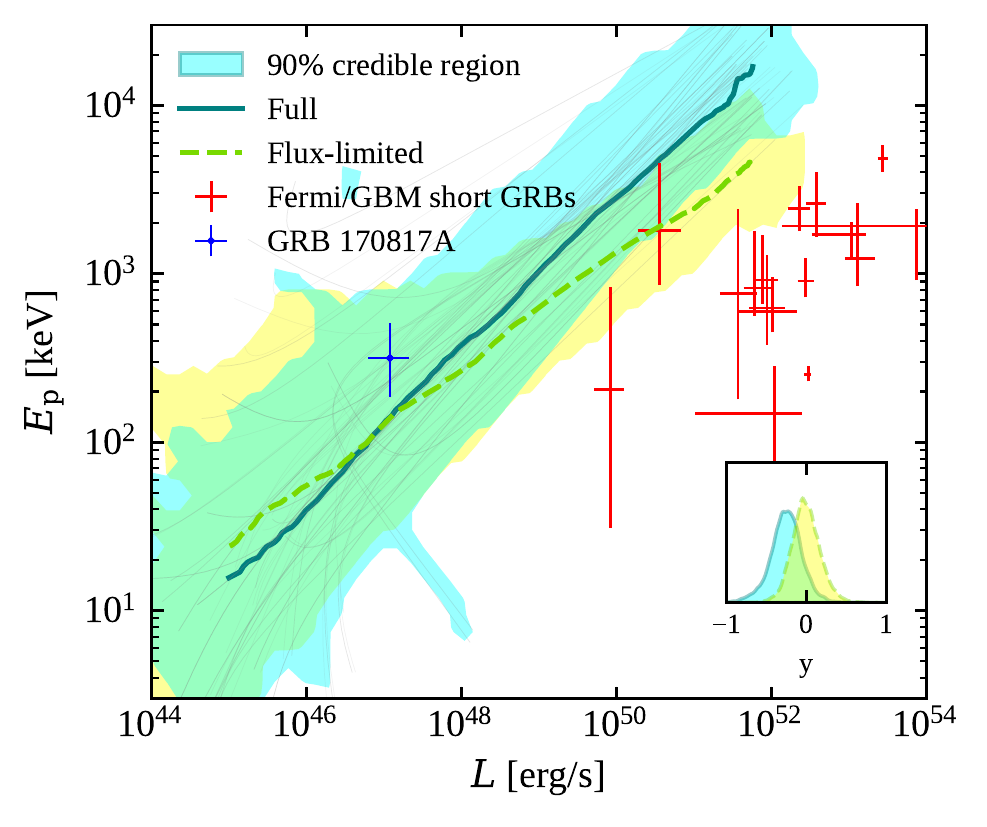}
 \caption{Average apparent jet structure in the $L-E_\mathrm{p}$ plane. The turquoise solid line connects the points $\left(\langle\bar L(\theta_\mathrm{v})\rangle,\langle\bar E_\mathrm{p}(\theta_\mathrm{v})\rangle\right)$ at varying viewing angles, where $\langle\cdot\rangle$ represents the median, from the full sample analysis. The turquoise shaded area encompasses the corresponding 90\% credible range. Grey lines show 100 posterior samples of the lines $\left(\bar L(\theta_\mathrm{v}),\bar E_\mathrm{p}(\theta_\mathrm{v})\right)$ from the same analysis.  The green dashed line and yellow area show the corresponding results for the flux-limited sample analysis. The crosses mark the positions of the \textit{Fermi}/GBM SGRBs with known redshift (all in red except for GRB~170817A, which is shown in blue). The inset shows the posterior probability density on the $y$ parameter that sets the slope of the correlation between the on-axis luminosity and the on-axis SED peak photon energy, with the results of the two analyses color coded as in the main panel. }
 \label{fig:Yonetoku}
\end{figure}

First of all, we focus on the apparent jet structure. Figure \ref{fig:tildeL_tilde_Ep} shows the constraint we obtained on the `average apparent jet structure' $\bar L=L_\mathrm{c}^\star\ell$ (central panel) and $\bar E_\mathrm{p}=E_\mathrm{p,c}^\star\eta$ (bottom panel) from the two analyses, with the insets showing the posterior distribution on the core dispersion parameters $A$ and $\sigma_\mathrm{c}$. In this and the following figures, solid lines refer to the full-sample analysis and dashed lines to the flux-limited sample analysis. The small panel at the top shows the posterior probability density on the logarithm of $\theta_\mathrm{c}$ (in grey) and of $\theta_\mathrm{w}$ (in green). The figure shows that the preferred apparent structure features a narrow core of $\sim 2-3 \,\mathrm{deg}$, outside of which the luminosity falls off approximately as $\theta_\mathrm{v}^{-4.7}$ and the SED peak photon energy as $\theta_\mathrm{v}^{-2}$. The posterior on the transition angle $\theta_\mathrm{w}$ rails against the upper boundary of its physical range, and the slopes $\beta_L$ and $\beta_{\Ep}$ at larger viewing angles are not well constrained by the available data (see Table \ref{tab:fit_results_table}). This indicates that the data are consistent with an apparent structure described by a single power law outside the core, which disfavours somewhat the presence of a distinct dissipation mechanism that dominates the gamma-ray emission at large viewing angles (e.g.\ cocoon shock breakout as proposed by \citealt{Gottlieb2018}). The structure is consistent by construction with the GRB~170817A luminosity and $\Ep$ at the relevant viewing angle, as shown by the green contours in the figure. The latter represents the 50\% and 90\% credible regions constructed using the viewing angle information from the GW analysis conditioned on the host galaxy distance (\S\ref{sec:viewing_angle_sample}) and our GRB~170817A spectral analysis at peak (Appendix \ref{sec:gbm_analysis}).




The jet structure functions can also be projected onto the $(L,\Ep)$ plane: Figure \ref{fig:Yonetoku} shows the median $(\bar L,\bar E_\mathrm{p})$ relation and the corresponding 90\% credible region for the two analyses. In both cases, the bulk of the SGRBs with known redshift (shown by red crosses) are close to the upper-right end of the relation, which indicates that they are observed close to on-axis in the model (more precisely, close to the edge of the core, see also \S\ref{sec:viewing_angles}).  Interestingly, the relation lies above most of the SGRBs in the known-redshift subsample, indicating that selection effects play a major role in how the plane is populated, according to the model. We expand on this in \S\ref{sec:Yonetoku_sel_effects}.

The inset in Figure \ref{fig:Yonetoku} shows the posterior probability density on the $y$ parameter that sets the correlation between the on-axis luminosity and peak SED photon energy. Despite our model allowing for such correlation, the posterior is fully compatible with $y=0$, that is, the absence of an intrinsic correlation between $L_\mathrm{c}$ and $E_\mathrm{p,c}$.

\subsection{Luminosity function}

\begin{figure}
 \centering
 \includegraphics[width=\columnwidth]{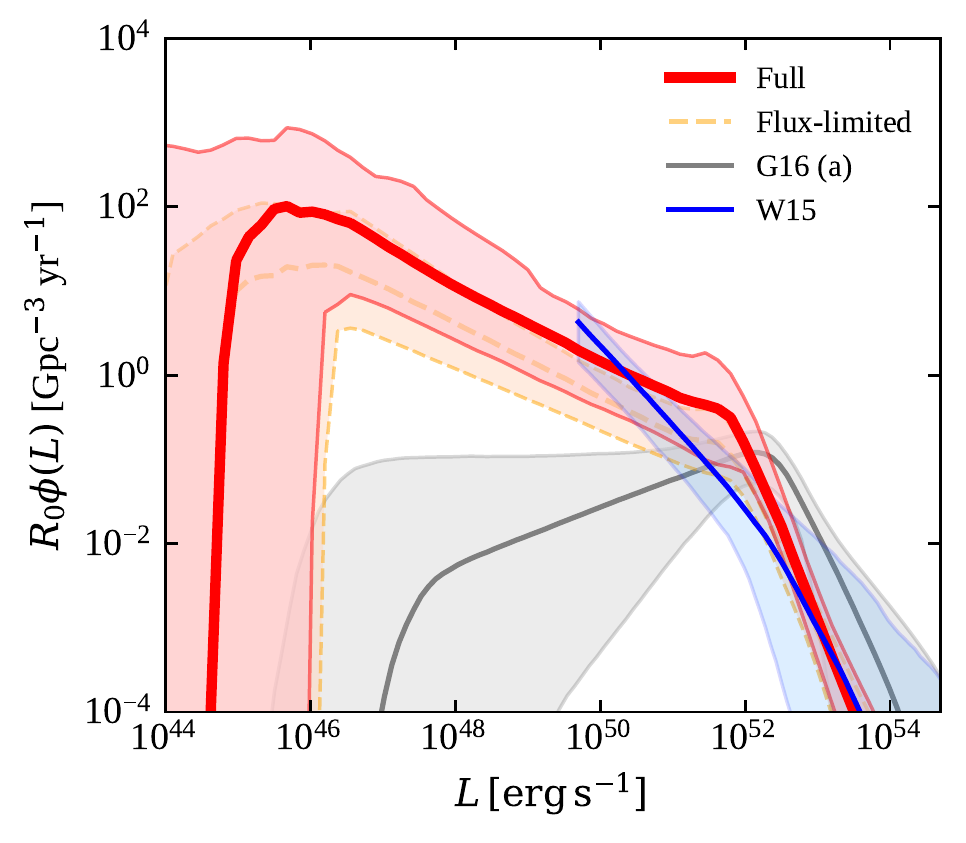}
 \caption{Luminosity function. The median of the posterior probability density of $R_0\phi(L)=\mathrm{d}R_0/\mathrm{d}\ln(L)$ (i.e.\ the local rate density per unit logarithm of the peak luminosity) from the full sample analysis is shown with a red solid line, while the orange dashed line refers to the flux-limited sample analysis. The shaded regions encompass the symmetric 90\% credible interval at each fixed $L$. We show for comparison the corresponding results from \citetalias{Wanderman2015} (blue) and \citetalias{Ghirlanda2016} (grey -- their fiducial model `a'). The credible region of \citetalias{Wanderman2015} is computed assuming uncorrelated errors on their parameters. }
 \label{fig:lum_func}
\end{figure}

Our analysis does not directly constrain the local rate density $R_0$ of SGRBs, because in our inference framework the latter is effectively only a normalization factor, as long as its prior is uniform in the logarithm (\citetalias{Mandel2019}; \citealt{Fischbach2018}). In order to derive the local rate density implied by our results, we required the observed rate of events with $p>p_\mathrm{lim}$ to be equal to\footnote{We are effectively neglecting here the Poisson error on the observed rate, which has a small impact on $R_0$ with respect to the uncertainty on $\lpop$.} $R_\mathrm{obs}(>p_\mathrm{lim,GBM})=N(>p_\mathrm{lim,GBM})/\eta_\mathrm{GBM}T_\mathrm{GBM}\approx 27.6\,\mathrm{yr^{-1}}$, where $N(>p_\mathrm{lim,GBM})=212$ is the number of SGRBs with $p>p_\mathrm{lim,GBM}$ in our sample, $\eta_\mathrm{GBM}=0.59$ is a factor that corrects for the accessible field of view and the duty cycle of GBM \citep{Burns2016}, and $T_\mathrm{GBM}=13\,\mathrm{yr}$ is the \textit{Fermi} mission duration at the time of the last SGRB in the observer-frame sample. Given $\lpop$, the local rate density is then
\begin{equation}
 R_0=\frac{R_\mathrm{obs}(>p_\mathrm{lim,GBM})}{\iiint \frac{\dot \rho/R_0}{1+z}\frac{\mathrm{d}V}{\mathrm{d}z} P(L,\Ep\,|\,\lpop) P_\mathrm{det,GBM}(L,\Ep,z)\,\mathrm{d}L\,\mathrm{d}\Ep\,\mathrm{d}z},
\end{equation}
where $P_\mathrm{det,GBM}$ is the flux-limited form from Eq.~\ref{eq:pdet_simple} and $\dot\rho(z)/R_0$ is independent of $R_0$ (see Eq.~\ref{eq:rhoz}). We applied the above expression to our population posterior samples $\lbrace\vec\lambda_{\mathrm{pop},k}\rbrace_{k=1}^{N_\mathrm{p}}$ to obtain an equal number of $\lbrace R_{0,k}\rbrace_{k=1}^{N_\mathrm{p}}$ samples. In turn, this allowed us to construct samples of the posterior distribution of the luminosity function, $\lbrace R_{0,k}\phi(L,\lambda_\mathrm{pop,k})\rbrace_{k=1}^{N_\mathrm{p}}$, where $\phi(L,\lambda_\mathrm{pop})=L\int P(L,\Ep\,|\,\lpop)\,\mathrm{d}\Ep$.

Figure~\ref{fig:lum_func} shows the luminosity function of SGRBs obtained in this way from the full-sample analysis (red solid) and the flux-limited sample analysis (orange dashed). The corresponding results from \citetalias{Wanderman2015} and \citetalias{Ghirlanda2016} (their fiducial model `a') are shown in blue and grey, respectively. The shape of the result is somewhat in between the \citetalias{Wanderman2015} and \citetalias{Ghirlanda2016} results, but the normalization of the high-luminosity end is in better agreement with \citetalias{Wanderman2015} than \citetalias{Ghirlanda2016}. The high-luminosity end slope ($1-A\sim 2$) is in agreement with both benchmark results, while the flattening below the break is reminiscent of that found by \citetalias{Ghirlanda2016}, but translated to a luminosity that is lower by almost one order of magnitude. At luminosities below $L_\mathrm{c}^\star\sim 3\times 10^{51}\,\mathrm{erg\,s^{-1}}$, the slope of the luminosity function is $\sim 0.4$, which is flatter than \citetalias{Wanderman2015} (but not by as much as \citetalias{Ghirlanda2016}) and more similar to the older results from \citet{Guetta2005,Guetta2006}. The low-luminosity end below $\sim 10^{49}\,\mathrm{erg\,s^{-1}}$ remains highly uncertain.

Overall, the luminosity function we recover bears some similarity with that obtained by \citet{Tan2020}, whose study also relies on a quasi-universal jet assumption and on a similar form of the `core' luminosity dispersion.

\subsection{Rate density}

\begin{figure*}
 \centering
 \includegraphics[width=\textwidth]{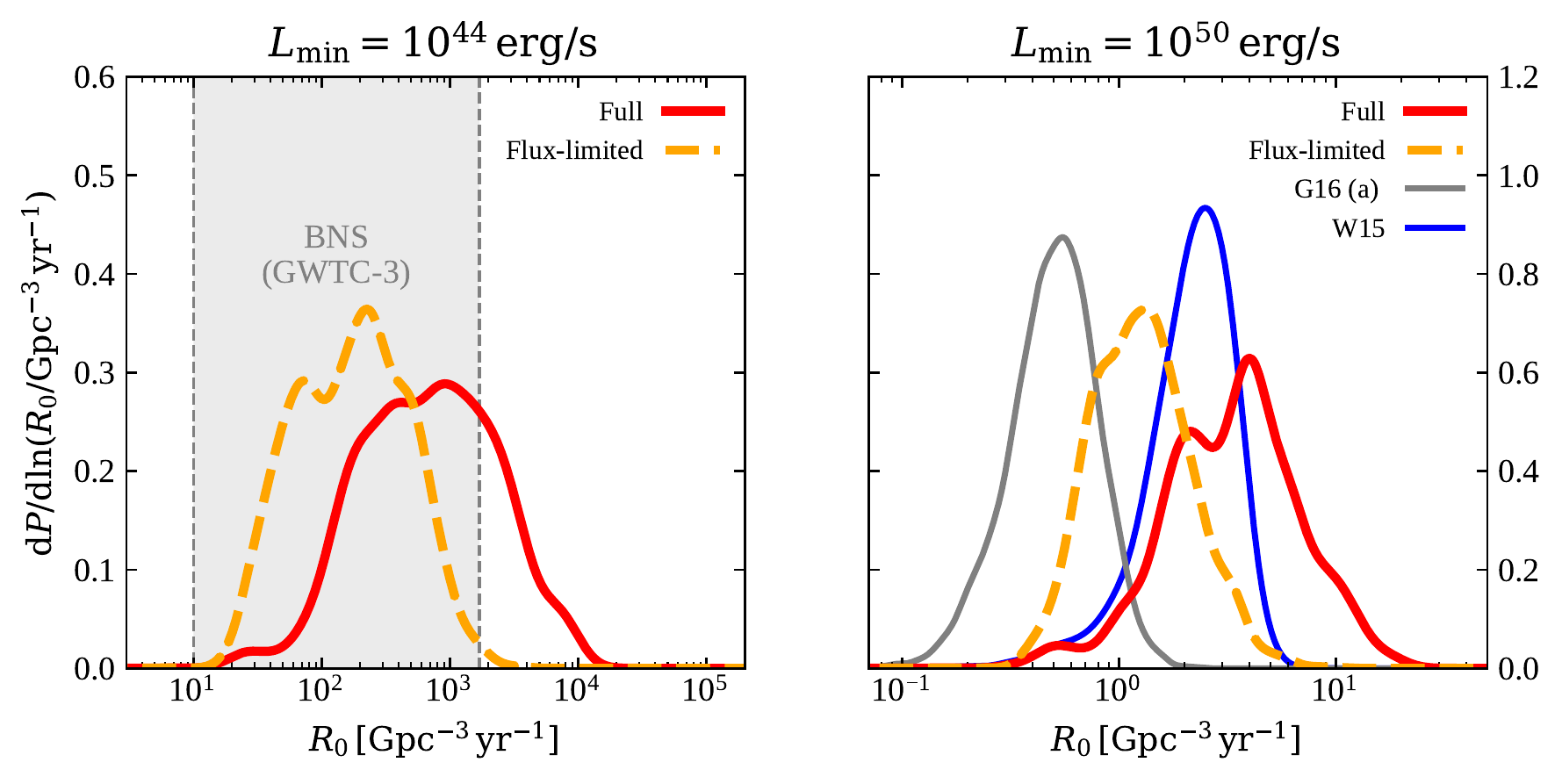}
 \caption{Local rate density. \textit{Left-hand panel}: the red solid (resp.\ orange dashed) line shows the posterior probability density on the total local rate density $R_0$ of SGRBs with a luminosity $L>10^{44}\,\mathrm{erg\,s^{-1}}$ (which encompasses all viewing angles) from our analysis using the full (resp.\ flux-limited) sample. The gray shaded band shows the constraint on the local rate density of binary neutron star mergers from GW observations, as derived in \citet{Abbott2021_GWTC3_pop}. \textit{Right-hand panel}: The red solid and orange dashed lines show the local rate density for events with a luminosity $L>10^{50}\,\mathrm{erg\,s^{-1}}$ in our two analysis setups. For comparison, we show the corresponding local rate densities from \citetalias{Wanderman2015} and \citetalias{Ghirlanda2016} above the same minimum luminosity.}
 \label{fig:R0}
\end{figure*}

The total local rate density $R_0$ of SGRBs (including all viewing angles) is not well-constrained by our analyses, because of the difficulty in determining the actual extent of the luminosity function with the available data. The full-sample analysis yields $R_0=740_{-630}^{+3870}\,\mathrm{Gpc^{-3}\,yr^{-1}}$ (median and symmetric 90\% credible interval), while the flux-limited sample analysis gives $R_0=180_{-145}^{+660}\,\mathrm{Gpc^{-3}\,yr^{-1}}$ (see Figure \ref{fig:R0}, left-hand panel). Both are compatible with the local binary neutron star (BNS) merger rate derived from gravitational wave observations \citep{Abbott2021_GWTC3_pop}, $R_\mathrm{0,BNS}=10-1700\,\mathrm{Gpc^{-3}\,yr^{-1}}$, and also with other recent estimates of the total, collimation-corrected SGRB rate based on different methods \citep[e.g.][see \citealt{MandelBroekgaarden:2021} for a review and further references]{RoucoEscorial2022}. The fact that the derived SGRB rate leans towards the high-end of the BNS merger rate uncertainty interval can be interpreted as an indication that the fraction of BNS mergers that yield a jet must be high, in agreement with the results of \citet{Salafia2022_fjet}, \citet{Sarin2022}, \citet{Beniamini2019}, and \citet{Ghirlanda2019}.

In order to compare our local rate density with those presented in the literature, we also computed the rate density of events above a minimum luminosity $L_\mathrm{min}=10^{50}\,\mathrm{erg\,s^{-1}}$. The right-hand panel in Figure \ref{fig:R0} shows the result from our two analyses, which yield $R_0(L>10^{50}\,\mathrm{erg\,s^{-1}})=3.6_{-2.5}^{+6.9}\,\mathrm{Gpc^{-3}\,yr^{-1}}$ (full sample) and $1.3_{-0.7}^{+2.0}\,\mathrm{Gpc^{-3}\,yr^{-1}}$ (flux-limited sample), compared with those obtained by integrating the \citetalias{Wanderman2015} and \citetalias{Ghirlanda2016} luminosity functions over the same luminosities. All results are in agreement with each other, placing the local rate density of luminous SGRBs around one event per $\mathrm{Gpc^3\,yr}$, with roughly one order of magnitude of uncertainty.

\begin{figure}
 \centering
 \includegraphics[width=\columnwidth]{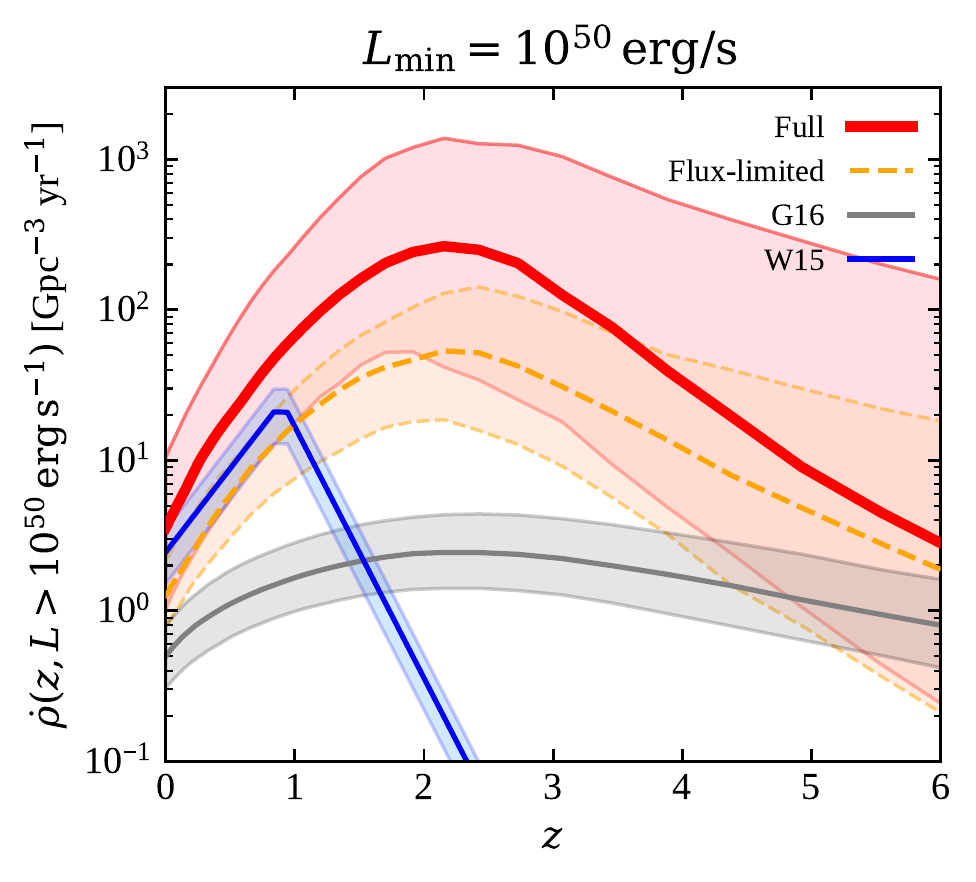}
 \caption{Rate density evolution. The figure shows the redshift evolution of the rate density from both our full sample (red solid lines) and flux-limited sample (orange dashed lines) analysis, limited to events above a minimum luminosity $L_\mathrm{min}=10^{50}\,\mathrm{erg\,s^{-1}}$. The corresponding evolutions from \citetalias{Wanderman2015} and \citetalias{Ghirlanda2016} are shown in blue and grey, respectively. The thick lines show the median of the posterior predictive distribution at each fixed redshift, while the shaded areas encompass the symmetric 90\% credible interval.}
 \label{fig:R(z)}
\end{figure}

Our model also constrains the evolution of the rate density with redshift, $\dot\rho(z)$, using the parametrization given in Eq.~\ref{eq:rhoz}. Figure \ref{fig:R(z)} shows the symmetric 90\% credible interval of the posterior probability distribution of $\dot \rho(z)$, at each fixed z, considering only events with $L\geq 10^{50}\,\mathrm{erg\,s^{-1}}$ (red solid: full-sample analysis; orange dashed: flux-limited sample analysis). Lowering the minimum luminosity would increase the uncertainty on the normalization, but leave the shape identical, as a consequence of our assumption of no evolution of the jet structure parameters with redshift.  For comparison, we show the corresponding results from \citetalias{Wanderman2015} and \citetalias{Ghirlanda2016}, where the shaded regions only account for the uncertainty on the local rate density, and thus represent an underestimate of the actual uncertainty in these models. The shape of the constraint is in qualitative agreement with the result of \citetalias{Ghirlanda2016}, who found an evolution that is compatible with the expectations from compact binary merger progenitors, while it is in strong disagreement with the sharp cut-off in the rate density at $z>0.9$ found by \citetalias{Wanderman2015}. We believe that the latter stems from a bias induced by the limited treatment of selection effects in that work, with a similar impact as that described in \citet{Bryant2021} for non-parametric methods.
We note, however, that three out of four SGRBs  with a photometric redshift in our rest-frame sample, namely 170127B, 180727A and 191031D, have median redshift larger than 1.9. If we remove the four SGRBs with a photometric redshift from our sample, we obtain generally similar results (with larger error bars) except for the redshift evolution, whose preferred low-redshift slope and peak become more consistent with the \cite{Madau2014} CSFR (but clearly with larger error bars: $a\sim 3\pm 1.5$, $z_\mathrm{p}\sim 1.9\pm 1$), which could reconcile the result with the expectations for BNS mergers with short delay times. Thus, the redshift distribution is sensitive to the reliability of these photometric redshifts.

The low-redshift scaling of the SGRB rate, $(1+z)^a$ with $a=4.6_{-0.8}^{+0.4}$, is steeper than that of the cosmic star formation rate (CSFR) at low redshift, $\mathrm{CSFR} \propto (1+z)^{2.7}$ \citep{Madau2014}, while the constraint on the peak of the SGRB rate $z_\mathrm{p}\sim 2.2_{-0.6}^{+0.8}$ indicates a preference for a rate that peaks at larger redshift than the CSFR (whose peak is at $z\sim 1.9$). This is difficult to explain with either very short delay times between star formation and BNS mergers (which would suggest that the SGRB rate should trace the CSFR) or long delay times, which would shift the peak of the SGRB to lower redshift than the CSFR peak (though the uncertainty on the SGRB redshift peak could allow for the long delay time interpretation).  This apparent discrepancy could point to a redshift-dependent evolution of the yield of merging binary neutron stars per unit star formation or a redshift-dependent evolution of the fraction of BNS mergers yielding SGRBs. Such effects could plausibly be caused by metallicity-dependent variations in stellar and binary evolution, including in NS masses.

\subsection{Comparison with the three reference samples}\label{sec:comparison_with_observations}

\begin{figure*}
 \centering
 \includegraphics[width=\textwidth]{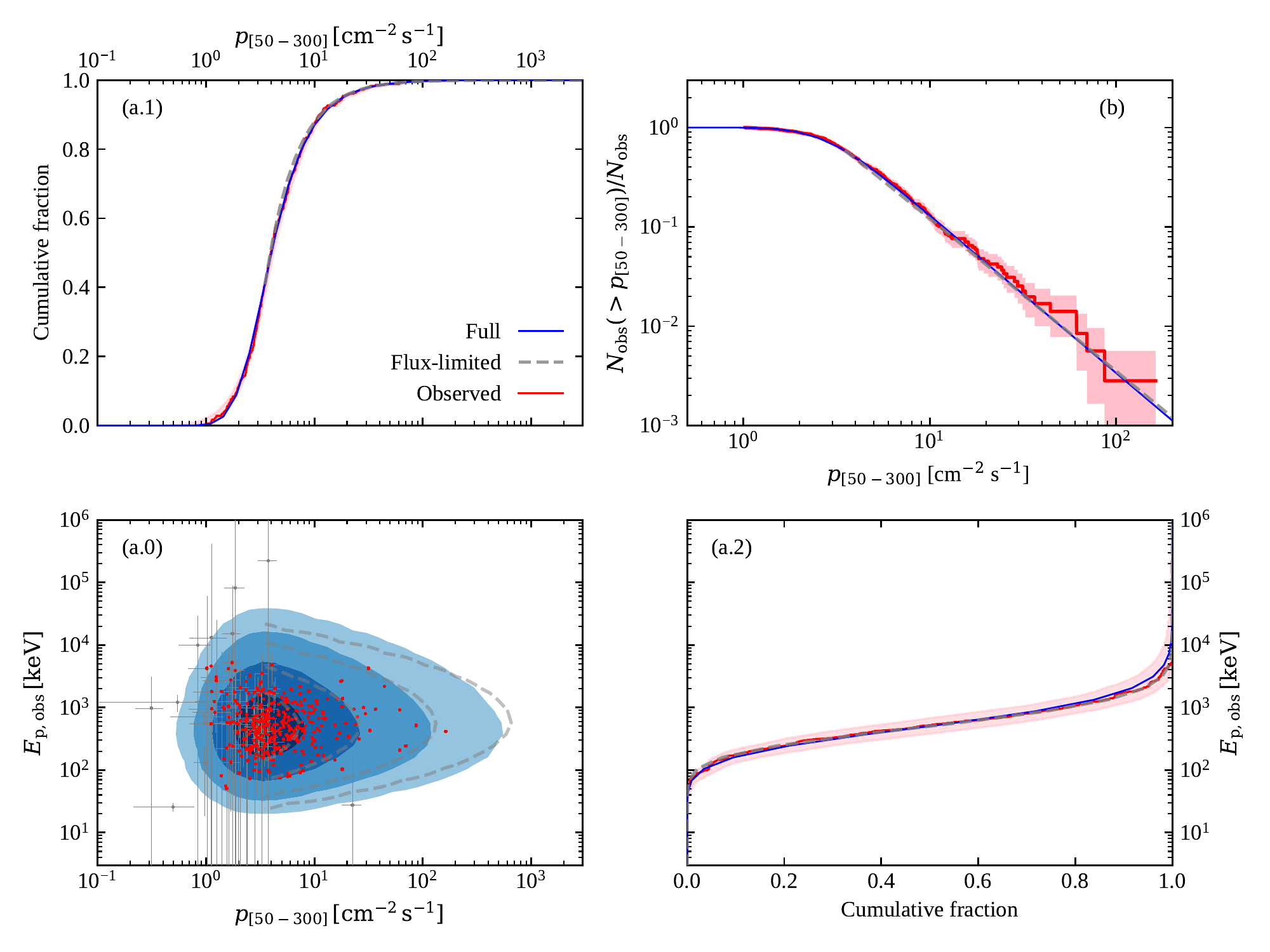}
 \caption{Observer-frame constraints and best-fit model predictions. Panel (a.0) shows the predicted distribution of \textit{Fermi}/GBM SGRBs on the $(p_{[50-300]},E_\mathrm{p,obs})$ plane for our best-fit model, with progressively lighter contours containing 50\%, 90\%, 99\% and 99\% of the events. Red dots show the observed data reported in the \textit{Fermi}/GBM catalog that pass our additional quality cuts, while grey points show the events that are discarded. We additionally show the error bars with thin grey lines for those events with a relative error larger than 50\% on either quantity, or both. Panels a.1 and a.2 show the predicted (solid blue line) and observed (solid red cumulative histogram, with the pink region showing the 90\% confidence region that stems from statistical uncertainties on spectral fitting parameters) cumulative distributions of $p_{[50-300]}$ (panel a.1) and $E_\mathrm{p,obs}$ (panel a.2) for events that pass the quality cuts. Panel (b) shows the inverse cumulative distribution of $p_{[50-300]}$, highlighting the behaviour at the high-flux end, which follows the expected $p_{[50-300]}^{-3/2}$ trend. Panel b uses the same conventions as panels a.1 and a.2, except the shaded pink region shows the one-sigma equivalent Poisson error.}
 \label{fig:obsframe}
\end{figure*}

\begin{figure*}
 \centering
 \includegraphics[width=\textwidth]{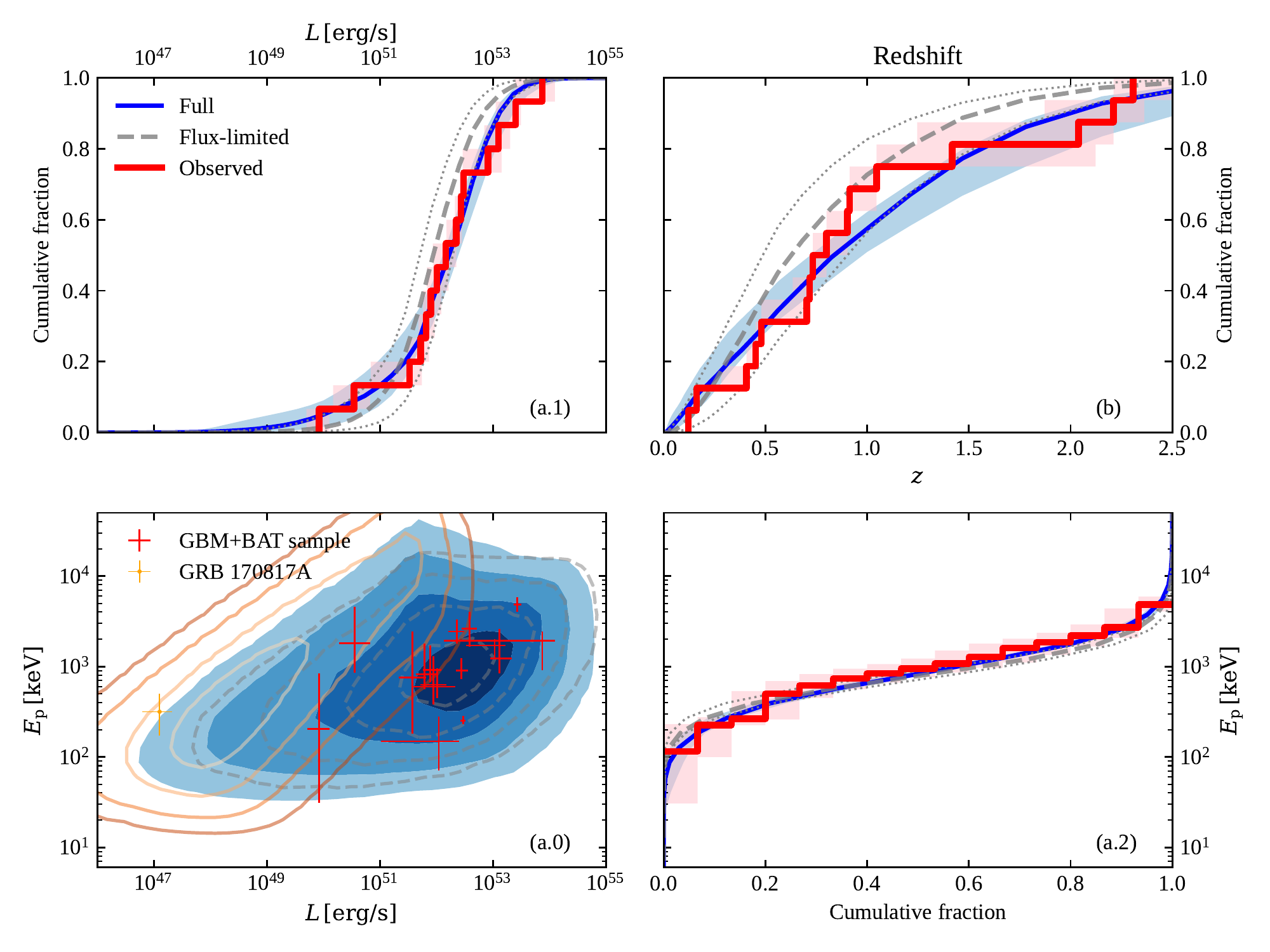}
 \caption{Rest-frame sample, viewing angle sample, and model predictions. This is similar to figure \ref{fig:obsframe}, but for rest-frame quantities $L$ and $\Ep$. Panel a.0 shows the predicted smallest regions in the $(L,E_\mathrm{p})$ plane where 50\%, 90\%, 99\% and 99.9\% of the SGRBs that are detected by \textit{Fermi}/GBM and by \textit{Swift}/BAT with $p_\mathrm{[15-150]}>p_\mathrm{lim,BAT}$ are located, according to our model, using the parameters at the median of the $\lpop$ posterior (filled blue contours: full sample analysis; dashed gray contours: flux-limited sample analysis). Red crosses show the SGRBs with a measured redshift in our rest-frame sample. Panels a.1, a.2 and b show the predicted (blue solid lines: full sample; grey dashed lines: flux-limited sample) and observed (red cumulative histograms, with pink shaded areas showing the 90\% confidence regions) cumulative distributions of luminosities (panel a.1), $E_\mathrm{p}$ (panel a.2), and redshifts (panel b). In panels a.1, a.2 and b, the 90\% credible bands stemming from the uncertainty on $\lpop$ are shown with blue shading  (full sample) and dotted lines (flux-limited sample). The additional orange contours in panel a.0 show the 50\%, 90\%, 99\% and 99.9\% containment regions for SGRBs with a BNS merger counterpart detected by \textit{Fermi}/GBM and by the aLIGO and Advanced Virgo detectors with O3 sensitivity. The orange cross shows the position of GRB~170817A on the plane. }
 \label{fig:restframe}
\end{figure*}

Figure \ref{fig:obsframe} compares the distributions of 64-ms peak photon flux $p_{[50-300]}$ and observed peak photon energy $E_\mathrm{p,obs} $ distributions predicted by our model (using the median of the $\lpop$ posterior distribution) with the observer-frame \textit{Fermi}/GBM sample. For the flux-limited sample analysis, we limit the comparison to the sub-sample of events with $p_{[50-300]}>p_\mathrm{lim,GBM}$. The figure demonstrates an excellent agreement both in the joint $p_{[50-300]}-E_\mathrm{p,obs}$ distribution and in the individual distributions. The fact that the shape of the low-end of the inverse cumulative $\log(N_\mathrm{obs})-\log(p_{[50-300]})$ distribution is well reproduced (panel b in the figure) indirectly demonstrates the ability of our \textit{Fermi}/GBM detection efficiency model to accurately reproduce the selection effects of the full sample.

In Figure \ref{fig:restframe} we also compare the distribution of $L$, $\Ep$ and $z$ of our rest-frame sample (\S\ref{sec:restframe_sample}), in red, with those predicted by the population model with the parameter constraints from the full sample analysis (blue) and flux-limited sample analysis (gray). In panels a.1, a.2 and b we show the 90\% credible bands that stem from the $\lpop$ uncertainty.
In this case, we apply the detection efficiency model described in Appendix \ref{sec:pdet_construction} to both the full sample and flux-limited sample analysis results in order to compute the predicted distributions.
In panel a.0 we additionally show the only event in our viewing angle sample, GRB~170817A (orange cross), along with the 50\%, 90\%, 99\% and 99.9\% containment contours (shades of orange) of the distribution of SGRBs detected by \textit{Fermi}/GBM and with a BNS merger counterpart detected by aLIGO and Advanced Virgo with the O3 sensitivity, as predicted by the population model with the best-fit parameters from the full sample analysis.

\section{Discussion}\label{sec:discussion}

\subsection{Jet intrinsic structures compatible with the derived apparent structure}\label{sec:intrinsic_structure_inference}

\subsubsection{Jet total luminosity and energy}
Keeping in mind that $L$ represents the luminosity at the peak of the light curve, the actual time-averaged gamma-ray luminosity can be written as $\langle L \rangle = \xi L$, where we take a reference value $\xi = 0.3\,\xi_{-0.5}$ which is the median of the peak flux to average flux ratios in the GBM sample. The prompt emission energy conversion efficiency is $\epsilon_\gamma = 0.1\,\epsilon_{\gamma,-1}$, where the reference value is based on the results of \citet{Beniamini2016}. Therefore the two factors compensate each other to some extent, and the jet core total (i.e.\ prior to dissipation that leads to gamma-ray emission) average isotropic-equivalent energy output rate is $\langle L_\mathrm{c,tot}\rangle=(\xi/\epsilon_\gamma)L_\mathrm{c}=9\times 10^{51}\xi_{-0.5}\epsilon_{\gamma,-1}^{-1}L_\mathrm{c,51.5}\,\mathrm{erg\,s^{-1}}$. If the jet duration is $T=1\,T_0\,\mathrm{s}$, the core isotropic-equivalent jet energy is then, by definition, $E_\mathrm{c,tot,iso}\sim \langle L_\mathrm{c,tot}\rangle T = 9\times 10^{51}\,\xi_{-0.5}\epsilon_{\gamma,-1}^{-1}L_\mathrm{c,51.5}T_0\,\mathrm{erg}$. This is compatible with typical estimates of the core isotropic-equivalent energy of the GRB~170817A jet, which fall in the range $10^{51}$ -- $3\times 10^{52} \mathrm{erg\,s^{-1}}$ (see e.g.\ figure S6 in \citealt{Ghirlanda2019}), and in line with cosmological SGRBs in general, which typically fall in a similar range \citep[e.g.][]{RoucoEscorial2022,Fong2015}. The total jet energy is on the order of $E_\mathrm{tot}\sim \theta_\mathrm{c}^2 E_\mathrm{c,tot,iso} \sim 2\times 10^{49}\, (\theta_\mathrm{c}/\mathrm{3\,\mathrm{deg}})^2\xi_{-0.5}\epsilon_{\gamma,-1}^{-1}L_\mathrm{c,51.5}T_0\,\mathrm{erg}$. Also this value is in line with those inferred from afterglow modelling of cosmological SGRBs \citep{RoucoEscorial2022}.

\subsubsection{Angular structure}

As discussed in \S\ref{sec:apparent_vs_intrinsic}, the relationship between the intrinsic jet structure and the apparent luminosity angular profile is not straightforward and dependent on the underlying dissipation and emission mechanism. Nevertheless, we can get some insight on the impact of our findings on the intrinsic structure of SGRB jets by adopting some simplifying assumptions: (i) the ratio $\xi$ of the peak luminosity to the average luminosity does not depend on the viewing angle (i.e.\ the light curve shape is preserved when changing the viewing angle),  and (ii) the observed duration of the emission is dominated by the central engine activity time, and hence it is also independent of the viewing angle. Under these assumptions, which likely hold only in a limited range of viewing angles close to the jet core, the peak luminosity scales with the viewing angle in the same way as the isotropic-equivalent energy (Eq.\ \ref{eq:Eiso_thv}). Let us further assume a power law profile for the jet total isotropic-equivalent energy $E_\mathrm{tot,iso}$ and bulk Lorentz factor $\Gamma$, that is let us set $E_\mathrm{tot,iso}(\theta)\propto (\theta/\theta_\mathrm{c})^{-s_E}$ and $\Gamma(\theta)\propto (\theta/\theta_\mathrm{c})^{-s_\Gamma}$ for $\theta>\theta_\mathrm{c}$.
It is likely that the gamma-ray efficiency $\epsilon_\gamma$ is also a function of the angle from the jet axis\footnote{In the internal shocks scenario, for instance, a large efficiency requires a large average bulk Lorentz factor and high Lorentz factor contrast between colliding shells. Since the bulk Lorentz factor is expected to decrease away from the core, the efficiency should decrease as well.}, so that we also set $\epsilon_\gamma(\theta) \propto (\theta/\theta_\mathrm{c})^{-s_\epsilon}$, with $s_\epsilon >0$. Therefore, the isotropic-equivalent energy radiated in gamma-rays at each angle goes as $E_\mathrm{rad,iso}(\theta) \propto (\theta/\theta_\mathrm{c})^{-s_E-s_\epsilon}$.

In principle, using Eq.~\ref{eq:Eiso_thv} one can derive the profile of the observed isotropic-equivalent energy $E_\mathrm{\gamma,iso}(\thv)$ (and hence peak luminosity $L$, given our assumptions) as a function of the viewing angle given the slopes $s_E$, $s_\epsilon$, $s_\Gamma$, the core bulk Lorentz factor $\Gamma_\mathrm{c}$ and the core half-opening angle $\theta_\mathrm{c}$. On the other hand, we can simplify the problem by noting that at relatively small viewing angles the emission is dominated by material along the line of sight, provided that $\Gamma_\mathrm{c}$ is relatively large, say $\Gamma_\mathrm{c}\gtrsim 100$. In this regime, $E_\mathrm{\gamma,iso}(\thv) \sim E_\mathrm{rad,iso}(\theta=\thv)$ \citep{Rossi2002}, thus $\alpha_L \sim s_E+s_\epsilon$ and hence $s_E\lesssim \alpha_L$. In the same regime, $\Ep(\thv)\sim\Gamma(\theta=\thv) E_\mathrm{p}^\prime(\theta=\thv)$, where $E_\mathrm{p}^\prime(\theta)$ is the comoving peak SED photon energy. The latter is likely positively correlated with $\Gamma$ (e.g.\ \citealt{Ghirlanda2018}), therefore $s_\Gamma\lesssim \alpha_{\Ep}$.

Using the upper end of the 90\% credible intervals for $\alpha_L$ and $\alpha_{\Ep}$ from our full sample analysis, these arguments therefore lead to the upper limits $s_E\lesssim 6$ and $s_\Gamma\lesssim 3$. We stress again that, given the assumptions, these results only hold for viewing angles close to the jet core.

While these limits clearly conflict with a `top-hat' jet structure (which would correspond to $s_E,s_\Gamma\to \infty$), they are still in agreement with the rather steep kinetic energy profiles and shallower Lorentz factor profiles found in studies of the GRB~170817A afterglow \citep[e.g.][]{Hotokezaka2019,Ghirlanda2019,Mooley2022}. The approximate $\theta^{-3}$ scaling of the jet isotropic-equivalent kinetic energy found in recent numerical simulations of SGRB jets \citep[e.g.][]{Gottlieb2020,Gottlieb2021,Gottlieb2022} is also compatible with these limits, even though it seems to conflict with the former findings based on the GRB~170817A afterglow.

\subsection{Viewing angles of \textit{Fermi}/GBM SGRBs with known redshift}\label{sec:viewing_angles}
Through our population model it is possible to derive a viewing angle probability for any SGRB using only the information on its luminosity and spectral peak energy. Here we focus on SGRBs with a measured redshift in our rest-frame sample. The posterior probability on the viewing angle $\thvi$ of the $i$-th SGRB in the sample (represented by data $d_i$ in the data vector $\vec d$) is
\begin{equation}
\begin{split}
 & P(\thvi\,|\,\vec d) = \iint P(\thvi\,|\,L_i,\Epi,\vec d) P(L_i,\Epi\,|\, \vec d)\,\mathrm{d}L_i\,\mathrm{d}\Epi =\\
 & \iint P(L_i,\Epi\,|\, d_i)\int P(\thvi\,|\,L_i,\Epi,\lpop)P(\lpop\,|\,\vec d)\,\mathrm{d}\lpop\mathrm{d}L_i\mathrm{d}\Epi\\
 & =\iint P(L_i,\Epi\,|\, d_i)\int \frac{P(L_i,\Epi\,|\,\thvi,\lpop)P(\thvi\,|\,\lpop)}{P(L_i,\Epi\,|\,\lpop)}\times \\
 & \times P(\lpop\,|\,\vec d)\,\mathrm{d}\lpop\,\mathrm{d}L_i\,\mathrm{d}\Epi\sim\\
 & \sim \frac{\sin\thvi}{N_\mathrm{s}N_\mathrm{p}}\sum_{j=1}^{N_\mathrm{s}}\sum_{k=1}^{N_\mathrm{p}}\frac{P(L_{i,j},E_{\mathrm{p},i,j}\,|\,\thvi,\vec \lambda_{\mathrm{pop},k})}{P(L_{i,j},E_{\mathrm{p},i,j}\,|\,\vec \lambda_{\mathrm{pop},k})},
\end{split}
\label{eq:viewing_angle_posterior}
\end{equation}
where the last equality follows from Monte-Carlo approximation of the integrals, with $\lbrace(L_{i,j},E_{\mathrm{p},i,j})\rbrace_{j=1}^{N_\mathrm{s}}$ being samples from the $P(L_i,\Epi\,|\, d_i)$ posterior obtained from the spectral analysis of the SGRB, and $\lbrace\vec \lambda_{\mathrm{pop},k}\rbrace_{k=1}^{N_\mathrm{p}}$ being samples from the population posterior.

\begin{figure*}
 \includegraphics[width=\textwidth]{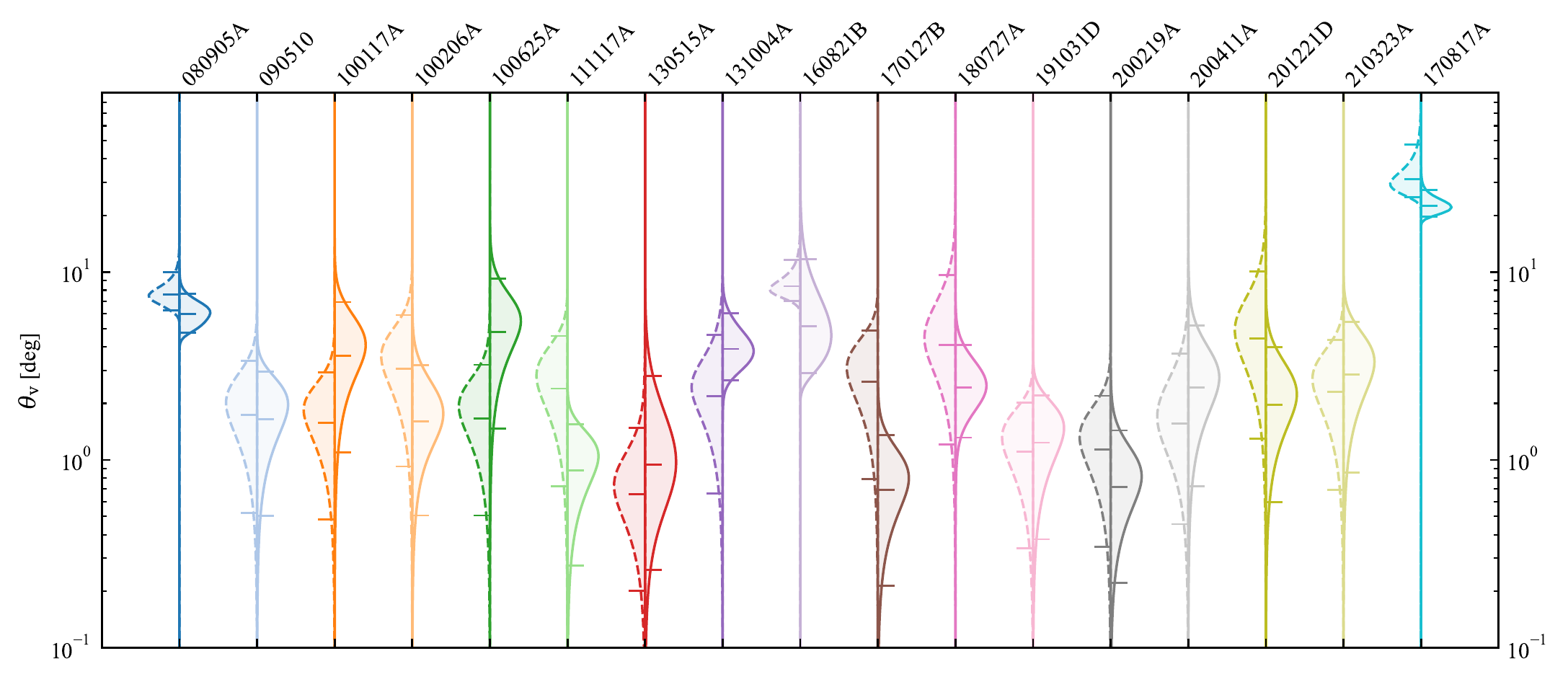}
 \caption{Viewing angle posterior probability densities for \textit{Fermi}/GBM SGRBs with known redshift. For each SGRB with known redshift in our sample, we show the posterior probability density $\mathrm{d}P/\mathrm{d}\ln\thvi=\thvi P(\thvi\,|\,\vec d)$ (see Eq.~\ref{eq:viewing_angle_posterior}) constructed using 300 population posterior samples from the full sample analysis (solid lines) or the flux-limited sample analysis (dashed lines). Tick marks indicate the 5$^\mathrm{th}$, 50$^\mathrm{th}$ and 95$^\mathrm{th}$ percentiles of each posterior probability. Individual SGRBs are assigned the same colors as in Fig.~\ref{fig:GBM_SGRBs_with_z}.\label{fig:viewing_angle_posteriors}}
\end{figure*}

Figure \ref{fig:viewing_angle_posteriors} shows the resulting population-informed viewing angle posterior probability densities for the SGRBs in our rest-frame sample. The constraints from the full sample and flux-limited sample analyses are in general agreement, with most jets likely viewed a few degrees from the jet axis. Focussing on the full sample analysis results, four SGRBs have population-informed viewing angles that are larger than 2 deg at 95\% credibility: GRB~080905A, GRB~131004A, GRB~160821B and, unsurprisingly, GRB~170817A. The median and 90\% credible interval of our population-informed viewing angle estimate for GRB~160821B is $\thv = 5_{-2}^{+7}\,\mathrm{deg}
$, which is compatible with the estimate $\thv=10_{-5}^{+14}\,\mathrm{deg}$ by    \citet{Troja2019_GRB160821B} based on afterglow modelling. The population-informed estimate for GRB~170817A is $\thv=23_{-3}^{+5}\,\mathrm{deg}$, in excellent agreement with afterglow-based estimates that include the information on the centroid proper motion from Very Long Baseline Interferometry imaging \citep[e.g.][]{Mooley2018,Hotokezaka2019,Ghirlanda2019,Mooley2022}, which find viewing angles in the range $15\lesssim\thv/\mathrm{deg}\lesssim 25$, and also with the estimate $\thv = 18 \pm 8\,\mathrm{deg}$ by \citet[][68\% credible interval]{Mandel:2018} based on a similar method as that employed in \S\ref{sec:viewing_angle_sample}, but which includes a marginalization over the cosmological parameters.

\subsection{The impact of selection effects on the $L$-$\Ep$ plane}\label{sec:Yonetoku_sel_effects}

\begin{figure*}
 \centering
 \includegraphics[width=0.6\textwidth]{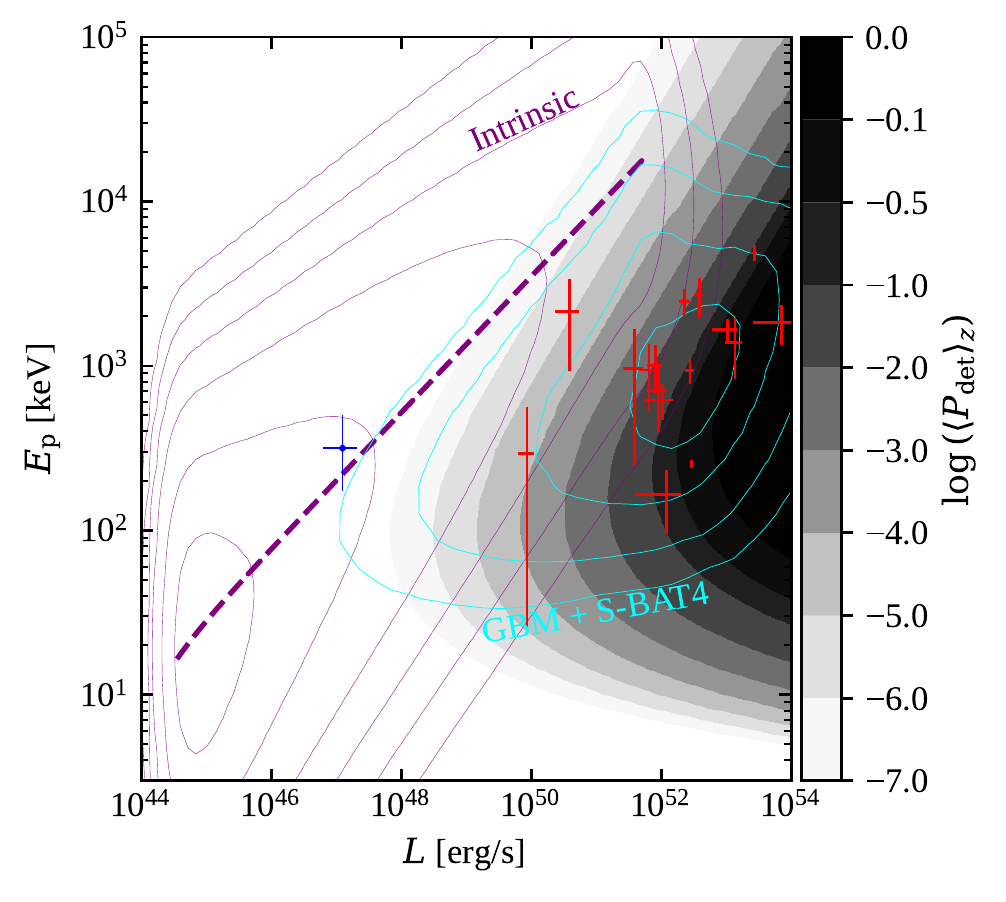}
 \caption{Visualization of the impact of selection effects on the $L$-$\Ep$ plane. The grey filled contours in the figure show the \textit{Fermi}/GBM SGRB detection efficiency averaged over redshift, assuming the redshift distribution to be described by our model with parameters corresponding to the median of the population posterior from the full sample analysis (values corresponding to different shades of grey are shown in the colorbar). Purple contours show the areas on the plane that contain 50\%, 90\%, 99\%, 99.9\%, 99.99\% and 99.999\% of SGRBs in the Universe according to the model. The thick dashed purple line shows the $\left(\bar L(\thv),\bar E_\mathrm{p}(\thv)\right)$ relation of the model. Cyan contours show the areas that contain 50\%, 90\%, 99\% and 99.9\% of SGRBs that are detected by GBM and have $p_{[15-150]}>p_\mathrm{lim,BAT}$, that is, that pass the selection criteria of our rest-frame sample. Red crosses show SGRBs with known $z$ in our rest-frame sample. GRB~170817A which is shown by a blue cross.}
 \label{fig:Pdet_Yonetoku}
\end{figure*}

Our result, shown in Figure \ref{fig:Yonetoku}, that the bulk of the SGRB population is located upwards of the apparent SGRB Yonetoku correlation came with a surprise to us. In order to demonstrate the role played by selection effects in this context, we show with grey shades in Figure \ref{fig:Pdet_Yonetoku} the detection efficiency that represents the selection effects acting on our rest-frame sample (\S\ref{sec:restframe_sample}), averaged over redshift, that is
\begin{equation}
 \left\langle P_\mathrm{det}\right\rangle_z(L,\Ep) = \int P_\mathrm{det,GBM}(L,\Ep,z)P_\mathrm{det,BAT}(L,\Ep,z)P(z\,|\,\lpop)\,\mathrm{d}z,
\end{equation}
where $P_\mathrm{det,GBM}(L,\Ep,z)$ is that from the full sample analysis (but the discussion remains unchanged when adopting that from the flux-limited sample analysis) and in this discussion we keep $\lpop$ fixed at the median of the full sample analysis posterior. The purple contours in the figure contain 50\%, 90\%, 99\%, 99.9\%, 99.99\% and 99.999\% of SGRBs in the Universe according to our model. For reference we also show the $\left(\bar L(\thv),\bar E_\mathrm{p}(\thv)\right)$ relation with a thick, purple dashed line. This `intrinsic' distribution of SGRBs in the $L$-$\Ep$ plane is distorted by selection effects into the distribution represented by cyan contours, that contain 50\%, 90\%, 99\% and 99.9\% of the SGRBs that pass the rest-frame sample selection criteria according to the model. The distribution represented by cyan contours can be understood as the product of the intrinsic (purple) distribution times the redshift averaged detection efficiency (grey filled contours).  As in previous figures, red crosses mark the positions of SGRBs with measured $z$ in our rest-frame sample, while the blue cross marks GRB~170817A. Focussing on SGRBs with $L>10^{50}\,\mathrm{erg\,s^{-1}}$, these reside in a part of the plane where the $\left\langle P_\mathrm{det}\right\rangle_z$ contours are roughly parallel and equally spaced. In other words, the gradient
 \renewcommand*{\arraystretch}{1.5}
\begin{equation}
   \vec\nabla_{\ln L, \ln\Ep}\ln \left\langle P_\mathrm{det}\right\rangle_z= \left(\begin{array}{c}
        \frac{\partial \ln \left\langle P_\mathrm{det}\right\rangle_z}{\partial \ln L}\\
        \frac{\partial \ln \left\langle P_\mathrm{det}\right\rangle_z}{\partial \ln \Ep}\\
       \end{array}\right)
\end{equation}
is roughly constant across the region occupied by the observed SGRBs in the plane, with a GBM detection efficiency that decreases by almost four orders of magnitude along the direction of this gradient. The variation in the density of the SGRBs in the rest-frame sample along the same direction, on the other hand, is clearly much less than four orders of magnitude. This suggests that the intrinsic density of SGRBs in this plane must increase steeply in the direction opposite to the $\left\langle P_\mathrm{det}\right\rangle_z$ gradient, in order for the variation in the intrinsic density of SGRBs to compensate the dramatic decrease in the detection efficiency. Hence, selection effects likely play a major role in shaping the observed $L$-$\Ep$ correlation in SGRBs. This conclusion and also, intriguingly, the slope and dispersion of the intrinsic $L$-$\Ep$ correlation we obtained, are in agreement with those found by \citet{Palmerio2021} for long GRBs (see section 5.1 in that paper). This might be taken as an indication of a common universal luminosity and $\Ep$ angular profile in the two populations.

On a different note, it is worth stressing that the existence of a tail of SGRBs with low luminosities $L\lesssim 10^{50}\,\mathrm{erg\,s^{-1}}$ but very high SED peak photon energies $\Ep\gtrsim 3\,\mathrm{MeV}$, predicted by our population model and visible in the figure, must be taken with a grain of salt. Our parametrization is constructed in such a way that the dispersion of $\Ep$ around the viewing-angle dependent average $\bar E_\mathrm{p}(\thv)$ is symmetric and identical at all viewing angles, so that the low-$L$, high-$\Ep$ tail is merely a result of the choice of parametrization, being unobservable with the current instrumentation and therefore not observationally constrained.

\subsection{Joint SGRB + GW detection predictions}

\begin{figure*}
 \centering
 \includegraphics[width=0.7\textwidth]{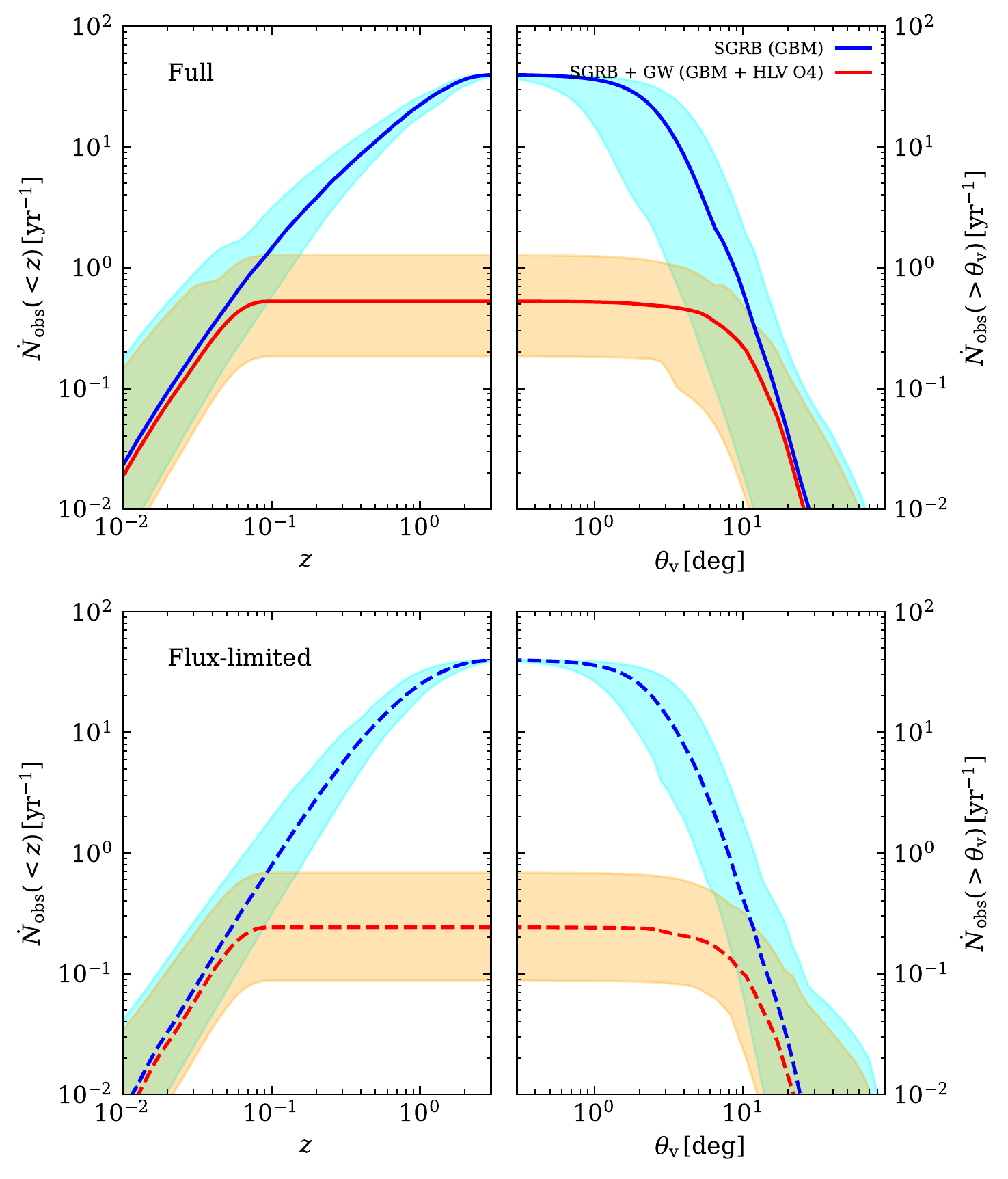}
\caption{Projected joint SGRB + GW rates for O4. The top (resp.\ bottom) panels refer to the results from the full (resp.\ flux-limited) sample analysis. Left-hand panels show the predicted rate $\dot N_\mathrm{obs}(<z)$ of SGRBs within a redshift $z$ detected by \textit{Fermi}/GBM alone (blue: median; cyan: 90\% credible range) or in coincidence with a network consisting of the two LIGO and the Virgo detectors with the projected O4 sensitivity (red: median; orange: 90\% credible range). The right-hand panels show the corresponding rates $\dot N_\mathrm{obs}(>\thv)$ for events seen at a viewing angle larger than $\thv$.}
\label{fig:SGRB+GW}
\end{figure*}

We can produce predictions for the rate of coincident detections of SGRBs and GWs by \textit{Fermi}/GBM and the ground-based GW detector network from the population model, leveraging the fact that it accounts for the jet luminosity as a function of the viewing angle. We focus here on a network consisting of the two Advanced LIGO detectors (Hanford and Livingston) plus the Virgo detector (a `HLV' network), and assume the projected sensitivities in the upcoming O4 observing run\footnote{\url{https://dcc.ligo.org/T2200043-v3/public}}. We assumed all SGRBs to be produced by BNS mergers with non-spinning components and the jets to be aligned with the orbital angular momentum. In practice, we constructed an HLV O4 GW detection efficiency $P_\mathrm{det,GW}(z,\thv)$ as a function of redshift and viewing angle as follows: we binned the simulated BNS mergers from \citet{Colombo2022} (selecting only those which produce a jet according to their criteria, which are $\sim 50\%$ in their population) in the $(\thv, z)$ plane and we computed the fraction $f_{\mathrm{GW},i,j}$ with a network signal-to-noise ratio (SNR)  $\rho_\mathrm{net}\geq 12$ in each bin (we assumed 100\% duty cycle for all detectors for simplicity). We then estimated $P_\mathrm{det,GW}(z,\thv)$ by linearly interpolating the $f_{\mathrm{GW},i,j}$'s on a grid with nodes corresponding to the centers of the bins. Samples of the cumulative joint detection rate were then computed based on 100 random population posterior samples as
\begin{equation}
\begin{split}
 & \dot N_\mathrm{GBM+GW}(<z,\vec\lambda_{\mathrm{pop}}) = \int_0^z \iiint\,\frac{\dot\rho(z',\vec\lambda_{\mathrm{pop}})}{1+z'}\frac{\mathrm{d}V}{\mathrm{d}z}P_\mathrm{det,GW}(z',\thv)\times \\
 & \times P(L,\Ep\,|\,\thv,\vec\lambda_{\mathrm{pop}})P_\mathrm{det,GBM}(L,\Ep,z')\sin\thv\,\mathrm{d}\thv\,\mathrm{d}L\,\mathrm{d}\Ep\,\mathrm{d}z'
 \end{split}
 \end{equation}
and similarly
 \begin{equation}
\begin{split}
 & \dot N_\mathrm{GBM+GW}(>\thv,\vec\lambda_{\mathrm{pop}}) = \int_{\thv}^{\pi/2} \iiint\,\frac{\dot\rho(z,\vec\lambda_{\mathrm{pop}})}{1+z}\frac{\mathrm{d}V}{\mathrm{d}z}P_\mathrm{det,GW}(z,\thv')\times \\
 & \times P(L,\Ep\,|\,\thv',\vec\lambda_{\mathrm{pop}})P_\mathrm{det,GBM}(L,\Ep,z)\sin\thv'\,\mathrm{d}L\,\mathrm{d}\Ep\,\mathrm{d}z\,\mathrm{d}\thv'.
 \end{split}
 \end{equation}
 Figure \ref{fig:SGRB+GW} shows the median and symmetric 90\% credible range over the population posterior samples of $\dot N_\mathrm{GBM+GW}(<z)$ (left-hand panels) and $\dot N_\mathrm{GBM+GW}(>\thv)$  (right-hand panels) using posterior samples from the full sample analysis (top panels) and flux-limited sample analysis (bottom panels). The plots also show the same result for the \textit{Fermi}/GBM detection only (i.e.\ setting $p_\mathrm{det\,GW}=1$) for comparison. All rates are normalized to the \textit{Fermi}/GBM observed SGRB detection rate of $40\,\mathrm{yr^{-1}}$ and hence include the limited field of view and duty cycle of the latter instrument. The total observed SGRB + GW rates are $\dot N_\mathrm{GBM+GW}=0.53_{-0.34}^{+0.75}\,\mathrm{yr^{-1}}$ for the full sample analysis and $\dot N_\mathrm{GBM+GW}=0.24_{-0.15}^{+0.44}\,\mathrm{yr^{-1}}$ for the flux-limited sample analysis (median and 90\% credible range). These predictions are in agreement with, but slightly more optimistic than, the rate $\dot N_\mathrm{SGRB+GW}=0.17_{-0.13}^{+0.26}\,\mathrm{yr^{-1}}$ estimated by \citet{Colombo2022}, and are generally in line with other recent predictions from the literature \citep[e.g.][]{Mogushi2019,Howell2019,Saleem2020,Yu2021,Patricelli2022}, including those from the LIGO-Virgo Collaboration \citep{Abbott2022_SGRB_search}.

\subsection{Effect of decreasing the lower bound of the prior on $L_\mathrm{c}^\star$}
As stated in \S\ref{sec:priors}, we chose the lower bound of the prior on the typical on-axis luminosity $L_\mathrm{c}^\star$ to be consistent with our assumption that the high end of the luminosty function is shaped by on-axis events, while intermediate and low luminosities are due to off-axis events and depend on the apparent structure. In order to investigate the effect of relaxing that assumption, we ran the full sample analysis one additional time with a much looser bound $L_\mathrm{c}^\star\geq 10^{48}\,\mathrm{erg\,s^{-1}}$. As expected (given the fact that the marginalized posterior on $L_\mathrm{c}^\star$ rails against the lower bound in the results described above), this results in a posterior that shows a mild preference for $L_\mathrm{c}^\star\sim 10^{50}\,\mathrm{erg\,s^{-1}}$, but with significant posterior support all the way down to $10^{48}\,\mathrm{erg\,s^{-1}}$ and up to $10^{52}\,\mathrm{erg\,s^{-1}}$, showing that the typical on-axis luminosity is poorly constrained by the available data within the proposed model. On the other hand, the lower $L_\mathrm{c}^\star$, the larger the local rate $R_0$ needed to reproduce the GBM observed SGRB rate: if we additionally require $R_0\leq 1700\,\mathrm{Gpc^{-3}\,yr^{-1}}$, to reflect the upper limit on the BNS merger rate from the GW population analysis \citep{Abbott2021_GWTC3_pop}, then the posterior becomes again consistent with that obtained with the original prior. We thus conclude that our prior, in addition to being a consequence of the assumed quasi-universal jet scenario, also has a similar impact as the requirement that the SGRB local rate does not exceed the BNS merger rate. This suggests that future improved constraints on the latter rate through GW observations will positively impact our ability to constrain the SGRB population properties, including their typical apparent jet structure.

\subsection{Difficulties in defining a `clean' sample of GRBs from compact binary mergers}
When selecting our sample, we focused on events with $T_{90}$ shorter than 2 seconds for simplicity, and for the practical reason that the spectral parameters (photon index, observed $E_\mathrm{p,obs}$, peak photon flux from the spectral analysis) in the \textit{Fermi}/GBM online catalog are given with 64-ms binning only for events with $T_{90}<2\,\mathrm{s}$. On the other hand, this selection may result in a sample that includes events from both the main progenitor classes, that is, compact binary mergers \textit{and} collapsars \citep[e.g.][]{Zhang2009,Bromberg2013}, with the contamination from the latter class being difficult to quantity. Even if we did not test explicitly for the dependence of our results on this choice, the fact that the high-end of the luminosity function we obtain agrees relatively well with that obtained by \citetalias{Wanderman2015} (who adopted a much more restrictive criterion in an attempt to select a sample of pure `non-collapsar' GRBs) lends support to the conclusion that any potential contamination from collapsars in our sample does not impact the results significantly, given the present uncertainties.

Conversely, some GRBs with $T_{90}>2$ s are now widely accepted as being the result of a compact binary merger rather than a collapsar (e.g.\,GRB211211A, \citealt{Rastinejad2022,Mei2022,Gompertz2023}). Hence, the definition of a clean sample of GRBs from compact binary mergers is not straightforward. We believe that the best approach to this kind of problem in the future will be that of modelling the entire GRB population as a mixture of the two classes, jointly fitting the two sub-populations in a hierarchical model, similarly to what is currently done for binary black hole merger GW population analyses \citep[e.g.][]{Bouffanais2019,Wong2021,Zevin2021}.

As a final note, we caution that the duration of an SGRB must eventually increase with the viewing angle: at large enough viewing angles, the observed duration of a single pulse can exceed the duration of the central engine activity \citep[e.g.][]{Salafia2016}. Therefore, our sample selection could contain a bias against events with a large viewing angle. In order to address this problem, the duration of the emission, and its dependence on the viewing angle, must be included in the model as well.

\subsection{Peak luminosity \textit{versus} average luminosity}
Population studies of GRBs, whose luminosity varies erratically without a clear-cut minimum variability time scale during the prompt emission, must confront the issue of \textit{defining} the relevant luminosity whose distribution is to be modelled. Given the stochastic nature of the light curves, the time-averaged luminosity is arguably the most relevant quantity from a physical point of view; on the other hand, the detectability (and hence the selection effects) depends more closely on the peak luminosity, so that from a practical point of view this is the most important quantity to be modelled, which has therefore become the standard in the field. To our knowledge, the relation between the two distributions, that is, the peak luminosity function $\phi(L)=\mathrm{d}P/\mathrm{d}\ln L$ and the average luminosity function $\Phi(\langle  L \rangle)=\mathrm{d}P/\mathrm{d}\ln \langle L \rangle$, 
has never been investigated. Quite clearly, since $L=\xi^{-1} \langle L \rangle$ and the factor $\xi^{-1}\geq 1$ can differ between two SGRBs with the same $\langle L \rangle $ due to the stochasticity in the light curve, then the $L$ distribution is necessarily broader than the $\langle L \rangle $ one.  In that sense, the `core luminosity dispersion' in our model, parametrized as in Eq.~\ref{eq:P(Lc)}, accounts at least partially for the dispersion due to these effects. At large viewing angles, on the other hand, the broadening of individual pulses is likely going to smooth out the light curve \citep{Salafia2016}, hence reducing this kind of scatter. Hence, another improvement over our approach could be that of including a viewing-angle-dependent scatter in $L$ (and similarly in $\Ep$), which may improve the recovery of the actual average luminosity and peak photon energy profiles $\ell$ and $\eta$.

\section{Summary and conclusions}

The observations of GW170817 and the GRB170817A afterglow provided clear support for the presence of a jet viewed off-axis and endowed with a non-trivial angular structure. The inferred properties of the core of that jet were found to be consistent with those typically derived from the afterglows of short gamma-ray bursts. This prompted the question whether jets underlying short gamma-ray bursts could be very similar to each other on average, with a large part of the diversity due to the geometric choice of a viewing angle rather than intrinsic variations in jet structure and energetics.  In this work, we showed that a good description of the observed short gamma-ray burst population can be obtained within such a scenario.  The implied typical jet properties are consistent with those inferred from the GRB~170817A afterglow and from the larger population of SGRBs with a known distance, adding stability to the foundations of a unification programme for short gamma-ray bursts under the quasi-universal jet scenario, and more generally to our physical understanding of these phenomena. 

The inferred jet structure features a $\sim 2\,\mathrm{deg}$ uniform core within which the observer sees a large typical SED peak photon energy ($\Ep \sim 5\,\mathrm{MeV}$) and luminosity ($L \gtrsim 5\times 10^{51}\,\mathrm{erg\,s^{-1}}$). Outside the core, the luminosity falls off with the viewing angle as a steep power law (slope $\alpha_L\sim 4.7$), while $\Ep$ decreases as a relatively shallower power law ($\alpha_{\Ep}\sim 2$). No evidence for a break in these power laws has been found with the present data and analysis approach. 
While we find no clear support for a correlation between the on-axis luminosity $L_\mathrm{c}$ and the on-axis peak SED photon energy $E_\mathrm{p,c}$, the combined viewing angle dependence of $L$ and $\Ep$ induces a correlation for events viewed outside the core, $L\propto E_\mathrm{p}^{0.4}$. In the observed sample, we find that this correlation is distorted by selection effects. 

The inferred local rate density of SGRBs (at all viewing angles) is compatible with that of binary neutron star mergers as inferred from gravitational-wave population studies,  suggesting these to be the dominant progenitors. The model shows a preference for a strong rate density evolution with redshift: the rate density steeply increases as $\dot\rho(z)\propto (1+z)^{4.6}$ at low redshifts, plateaus toward a maximum near $z\sim 2.2$, and declines at higher redshifts, where it is poorly constrained. These results, on the other hand, may be driven by the rather large redshifts of three out of four SGRBs with a photometric redshift in our rest-frame sample: if photometric redshifts are excluded from the analysis, the redshift evolution becomes consistent with that of the cosmic star formation rate, and hence with progenitor binaries that merge rapidly after formation.  Based on the model and on the projected sensitivity of the aLIGO and Advanced Virgo network, we predict around 0.2 to 1.3 joint SGRB and GW detections per year during the O4 observing run.

Through the population model, it is possible to derive a population-informed viewing angle estimate for every SGRB whose intrinsic luminosity and peak photon energy are reasonably constrained. The estimates obtained for SGRBs with a known redshift in our sample indicate that most of them are viewed close to the edge of the core (either just within the core or slightly outside), with a few exceptions with a somewhat larger viewing angle. The largest viewing angle is clearly that of GRB~170817A, for which we estimate $\theta_\mathrm{v}=23_{-3}^{+5}\,\mathrm{deg}$, in excellent agreement with the estimates based on the afterglow and the VLBI superluminal motion.

A unification of SGRBs under a quasi-universal jet scenario would call for a relatively narrow progenitor parameter space, which can eventually help in pinpointing the long debated jet launching mechanism and the nature of the central engine. As demonstrated by the amount of information contained in the single GW170817 event, future multi-messenger observations of binary neutron star mergers and their jets will be of utmost importance in order for this programme to be successful.

\begin{acknowledgements}
The authors acknowledge Paolo D'Avanzo for support in building the S-BAT4ext sub-sample used in this work, and Ruben Salvaterra for insightful comments that helped deepening our understanding of some of the results. 
OS thanks Riccardo Buscicchio for many amusing and illuminating discussions. 
IM is a recipient of the Australian Research Council Future Fellowship FT190100574.
The color set used to identify SGRBs in the rest-frame sample is from \url{colorbrewer2.org} \citep{colorbrewer}.
\end{acknowledgements}

\bibliographystyle{aa}
\footnotesize
\bibliography{references}
\normalsize

\begin{appendix}

\section{Additional details on the results}

\subsection{Corner plot}

\begin{figure*}
 \includegraphics[width=\textwidth]{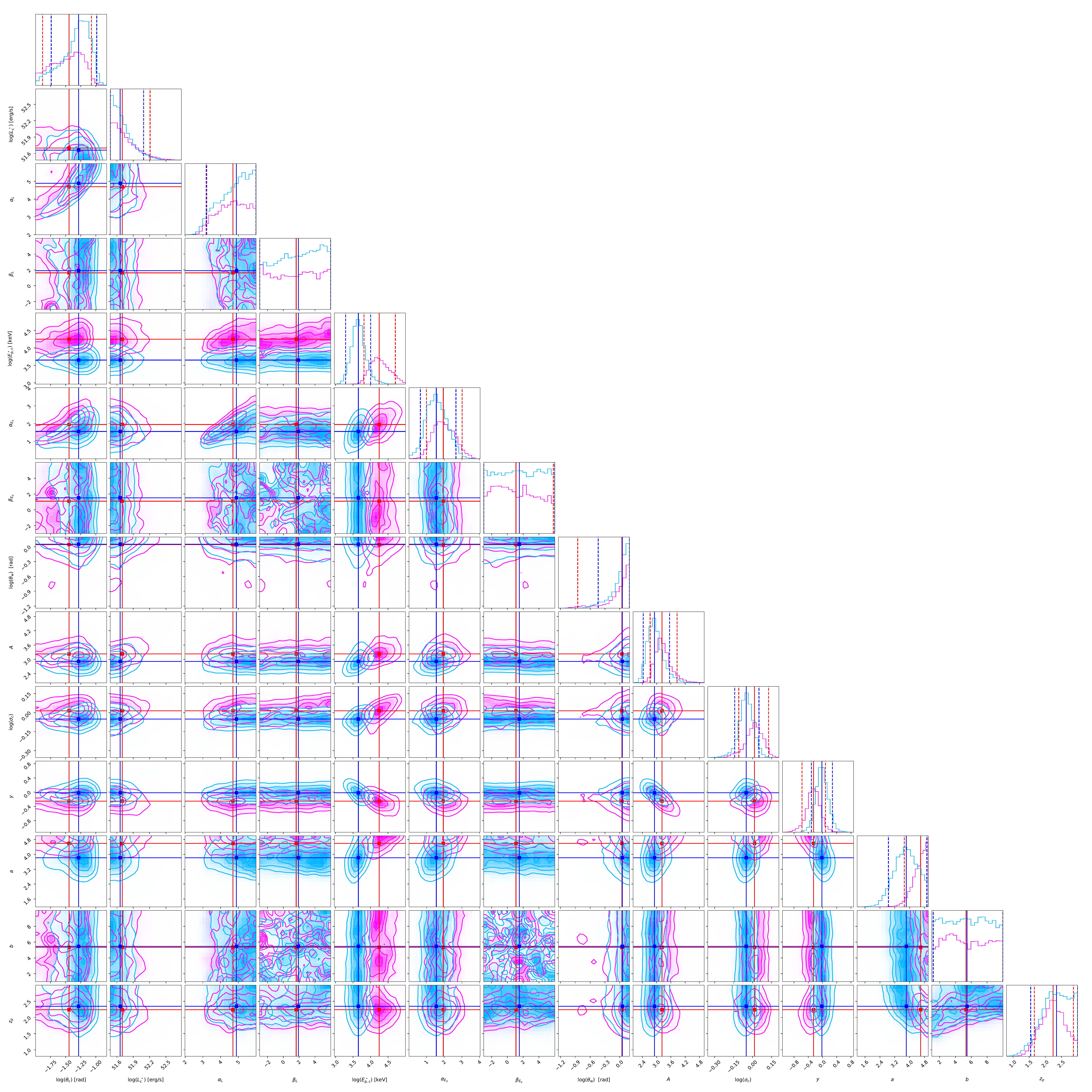}
 \caption{Corner plot of the posterior probability densities from the two analyses. The full sample analysis is shown in magenta, while the flux-limited sample analysis is shown in blue. The histograms on the diagonal represent the marginalized posterior probability densities constructed from the posterior samples, with the solid vertical lines marking the medians and the vertical dashed lines delimiting the symmetric 90\% credible interval (i.e.\ the 5$^\mathrm{th}$ and the 95$^\mathrm{th}$ percentiles -- note that these are not shown if they differ by less than one bin size from the nearest edge of the allowed range). The contours in the remaining panels show the one, two, three and four sigma credible areas from the two-dimensional marginalized joint posterior probability densities, with the dots showing the intersections of the medians of the corresponding one-dimensional marginalized posterior probability densities.}
 \label{fig:corner_full}
\end{figure*}

Figure \ref{fig:corner_full} shows the full, 14-parameter corner plot of both the full sample analysis (magenta) and the flux-limited sample analysis (light blue).

\section{Spectral analysis of \textit{Fermi}/GBM short gamma-ray burst with known redshift}\label{sec:gbm_analysis}
For each Fermi/GBM short GRB with known redshift in our sample, we analyzed the spectrum at the peak flux of the lightcurve binned with a 64-ms timescale. As starting time of the interval selected for the spectral analysis, we used the one reported in the GBM Catalogue (referred to as \textit{'Flux\_64\_Time'}\footnote{\url{https://heasarc.gsfc.nasa.gov/W3Browse/fermi/fermigbrst.html}}).

The spectral data ﬁles and corresponding latest response matrix ﬁles (rsp2) were obtained from the online HEASARC archive. We used the public software GTBURST to extract the spectral data. As part of the standard procedure, we selected the spectral data of the two most illuminated NaI detectors with a viewing angle smaller than 60° and the most illuminated BGO detector. In particular, we considered the energy channels in the range 10–900 keV for the NaI detectors, and 0.3–40 MeV for the BGO detector. We used intercalibration factors among the detectors, scaled to the most illuminated NaI and free to vary within 30\%. To model the background, we manually selected time intervals before and after the burst and modelled them with a polynomial function whose best-fitting order is automatically found by GTBURST. 

The spectral analysis has been performed with the public software XSPEC (v. 12.12.1). In the ﬁtting procedure, we used the PG-Statistic, valid for Poisson data with a Gaussian background.
Each peak flux spectrum was analyzed with a cutoff powerlaw model, typically used for the analysis of short GRBs spectra, and it consists of three parameters: the low-energy photon index $\alpha$, the characteristic energy $E_{\rm cut}$ (from which the peak energy of the spectrum can be derived as $E_{\rm p}=E_{\rm cut}(2-\alpha)$) and the normalization.
To obtain the luminosity (and its uncertainties) directly from the fit, the cutoff powerlaw model is multiplied by the \textsc{clumin} function available in XSPEC\footnote{For SGRBS with a photometric redshift, we used the \textsc{cflux} multiplicative model instead.}, which computes the luminosity of a specific model component, provided the redshift value of the source. The source-frame energy band over which luminosity is calculated is 1 keV-10 MeV.  Since the \textsc{clumin} model is used, the normalization of the cutoff powerlaw model has been kept fixed to 1 for all the spectra analyzed.
The parameters left free to vary in each fit are the following: the luminosity in the 1 keV-10 MeV energy range, the low-energy photon index $\alpha$ and the characteristic energy $E_{\rm cut}$, alongside the two inter-calibration factors for the GBM detectors.
Best-fit values and confidence ranges on these parameters have been derived within XSPEC, through the built-in Markov chain Monte Carlo algorithm (using the \textsc{chain} command). The 50\% and 90\% contours for $L_{\rm peak,iso}$ and $E_{\rm p}$ derived from the spectral analysis of each short GRB analyzed in this work are reported in Figure \ref{fig:GBM_SGRBs_with_z}. 

\section{\textit{Fermi}/GBM short gamma-ray burst detection efficiency}\label{sec:pdet_construction}

In order to construct a reliable estimate of the SGRB detection efficiency of \textit{Fermi}/GBM, a detailed simulation of its response to an event of that class is needed. The onboard trigger algorithms of GBM, described for example in \citealt{vonKienlin2020} (\citetalias{vonKienlin2020} hereafter), monitor the counts (binned over a given timescale $\Delta t$) recorded in a subset of the 8 energy channels of the NaI detectors\footnote{Algorithms that analyze the BGO detector data stream exist as well, but most of them require a trigger in the NaI detectors to be activated, or they have very restrictive triggering conditions, so they are effectively redundant and we neglect them here \citepalias{vonKienlin2020}.}. Each algorithm looks for an excess in the background-subtracted, binned counts $C_{\Delta t}(t)-B_{\Delta t}(t)$ (where $B_{\Delta t}$ is an estimate of the counts due to the background), over a certain multiple $n_\sigma$ of the standard deviation $\sigma_\mathrm{bkg,\Delta t}(t)$ of the counts recorded in a time interval immediately preceding $t$. Multiple algorithms operating on the same channels and with the same binning timescale are run in parallel, differing from each other only by small time offsets, which improves the triggering efficiency by minimizing cases with sub-optimally placed bins. A burst trigger is initiated whenever an excess is recorded at a consistent time in at least two of the NaI detectors by any of the running algorithms.

Since a sufficient condition for a trigger is that the \textit{peak} counts exceed the threshold, it is enough to simulate the peak counts generated by a putative source. On the other hand, since different algorithms operate with different binning timescales $\Delta t$, and since SGRBs are highly variable, it is necessary that the peak counts generated by the same source, but measured with a different $\Delta t$, reflect the actual peak count ratios induced by the temporal behavior of the events of that class. The simulated counts must then be compared with realistic background counts, and this must be done in each of the NaI detectors and consistently for each of the triggering algorithms. In what follows, we describe our approach to these problems.

\subsection{NaI background}\label{sec:NaI_bkg}
\begin{figure}
 \centering
 \includegraphics[width=0.8\columnwidth]{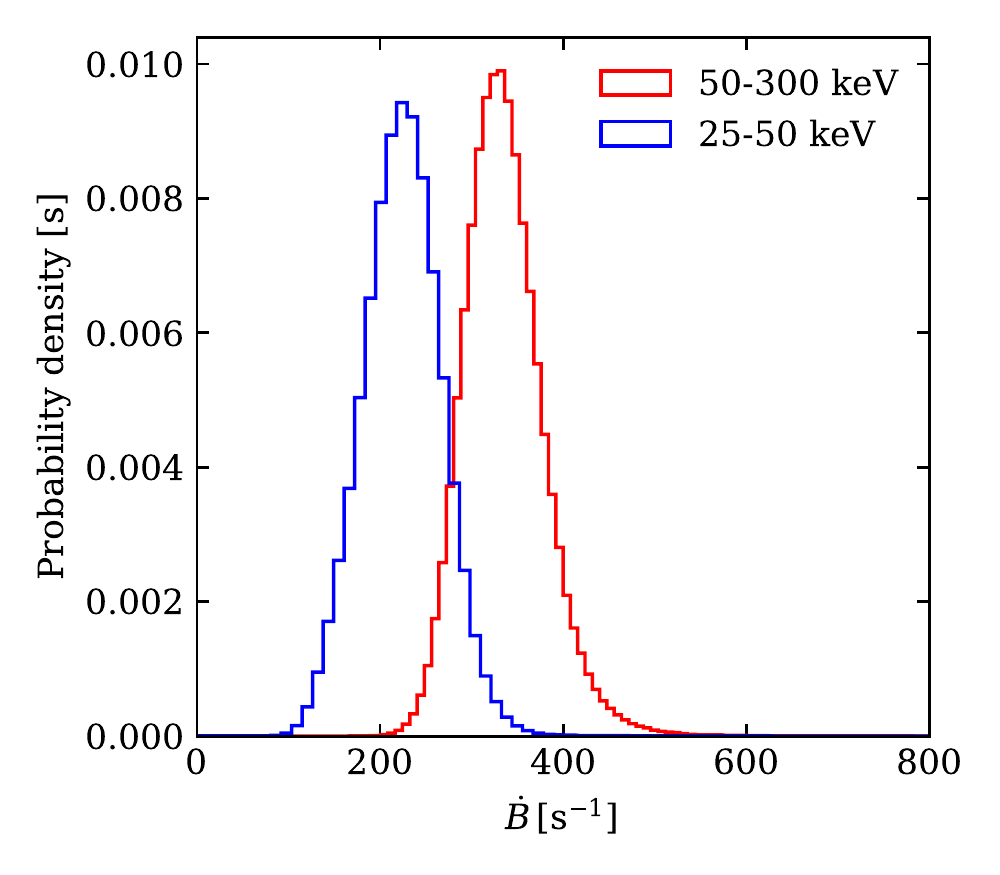}
 \caption{Probability density of the NaI background count rates in the 25-50 and 50-300 keV bands, constructed by sampling GBM \texttt{ctime} day data from days 210530 and 220928 at random times, combined from all 12 detectors. }
 \label{fig:bkg_count_rate_pdfs}
\end{figure}

In order to simulate a realistic background in each of the channels of the NaI detectors, we extracted the observed count rates from publicly available GBM daily data\footnote{\url{https://heasarc.gsfc.nasa.gov/W3Browse/fermi/fermigdays.html}}, at a randomly sampled time, and multiplied them by $\Delta t$ to get the counts. In particular, we considered \texttt{ctime} data relative to two selected days without triggers, 210530 and 220928, and extracted the count rates of the 12 NaI detectors in the 8 energy channels at random times, excluding periods when the detectors were turned off. The resulting probability density distributions of count rates in the 50-300 keV band (channels 3-4, combined from all 12 detectors) is shown in Figure \ref{fig:bkg_count_rate_pdfs} (red histogram), along with that for the 25-50 keV band (channel 2), for comparison.

\subsection{Detector response to an SGRB}

We assumed the photon spectrum (number of SGRB photons per unit time, per unit area, per unit photon energy as measured at the \textit{Fermi}/GBM position) of every SGRB to be time-independent and well described by the same cut-off power law as used in the population modelling, namely
\begin{equation}
 \frac{\mathrm{d^3N}}{\mathrm{d}t\,\mathrm{d}A\,\mathrm{d}E}\propto E^\alpha \exp\left(-(2+\alpha)\frac{E}{E_\mathrm{p,obs}}\right),
\end{equation}
which corresponds to the \textsc{comp} model commonly used in \textit{Fermi}/GBM data analyses \citepalias{vonKienlin2020}. 
Let us indicate with $f(E)$ the above spectrum normalized to unity in the $50-300$ keV band, that is,
\begin{equation}
f(E,\alpha,E_\mathrm{p,obs})=\frac{\mathrm{d^3N}/\mathrm{d}t\,\mathrm{d}A\,\mathrm{d}E}{\int_{50\,\mathrm{keV}}^{300\,\mathrm{keV}}(\mathrm{d^3N}/\mathrm{d}t\,\mathrm{d}A\,\mathrm{d}E)\,\mathrm{d}E}.
\end{equation}
The expected counts in the $j$-th channel ($j=1,...,8$) of the $k$-th NaI \textit{Fermi}/GBM detector ($k=1,...,12$), with a binning timescale $\Delta t$, produced by an SGRB with a peak photon flux $p_{[50-300]}$ (with 64 ms binning) and spectral parameters $\alpha$ and $E_\mathrm{p,obs}$, located at a sky position $(\theta,\phi)$ (polar and azimuthal angles with respect to the spacecraft $z$-axis), were computed as
\begin{equation}
\begin{split}
 & \bar C_{j,k,\Delta t,\mathrm{SGRB}}(\theta,\phi,p_{64},\alpha,E_\mathrm{p,obs})=\\
 & R_\mathrm{pflx}(\Delta t)p_{[50-300]}\Delta t\int_{E_{\mathrm{min},j}}^{E_{\mathrm{max},j}}f(E,\alpha,E_\mathrm{p,obs})A_\mathrm{eff}(\Theta_{k},E)\,\mathrm{d}E,
\end{split}
\end{equation}
where $E_{\mathrm{min},j}$ ($E_{\mathrm{max},j}$) is the lower (upper) photon energy bound of channel $j$ as given in Figure 8 of \citealt{Meegan2009} (\citetalias{Meegan2009} hereafter), $A_\mathrm{eff}$ is the NaI detector effective area at incidence angle $\Theta_k(\theta,\phi)$ measured with respect to the vector normal to the detector's entrance window (see Appendix~\ref{sec:eff_area}), and $R_\mathrm{pflx}(\Delta t)$ is the ratio of the count rate peak flux measured with $\Delta t$ binning to that measured with $64$ ms binning. The incidence angle  can be computed knowing the orientation of the $k$-th NaI detector with respect to the spacecraft $z$-axis \citepalias[Table 1 of][]{Meegan2009}.

\begin{figure}
\centering
 \includegraphics[width=0.8\columnwidth]{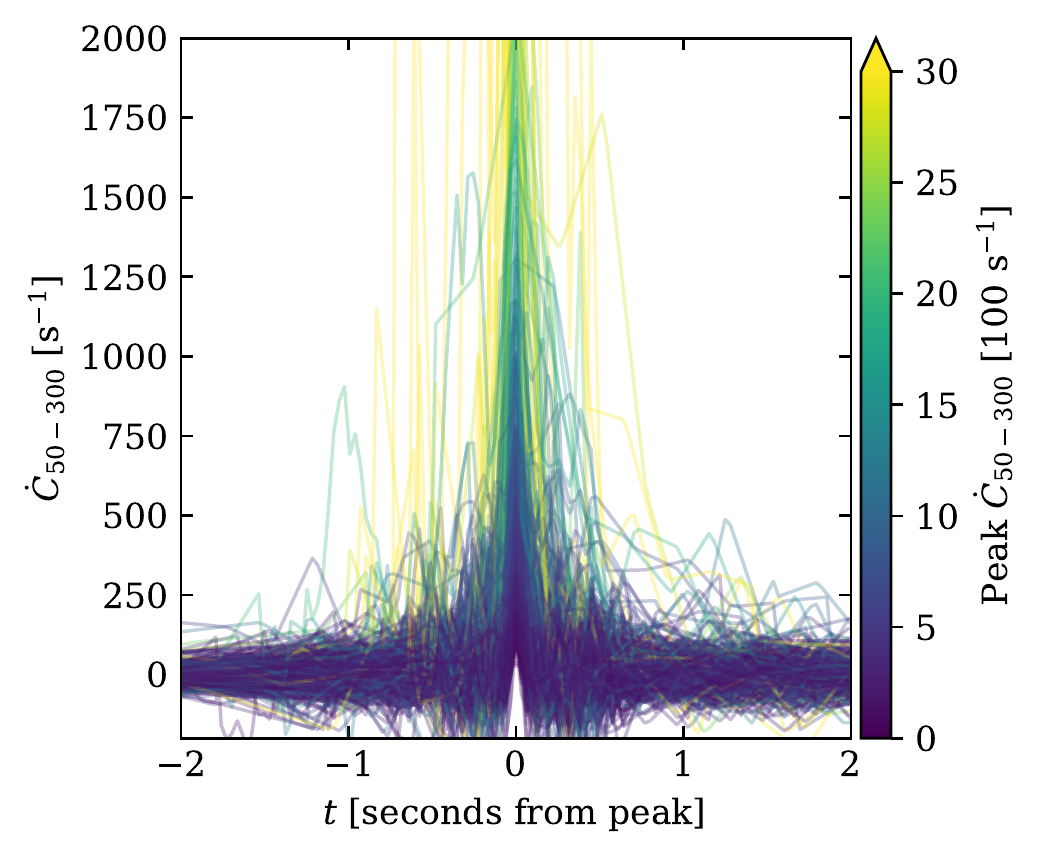}
 \caption{Background-subtracted count rate light curves in the 50-300 keV band for SGRBs observed by \textit{Fermi}/GBM. Each light curve is plotted with a color that depends on its peak count rate, as shown in the colorbar on the right. More details are given in the text.}
 \label{fig:C50-300}
\end{figure}

\begin{figure}
\centering
 \includegraphics[width=0.8\columnwidth]{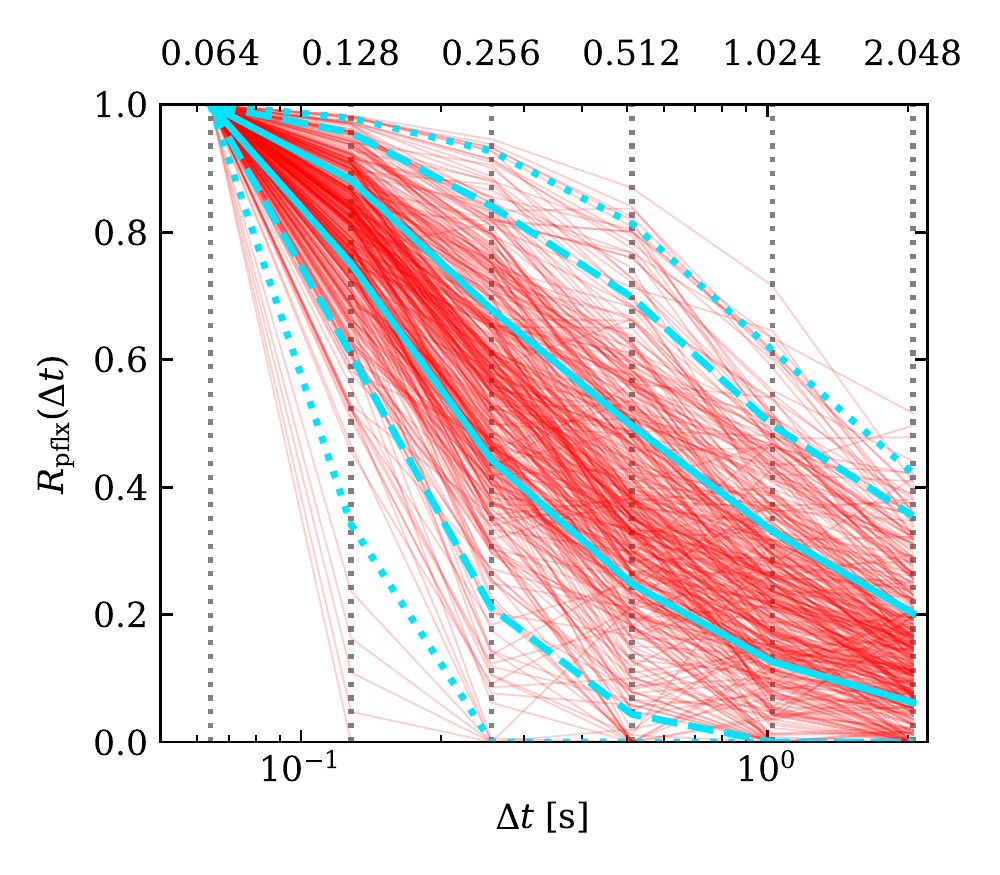}
 \caption{Ratio of SGRB count rate peak flux measured with a $\Delta t$ binning over that measured with a 64 ms binning. This is constructed using \textit{Fermi}/GBM \texttt{trigdata} files as explained in the text. Each thin red line corresponds to a single observed SGRB, while the light blue lines mark the 1$^\mathrm{st}$, 5$^\mathrm{th}$, 25$^\mathrm{th}$, 75$^\mathrm{th}$, 95$^\mathrm{th}$ and 99$^\mathrm{th}$ percentiles.}
 \label{fig:Rpflx}
\end{figure}

The ratio $R_\mathrm{pflx}(\Delta t)$ was constructed by extracting the count rate light curves in the 50-300 keV band (corresponding to channels 3 and 4, where the effective area is maximal) from the publicly available \texttt{trigdata} files\footnote{The data were obtained from \url{https://heasarc.gsfc.nasa.gov/W3Browse/fermi/fermigtrig.html}, where 527 \texttt{trigdata} files for bursts with $T_{90}<2\,\mathrm{s}$ were available. Some of the events had apparently incorrect background estimates, which were much larger than the actual counts recorded, resulting in negative background-subtracted counts at peak. Others had corrupted \texttt{trigdata} files. After removing these problematic events, we were left with 449 valid SGRBs.} of all \textit{Fermi}/GBM SGRBs. For each burst, we extracted the background-subtracted count rate light curves $\dot C_{50-300}(t)=\sum_{j=3}^4\sum_{k\in T}(\dot C_{j,k}(t)-\dot B_{j,k})$ in channels $j=3,4$, where $k\in \mathrm{T}$ means that we summed only over detectors for which an excess was found, $\dot C_{j,k}$ represents the count rate and $\dot B_{j,k}$ the estimated background count rate (as reported in the \texttt{trigdata} file). Figure \ref{fig:C50-300} shows the count rate light curves obtained in this way. We then computed
\begin{equation}
 R_\mathrm{pflx}(\Delta t)=\frac{64\,\mathrm{ms}}{\Delta t}\frac{\int_{-\Delta t/2}^{+\Delta t/2}\dot C_{50-300}(t)\,\mathrm{dt}}{\int_{-32\,\mathrm{ms}}^{+32\,\mathrm{ms}}\dot C_{50-300}(t)\,\mathrm{dt}},
\end{equation}
where  the time coordinate here is defined in such a way that $t=0$ corresponds to the maximum of $\dot C_{50-300}$. The resulting ratios for all the SGRBs analyzed are shown in Figure \ref{fig:Rpflx}. For each simulated SGRB, we randomly picked one of the $R_\mathrm{pflx}(\Delta t)$ curves shown there.

For each considered detection algorithm, we computed the background-subtracted counts as $C_{k,\Delta t}-B_{k,\Delta t}=\mathcal{P}(\sum_j\bar C_{j,k,\Delta t,\mathrm{SGRB}}+\sum_j\bar C_{j,k,\Delta t,\mathrm{bkg}})-\sum_j\bar C_{j,k,\Delta t,\mathrm{bkg}}$, where $j$ runs over the channels monitored by the algorithm, $\bar C_{j,k,\Delta t,\mathrm{bkg}}$ represents a randomly picked matrix of NaI background counts as explained in Appendix~\ref{sec:NaI_bkg}, and $\mathcal{P}(n)$ represents a random sample from a Poisson distribution with mean $n$. We then set $\sigma_{\mathrm{bkg},\Delta t,k}=\sqrt{\sum_j\bar C_{j,k,\Delta t,\mathrm{bkg}}}$ (which is a good approximation given the background count distributions shown in Fig.~\ref{fig:bkg_count_rate_pdfs}) and considered a trigger in the $k$-th detector if $(C_{k,\Delta t}-B_{k,\Delta t})>n_\sigma \sigma_{\mathrm{bkg},\Delta t,k}$. The simulated SGRB was marked as detected whenever at least two NaI detectors had at least one algorithm that triggered.

\subsection{Detection algorithms}

\begin{table}
\caption{Trigger algorithms considered in our framework. Each row lists the identification number, monitored channels, binning timescale, ratio of threshold to background standard deviation and corresponding \textit{Fermi}/GBM algorithm numbers for one of the triggering algorithms implemented in our framework. }\label{tab:trig_algs}
 \begin{tabular}{cccccc}
  Alg. & Channels & Band [keV] & $\Delta t$ [s] & $n_\sigma$ & GBM algs.$^a$ \\
  \hline
  1 & 3-4 & 50-300 & 0.064 & 5.0 & 4-7\\
  2 & 3-4 & 50-300 & 0.128 & 5.0 & 8-9\\
  3 & 3-4 & 50-300 & 0.256 & 4.5 & 10-11\\
  4 & 3-4 & 50-300 & 0.512 & 4.5 & 12-13\\
  5 & 3-4 & 50-300 & 1.024 & 4.5 & 14-15\\
  6 & 3-4 & 50-300 & 2.048 & 4.5 & 16-17\\
  \hline
 \end{tabular}\\
\footnotesize{$^a$identification numbers of the \textit{Fermi}/GBM algorithms \citepalias{vonKienlin2020} that are represented by each single algorithm in our framework.}
\end{table}

In our framework, each detection algorithm is characterized by a range of monitored channels, a binning timescale and a factor $n_\sigma$ that sets the threshold for triggering. When compared to the actual triggering algorithms running onboard \textit{Fermi}/GBM \citepalias{vonKienlin2020}, we do not consider a time offset, as our definition of $R_\mathrm{pflx}(\Delta t)$ essentially corresponds to always taking the offset that maximizes the peak count rates for each timescale. Hence, each of our algorithms effectively covers all the GBM algorithms with the same channels and binning timescales, but different offsets. Table \ref{tab:trig_algs} reports the 6 algorithms that we implemented in our framework. These are limited to the 50-300 keV band because, by inspection of the \texttt{trigdata} files, we verified that essentially all the SGRBs detected by GBM triggered one of the algorithms operating in that band. We implemented all algorithms that operate on timescales $\Delta t\geq 64$ ms, because \texttt{trigdata} files do not provide count rates with a finer resolution. We do not consider this as a limiting factor, since the variability of the SGRB light curves over such short timescales is hardly important to our purposes.

\subsection{NaI detector effective area}\label{sec:eff_area}

\begin{figure*}[t]
 \includegraphics[width=\textwidth]{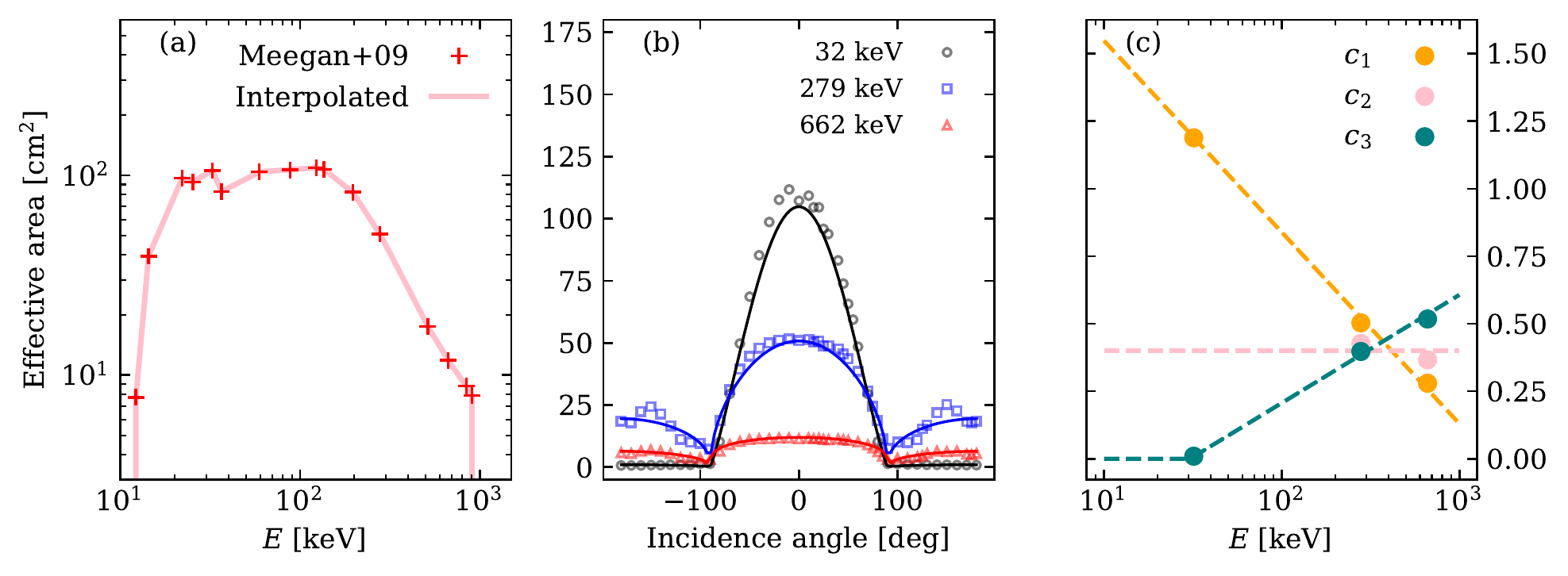}
 \caption{NaI detector effective area model. Panel (a) shows the NaI detector effective area measurements (red crosses), for zero incidence angle, reported in \citetalias{Meegan2009}. The pink line shows our adopted interpolation. Panel (b) shows the measured effective area \citepalias{Meegan2009} for different photon incidence angles at three reference photon energies (black circles: 32 keV; blue squares: 279 keV; red triangles: 662 keV). The solid lines show our best-fitting model. Panel (c) shows the best-fit values of parameters $c_1$, $c_2$ and $c_3$ of our model (Eq.~\ref{eq:Aeff_ansatz}) for the angular dependence of the effective area at the three reference energies (colored circles). The dashed lines show our assumed behavior of $c_i$ as a function of the photon energy. }
 \label{fig:eff_area}
\end{figure*}

We decomposed the effective area of a NaI detector into its value for zero incidence angle $A_\mathrm{eff}(0,E)$, times a dimensionless function that captures its variation with the incidence angle, $A_\mathrm{eff}(\Theta,E)\equiv A_\mathrm{eff}(0,E)\mathcal{A}(\Theta,E)$. We computed $A_\mathrm{eff}(0,E)$ by linearly interpolating the measurements reported Figure 11 of \citetalias{Meegan2009} on a log-log plane, as shown in panel (a) of Fig.~\ref{fig:eff_area}. To model $\mathcal{A}(\Theta,E)$ we considered the measurements reported in Figure 12 of \citetalias{Meegan2009} (shown in panel (b) of Figure \ref{fig:eff_area}), relative to the three reference photon energies $E_\mathrm{ref,1}=32$ keV, $E_\mathrm{ref,2}=279$ keV and $E_\mathrm{ref,3}=662$ keV, and normalized these data to the peak (which corresponds to $\Theta=0$ in each case). We assumed an ansatz analytical form
\begin{equation}
 \tilde{\mathcal{A}}(\Theta,c_1,c_2,c_3) = \left\lbrace\begin{array}{lr}
                                    |\cos(\Theta)|^{c_1} & \cos(\Theta)\geq 0\\
                                    c_3|\cos(\Theta)|^{c_2} & \cos(\Theta)< 0\\
                                   \end{array}
\right.,\label{eq:Aeff_ansatz}
\end{equation}
and fit it to the data at each reference photon energy by minimizing the sum squares of the residuals, obtaining the values reported in Table~\ref{tab:incidence-angle-eff-area} (the best-fit models are shown by solid lines in panel (b) of Fig.~\ref{fig:eff_area} -- we note that $c_2$ remains unconstrained at $E_\mathrm{ref,1}$, because $c_3\sim0$). In order to extend the model to other photon energies, we then assumed a linear dependence of the parameters $c_1$, $c_2$ and $c_3$ on the natural logarithm of the photon energy normalized to 32 keV, $\epsilon=\ln(E/32\,\mathrm{keV})$, namely $c_i=q_i+m_i\epsilon$, and obtained a good description of the data with $q_1=1.19$, $m_1=-0.308$, $q_2=0.4$, $m_2=0$, $q_3=0.00938$ and $m_3=0.173$, as shown in panel (c) of Fig.~\ref{fig:eff_area} (in the case of $c_3$, since it is naturally positive definite, we took $c_3=\max(q_3+m_3\epsilon,0)$). We therefore used $\mathcal{A}(\Theta,E)=\tilde{\mathcal{A}}(\Theta,q_1+m_1\epsilon,q_2+m_2\epsilon,\max(q_3+m_3\epsilon,0))$.

\begin{table}
\caption{Best-fit parameter values for the ansatz function, Eq.~\ref{eq:Aeff_ansatz}, fitted to the incidence-angle-dependent effective area data from \citetalias{Meegan2009} relative to three different reference photon energies.}
\centering
\begin{tabular}{c|ccc}
~ & \multicolumn{3}{c}{$E_\mathrm{ref}$}\\
 Param. & $32$ keV &  $279$ keV & $663$ keV \\
 \hline
 $c_1$ & 1.19 & 0.503 & 0.280 \\
 $c_2$ & - & 0.427 & 0.366 \\
 $c_3$ & 0.00938 & 0.397 & 0.518 \\
\end{tabular}
\label{tab:incidence-angle-eff-area}
\end{table}

\subsection{Resulting detection efficiency}
Within the framework described in the preceding sections, we computed the detection efficiency $\eta_\mathrm{det,3D}(p_{[50-300]},E_\mathrm{p,obs},\alpha)$ at a number of points on a $(p_{[50-300]},E_\mathrm{p,obs},\alpha)$ three-dimensional grid. For each point of the grid, we simulated a large number of SGRBs with isotropic $(\theta,\phi)$ positions in the \textit{Fermi} sky, with randomly sampled NaI backgrounds, by repeatedly following the procedure outlined above. Finally, we estimated $\eta_\mathrm{det,3D}(p_{[50-300]},E_\mathrm{p,obs},\alpha)$ as the fraction of simulated SGRBs that yielded a detection over the total.

\begin{figure}
 \includegraphics[width=\columnwidth]{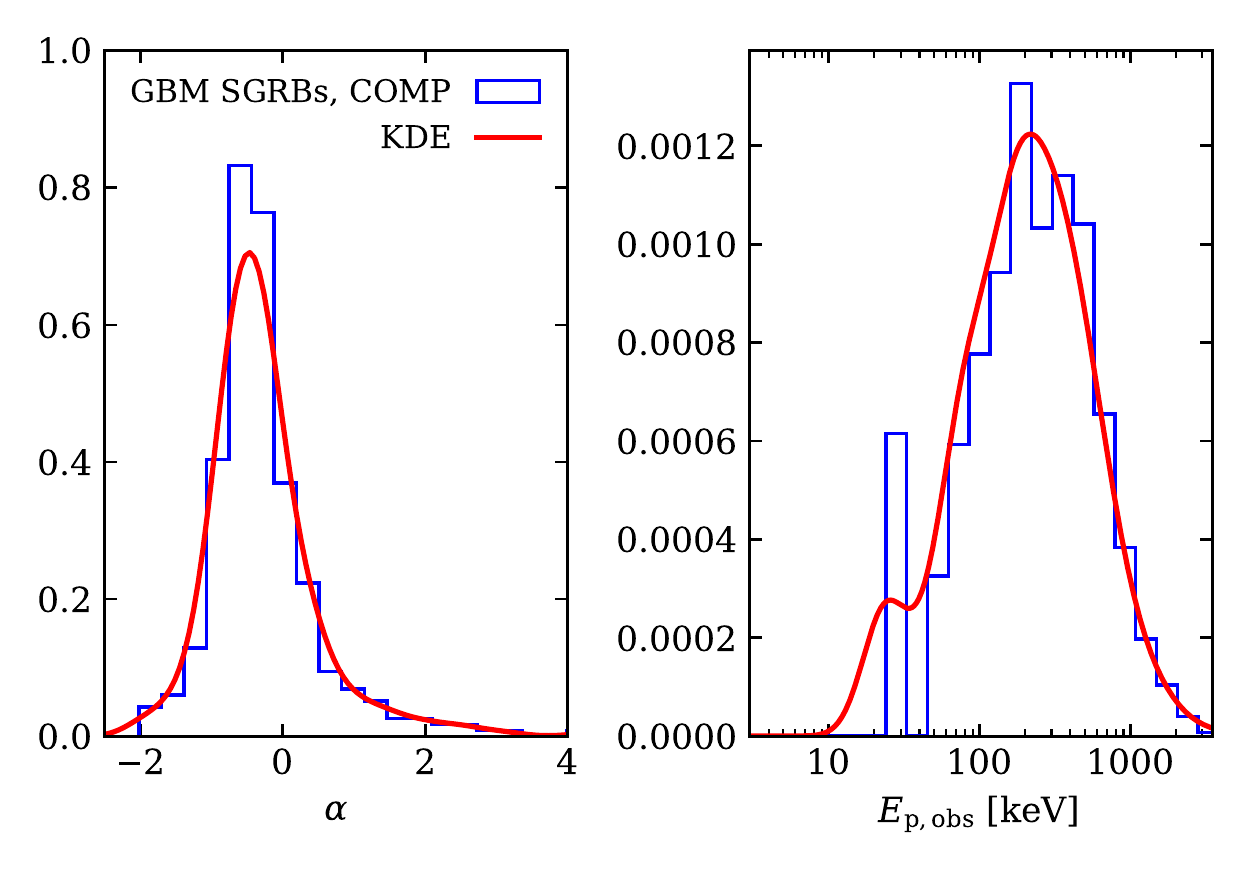}
\caption{Distributions of the low-energy photon index $\alpha$ and peak SED photon energy $E_\mathrm{p,obs}$ within the observed \textit{Fermi}/GBM sample of SGRBs successfully modelled with the \textsc{comp} model at peak flux. Blue histograms show density estimates constructed by binning the best-fit values reported in the GBM catalog. Red lines show the corresponding kernel density estimates. }
\label{fig:alpha_ep_obs_distribs}
\end{figure}

\begin{figure*}
 \includegraphics[width=\textwidth]{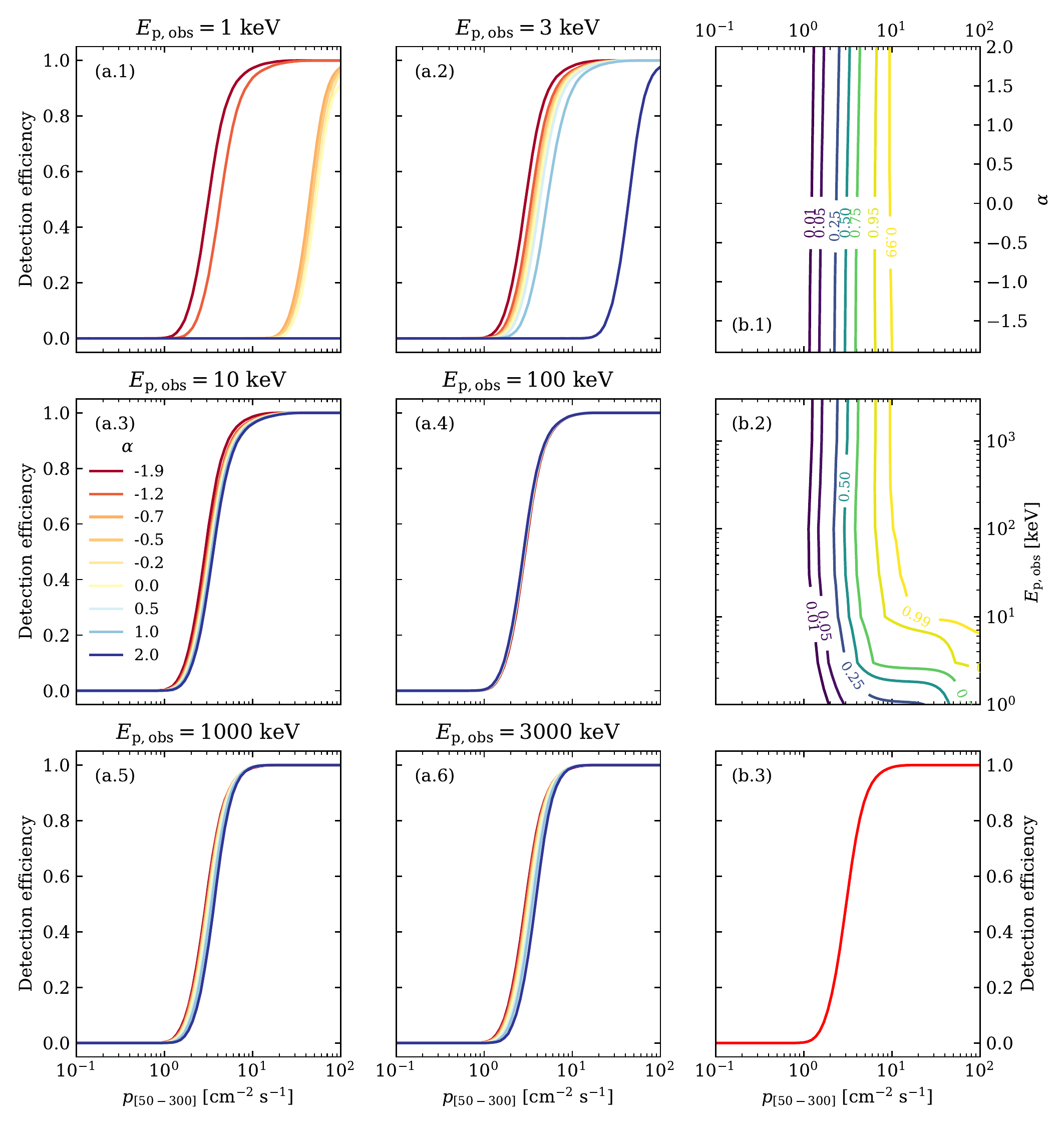}
\caption{\textit{Fermi}/GBM detection efficiency for SGRBs. Panels (a) show the detection efficiency $\eta_\mathrm{det,3D}$ for a fixed value of $E_\mathrm{p,obs}$ (reported on top of each panel) and different photon indices $\alpha$ (values shown in the legend), as a function of the 64-ms peak photon flux $p_{[50-300]}$. Panels (b) are obtained by averaging $\eta_\mathrm{det,3D}$ over the observed distribution of $E_\mathrm{p,obs}$ (panel b.1), that of $\alpha$ (panel b.2), or both (panel b.3). In panels b.1 and b.2, the solid lines are contours of constant averaged $\eta_\mathrm{det,2D}$, with the values reported along each line.}
\label{fig:pdet_gbm}
\end{figure*}

Panels (a) in Figure~\ref{fig:pdet_gbm} show the resulting detection efficiency as a function of $p_{[50-300]}$ for a number of values of $E_\mathrm{p,obs}$ and $\alpha$. Panels (b) additionally show the detection efficiency averaged over the observed distribution of $E_\mathrm{p,obs}$ (panel b.1), $\alpha$ (panel b.2) or both (panel b.3). The averaging is done by constructing kernel density estimates $w(E_\mathrm{p,obs})$ and $w(\alpha)$ of the distributions of these quantities from the best-fit values reported in the GBM spectral catalog, limiting to the events for which a valid value for the corresponding \textsc{comp} model parameter was available. The distributions are shown in Figure~\ref{fig:alpha_ep_obs_distribs}. The averaged $\eta_\mathrm{det,3D}$ is then obtained as
\begin{equation}
 \begin{array}{l}
\eta_\mathrm{det,2D,\alpha}(\alpha,p_{[50-300]})=\int_0^\infty \eta_\mathrm{det,3D}(\alpha,E_\mathrm{p,obs},p_{[50-300]})w(E_\mathrm{p,obs})\,\mathrm{d}E_\mathrm{p,obs},\\
\eta_{\mathrm{det,2D},E_\mathrm{p,obs}}(E_\mathrm{p,obs},p_{[50-300]})=\int_0^\infty \eta_\mathrm{det,3D}(\alpha,E_\mathrm{p,obs},p_{[50-300]})w(\alpha)\,\mathrm{d}\alpha,\\
\eta_\mathrm{det,1D}(p_{[50-300]})=\int_0^\infty \eta_{\mathrm{det,2D},E_\mathrm{p,obs}}(E_\mathrm{p,obs},p_{[50-300]})w(E_\mathrm{p,obs})\,\mathrm{d}E_\mathrm{p,obs}.\\
 \end{array}
\end{equation}

\begin{figure*}
    \centering
    \includegraphics[width=\textwidth]{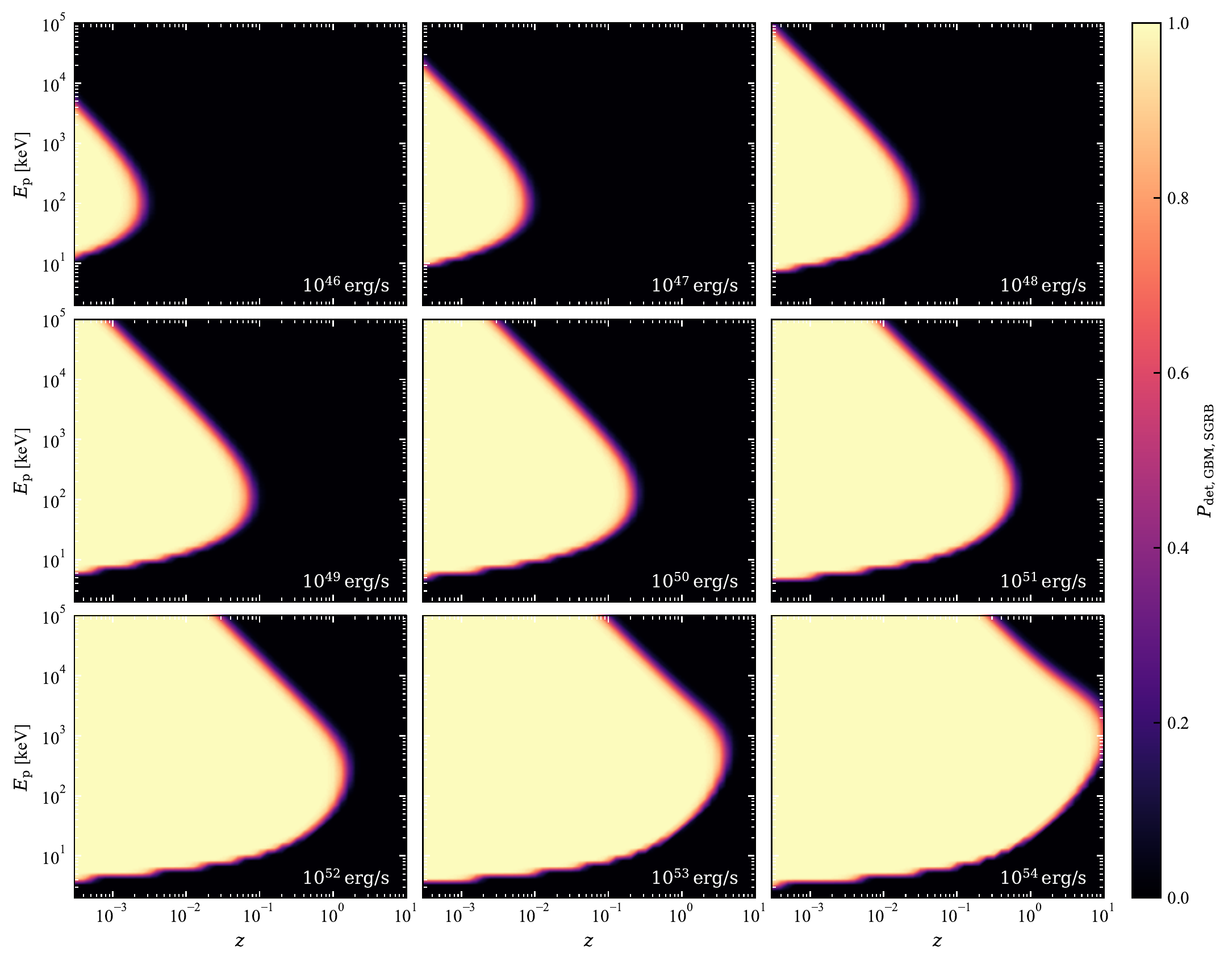}
    \caption{Fermi/GBM triggering efficiency for SGRBs as a function of their isotropic-equivalent peak luminosity $L$ (annotated within each panel), SED peak photon energy $E_\mathrm{p}$ and redshift $z$, assuming a low-energy photon index $\alpha=-0.4$. }
    \label{fig:Pdet_GBM_SGRB}
\end{figure*}

The figure demonstrates that the dependence of the detection efficiency on the low-energy photon index is essentially negligible, except for extreme cases where $E_\mathrm{p,obs}$ lies well below the 50-300 keV band. For these cases, on the other hand, the values $p_{[50-300]}$ reported on the x-axis would correspond to unrealistically large \textit{bolometric} fluxes. The dependence on $E_\mathrm{p,obs}$ is relevant only for very low values of this quantity. For the purposes of our study, we define the detection probability, expressed as a function of the source intrinsic parameters $\lsource=(L,\Ep,z)$, as
\begin{equation}
 P_\mathrm{det,GBM}(L,\Ep,z)=\eta_{\mathrm{det,2D},E_\mathrm{p,obs}}\left(\Ep/(1+z),p(L,\Ep,z)\right),
 \label{eq:Pdet_GBM_SGRB}
\end{equation}
where the peak photon flux $p$ is computed as defined in Equation~\ref{eq:pflux}, setting $\alpha=-0.4$. In practice, this is obtained by linear interpolation of $\eta_{\mathrm{det,2D},E_\mathrm{p,obs}}$ over its grid. The result, for a number of fixed peak luminosities, is shown in Figure~\ref{fig:Pdet_GBM_SGRB}.

\end{appendix}

\end{document}